\documentclass[letterpaper,twocolumn,10pt]{article}
\usepackage{usenix-2020-09}

\makeatletter
\g@addto@macro\UrlBreaks{%
  \do\a\do\b\do\c\do\d\do\e\do\f\do\g\do\h\do\i\do\j\do\k\do\l\do\m%
  \do\n\do\o\do\p\do\q\do\r\do\s\do\t\do\u\do\v\do\w\do\x\do\y\do\z%
  \do\A\do\B\do\C\do\D\do\E\do\F\do\G\do\H\do\I\do\J\do\K\do\L\do\M%
  \do\N\do\O\do\P\do\Q\do\R\do\S\do\T\do\U\do\V\do\W\do\X\do\Y\do\Z%
  \do\0\do\1\do\2\do\3\do\4\do\5\do\6\do\7\do\8\do\9%
  \do\.\do\-\do\_\do\/}%
\makeatother

\usepackage{amsmath}
\usepackage{amssymb}
\usepackage[numbers,sort&compress]{natbib}
\usepackage{listings}
\usepackage{booktabs}
\usepackage[ruled,vlined,linesnumbered]{algorithm2e}
\usepackage{cleveref}
\usepackage{subcaption}
\usepackage{paralist}
\usepackage{multirow}
\usepackage{tikz}
\usetikzlibrary{arrows.meta,positioning,calc,fit,backgrounds,shapes.geometric}

\usetikzlibrary{positioning,arrows.meta}

\definecolor{pybg}{RGB}{248,248,248}
\definecolor{pykw}{RGB}{0,92,197}
\definecolor{pystr}{RGB}{163,21,21}
\definecolor{pycom}{RGB}{0,128,0}

\lstdefinestyle{mypython}{
  language=Python,
  basicstyle=\ttfamily\scriptsize,
  keywordstyle=\color{pykw}\bfseries,
  stringstyle=\color{pystr},
  commentstyle=\color{pycom}\itshape,
  showstringspaces=false,
  keepspaces=true,
  columns=fullflexible,
  breaklines=true,
  frame=none,
  aboveskip=0pt,
  belowskip=0pt,
  xleftmargin=0pt,
  xrightmargin=0pt
}

\newsavebox{\helperbox}

\definecolor{codegreen}{rgb}{0,0.5,0}
\definecolor{codegray}{rgb}{0.5,0.5,0.5}
\definecolor{codepurple}{rgb}{0.58,0,0.82}
\definecolor{backcolour}{rgb}{0.97,0.97,0.97}

\lstdefinestyle{cstyle}{
  language=C,
  backgroundcolor=\color{backcolour},
  basicstyle=\ttfamily\small,
  keywordstyle=\color{blue}\bfseries,
  commentstyle=\color{codegreen},
  stringstyle=\color{codepurple},
  numberstyle=\tiny\color{codegray},
  numbers=left,
  numbersep=5pt,
  breaklines=true,
  frame=single,
  captionpos=b,
  tabsize=4,
  showstringspaces=false,
  columns=flexible,
  xleftmargin=1.5em,
  framexleftmargin=1em,
}

\lstdefinestyle{ruststyle}{
  language=C, % no built-in Rust; C is close enough for keywords
  backgroundcolor=\color{backcolour},
  basicstyle=\ttfamily\scriptsize,
  keywordstyle=\color{blue}\bfseries,
  commentstyle=\color{codegreen},
  stringstyle=\color{codepurple},
  numberstyle=\tiny\color{codegray},
  numbers=left,
  numbersep=5pt,
  breaklines=true,
  breakatwhitespace=false,
  frame=single,
  captionpos=b,
  tabsize=4,
  showstringspaces=false,
  columns=flexible,
  xleftmargin=1.5em,
  framexleftmargin=1em,
  morekeywords={fn,let,mut,unsafe,pub,static,use,match,Ok,Err,
                Result,Option,Some,None,impl,struct,enum,mod,
                crate,self,super,where,trait,type,as,const,
                loop,return,if,else,for,in,while,break,continue,
                u8,u16,u32,u64,i8,i16,i32,i64,bool,usize,isize},
  morecomment=[l]{//},
  morecomment=[s]{/*}{*/},
}

\lstdefinestyle{pythonstyle}{
  language=Python,
  backgroundcolor=\color{backcolour},
  basicstyle=\ttfamily\small,
  keywordstyle=\color{blue}\bfseries,
  commentstyle=\color{codegreen},
  stringstyle=\color{codepurple},
  numberstyle=\tiny\color{codegray},
  numbers=left,
  numbersep=5pt,
  breaklines=true,
  frame=single,
  captionpos=b,
  tabsize=4,
  showstringspaces=false,
  columns=flexible,
  xleftmargin=1.5em,
  framexleftmargin=1em,
}

\lstdefinestyle{promptstyle}{
  basicstyle=\ttfamily\scriptsize,
  backgroundcolor=\color{backcolour},
  breaklines=true,
  frame=single,
  captionpos=b,
  tabsize=2,
  showstringspaces=false,
  columns=flexible,
  xleftmargin=1.5em,
  framexleftmargin=1em,
  numbers=none,
}

\date{}
\title{\Large \bf \textsc{Heimdall}: Formally Verified Automated Migration of Legacy eBPF Programs to Rust}

\author{
\begin{tabular}{ccc}
{\rm Vishnu Asutosh Dasu} & {\rm Monika Santra} & {\rm Md Rafi Ur Rashid}\\
Pennsylvania State University & Pennsylvania State University & Pennsylvania State University\\
vdasu@psu.edu & monikas@psu.edu & rafiurrashid@psu.edu\\[1.5ex]
{\rm Ashish Kumar} & {\rm Saeid Tizpaz-Niari} & {\rm Gang Tan}\\
Pennsylvania State University & University of Illinois Chicago & Pennsylvania State University\\
azk640@psu.edu & saeid@uic.edu & gtan@psu.edu
\end{tabular}
}

\newcommand{\toolname}{\textsc{Heimdall}}

\newtheorem{definition}{Definition}[section]
\crefname{definition}{Definition}{Definitions}
\Crefname{definition}{Definition}{Definitions}

\begin{document}

\maketitle

%% ----------------------------------------------------------------
%%  Abstract
%% ----------------------------------------------------------------
\begin{abstract}
Extended Berkeley Packet Filter (eBPF) programs are kernel
extensions used for networking, observability, and security
enforcement in the Linux kernel. The in-kernel eBPF verifier
checks low-level memory safety and termination on eBPF programs, but it does not enforce
many higher-level source-level properties, such as initialization
discipline, schema consistency, or error handling.
We document nine classes of source-level bugs that compile, pass
the kernel verifier, and can silently corrupt data,
leak kernel memory to userspace, or yield incorrect enforcement outcomes.
To harden such verifier-accepted buggy programs and support safe migration, we present \toolname, an automated pipeline that uses
large language models to translate legacy libbpf C programs
to Aya Rust. \toolname~iteratively repairs compilation and
kernel-verifier failures, rejects unsafe escape hatches in Rust-Aya with a static analysis safety engine, and proves per-program equivalence to the original via symbolic execution and Z3-based equivalence checking.
Across 115~eBPF programs,~\toolname~generates
109~formally proven-equivalent translations (94.8\%).
In the process,~\toolname~identifies nine bugs in real-world eBPF programs and fixes all of them, two of which leak randomized kernel addresses to userspace and break KASLR. Eight of the nine have also been acknowledged and fixed upstream by the developers.
\toolname~is the first system to automate memory-safe-language migration of production eBPF programs with per-program formal guarantees that the migration preserves observable behavior.

\end{abstract}

\section{Introduction}
\label{sec:intro}

The extended Berkeley Packet Filter (eBPF)~\cite{ebpfio,gregg2019bpf} lets developers extend the Linux
kernel without modifying its source. A small program is compiled, loaded through
the \texttt{bpf()} system call, vetted by an in-kernel verifier for properties
such as memory safety and bounded execution, and then runs in response to system
calls, network packets, or function entries. eBPF programs now underpin
high-performance networking~\cite{xdp}, continuous
profiling~\cite{parca-agent}, security policy
enforcement~\cite{ebpfio}, and system
observability~\cite{bcc,libbpf-tools} across virtually every modern Linux
deployment. The eBPF verifier is treated as the admission boundary for the Just-In-Time (JIT) compiled bytecode that enters the kernel. Passing it establishes that a program satisfies
the kernel's admission policy. However, it does not establish that the
program is free of defects. 
% Verifier-accepted programs can still leak
% sensitive data to userspace or corrupt the behavior they were deployed to enforce.

Across widely deployed libbpf-tools utilities~\cite{libbpf-tools}, we found verifier-accepted
programs that export uninitialized stack memory to userspace and programs
that use a kernel-read helper with the full capacity of the destination buffer instead of the length that actually needs to be filled, copying adjacent kernel memory past the data
of interest. Neither pattern is a memory-safety violation the verifier is meant to reject.
The write stays within the program's own buffer, and these defects violate obligations at the source-code level while the verifier checks at the bytecode level. The missing initialization is one the verifier deliberately waives for privileged loaders, and the oversized read cannot be inferred from the bytecode alone.
Yet in both cases the exported bytes carry kernel-text and module
pointers, and by loading each program unmodified and reading the data it
exposes, we recovered the slide of Kernel Address Space Layout Randomization
(KASLR) and the live kernel-text base. 

Migration to a memory-safe language such as Rust and enforcing type safety at the source-code level can close these gaps~\cite{sok-ebpf-oakland25}, 
but manually rewriting the existing eBPF library is impractical at
scale. Aya~\cite{aya} is a natural target; it is an eBPF
library built purely in Rust, providing safe and idiomatic wrappers to the BPF
application binary interface (ABI), supporting the BPF Type Format, and requiring no C toolchain. Its typed interface encodes obligations that compilation otherwise erases---required struct
fields, error-valued helper results, pointee types, and hook-specific action
namespaces---so the Rust compiler checks them. 
Large language model (LLM) agents make automated migration to Aya plausible, yet compilation,
kernel acceptance, and finite tests establish neither semantic preservation nor
use of Aya's safest abstractions. Automating this workflow faces three concrete challenges.
% Heimdall is a migration and review tool for
% maintainers of trusted production eBPF programs; it neither replaces the kernel
% verifier nor mediates arbitrary eBPF loading. Our central contribution is
% \toolname{}, which combines autonomous translation, Aya-specific safety
% enforcement, bytecode-level equivalence, and counterexample triage in one
% workflow, turning semantic equivalence from a pass/fail check into a mechanism
% that both repairs mistranslations and exposes verifier-accepted bugs in legacy C.

% Even with
% the kernel’s \texttt{BPF\_PROG\_TEST\_RUN} facility, comparing side effects
% across two program versions requires custom per-program har-
% nesses, and subtle semantic divergences (endianness conversions,
% sign extension, map-access patterns) escape any finite test suite by
% construction.

First, semantic equivalence is desired between the original C eBPF and Aya Rust code. The kernel’s \texttt{BPF\_PROG\_TEST\_RUN} facility requires custom per-program harnesses and is insufficient to claim equivalence since subtle semantic divergences escape any finite test suite by construction~\cite{llm-translation-formal}.
Comparing C and Rust source would introduce ownership,
integer-promotion, and panic differences erased by compilation. Since both libbpf C and Aya Rust code are compiled to eBPF bytecode, it is possible to check semantic equivalence at the bytecode level via symbolic execution. However, no existing symbolic-execution framework supports realistic eBPF programs at the bytecode level; such a framework needs to support symbolic hash-map keys, arrays with per-index states, etc.

% GT: summarize the paragraph in the last sentence
%When performing symbolic execution using opaque packet and tracing contexts, their important outputs are usually side effects: return values, map state, and events emitted to userspace.  Hash-map keys can be symbolic, so writes may alias earlier entries and last-write-wins behavior must track both value and key presence; arrays require per-index state, and emitted records must be compared in sequence. 

Second, the output must use safe and idiomatic Aya. 
Code in Rust can compile and
pass the eBPF verifier while bypassing Aya's idiomatic typed APIs through raw fixed-size reads,
discarded helper \texttt{Result}s, direct generated-helper calls, writable
\texttt{static mut} configuration, raw-pointer mutation, wrong-hook action
constants, or unnecessarily broad \texttt{unsafe} blocks. An explicit
Aya-specific policy must reject these patterns while permitting the narrowly
scoped operations required by approved Aya interfaces.

Third, a literal translation can faithfully preserve a C defect, while the canonical Aya Rust API can repair it and thereby produce a C--Rust behavioral difference. A counterexample to equivalence therefore says only that the programs differ, but cannot tell whether the Rust translation might be wrong, or the Rust translation may have fixed a bug in the C source. Forcing Rust code to fix the difference can reintroduce the C bug, while accepting every divergence can hide a mistranslation. The pipeline must distinguish translation regressions from C bugs without requiring safe Rust to reproduce bugs.

\toolname{} addresses these challenges with a six-stage pipeline. Given a libbpf C source,
entry symbol, and map schema, it (1) generates an Aya Rust translation,
(2) repairs compiler and kernel-verifier failures, (3) applies the safe-Aya
policy, (4) symbolically executes the C and Rust bytecode, (5) asks Z3 whether
their return values, map state, and output events can differ, and (6) triages
any counterexample. An \texttt{UNSAT} result yields \emph{Verified} under the
conditional-equivalence model. A \texttt{SAT} result enters a bug-triage stage, where a
translation-mismatch verdict returns automatically to counterexample-guided
repair, whereas an LLM \emph{Bug Found} claim halts the translation and is reported. 

Across 115~real-world eBPF programs, \toolname{} produces
\textbf{109~formally verified translations (94.8\%)}. Three are partially verified, with a
subset of entry points proven equivalent, and three exceed symbolic-execution
scalability before reaching a verdict. Its bug triage stage surfaced nine verifier-accepted bugs across nine programs. Two of the nine defeat KASLR, and eight have been acknowledged and fixed upstream.

% Compilation and kernel verification are
% poor proxies for trustworthy migration: compile-only LLM baselines pass all
% programs yet clear kernel verification, safety analysis, and equivalence together for only a fraction, whereas \toolname{} verifies nearly all of them.

\paragraph{Contributions.}
This paper makes the following contributions:
\begin{itemize}

  \item \textbf{An eBPF symbolic-execution backend for angr.}
  We extend angr to support eBPF ISA, including
  helpers, maps, relocations, atomic operations, BPF-to-BPF calls, and symbolic
  map keys with an if-then-else-chain encoding that captures last-write-wins
  value and key-presence semantics.

\item \textbf{Conditional equivalence with counterexample triage.}
  \toolname{} turns semantic equivalence from a pass/fail translation check into
  a dual-purpose mechanism: a counterexample either guides Rust repair or identifies a verifier-accepted bug in the C reference,
  proving equivalence on executions that do not trigger the declared safety
  failures.

  \item \textbf{A safe-Aya static analyzer.}
  The analyzer rejects verifier-accepted Rust translations that bypass Aya's
  typed surface, including raw fixed-size reads, discarded helper results, and gratuitous \texttt{unsafe}, while permitting
  the narrowly scoped operations that Aya requires.
  
  \item \textbf{An automated six-stage C-to-Rust pipeline for eBPF.}
  \toolname{} integrates LLM translation, compiler and verifier repair,
  Aya-specific safety analysis, symbolic execution, equivalence checking,
  counterexample-guided repair, and bug discovery.

  \item \textbf{Production security and correctness findings.}
  We identify, repair, and re-verify nine verifier-accepted bugs across nine
  production programs, including two disclosures of kernel-text addresses that
  defeat KASLR. Eight bugs have been acknowledged and fixed upstream.

  \item \textbf{A full-scale evaluation and public benchmark.}
  We evaluate \toolname{} on 115~real-world eBPF programs, producing
  109~formally verified translations, substantially outperforming compile-only
  LLM baselines. Additionally, we release all the artifacts.
\end{itemize}

%%% Local Variables:
%%% mode: LaTeX
%%% TeX-master: "main"
%%% End:

\section{Threat Model}
\label{sec:threat_model}

% \toolname~helps eBPF developers
% harden their programs against errors that the kernel verifier, by design or by defect, does not catch. As a result, these errors can inadvertently compromise the system security (e.g., breaking KASLR) if not fixed. Our goal is to improve the security of the systems on which eBPF programs are deployed (e.g., preventing accidental stack leaks) and the reliability of eBPF programs themselves, which are often deployed in security and network monitoring applications. 
\toolname~helps developers harden their eBPF programs against errors that the kernel verifier, by design or by defect, does not catch. Since the verifier is the kernel's only mechanism to enforce safety, these errors can compromise the system's security (e.g., breaking KASLR) if not fixed. Our goal is to fix these errors to improve the security of the systems on which eBPF programs are deployed and the reliability of eBPF programs themselves, which are often deployed in security and network monitoring applications. 
We consider programs that execute in the kernel on untrusted runtime
inputs, including network packets, system call arguments, filenames, and
potentially attacker-influenced map state. 
The developer seeks to write a correct
program rather than subvert our tooling, and we do not consider programs crafted to deliberately break our tooling (e.g., prompt
injections placed as source-code comments). The language model that performs
the translation is untrusted. Every translation undergoes symbolic equivalence
checking against the original program, so a mistranslation is rejected rather
than certified. The soundness of a \emph{Verified} verdict, therefore, relies on our
symbolic-execution engine and equivalence checker, which together form the
trusted computing base (TCB) of \toolname. We validate the TCB in \Cref{sec:differential-testing}. 
% Safety
% enforcement additionally trusts \texttt{rustc} and Aya's typed APIs, and the
% unmodified kernel verifier remains the final admission check on every
% translated program at load time. 
\section{Motivation}
\label{sec:motivation}

Passing the kernel verifier establishes that an eBPF program satisfies the
kernel's admission policy, not that the program is free of defects.  
% Verified
% programs can still publish uninitialized memory, misuse helpers, and confuse
% hook types.  
\Cref{sec:layered-verification} explains why closing these gaps requires
source-level checks working with, not replacing, the kernel's bytecode
verifier.  \Cref{sec:verifier-gaps} grounds the argument in bugs that
\toolname{} exposed in real-world programs.

\subsection{Layered Verification for Safer eBPF Programs}
\label{sec:layered-verification}

\begin{figure}[t]
\centering
\resizebox{\columnwidth}{!}{%
\begin{tikzpicture}[
    >=Stealth,
    node distance=1.6cm,
    procbox/.style={
      draw, rounded corners=3pt, fill=blue!8,
      font=\normalsize, inner sep=6pt, minimum width=2.6cm,
      minimum height=1.0cm, align=center, line width=0.7pt,
    },
    toolbox/.style={
      draw, rounded corners=3pt, fill=green!12,
      font=\normalsize, inner sep=6pt, minimum width=2.6cm,
      minimum height=1.0cm, align=center, line width=1.0pt,
    },
    libbox/.style={
      draw, rounded corners=3pt, fill=orange!12,
      font=\normalsize, inner sep=6pt, minimum width=2.8cm,
      minimum height=1.0cm, align=center, line width=0.7pt,
    },
    arr/.style={->, thick, color=black!60},
    edgelab/.style={font=\small, align=center},
    layerlab/.style={font=\small\itshape, color=black!60, align=center},
  ]

  %% --- main horizontal flow ---
  \node[toolbox] (tool) {\toolname};
  \node[procbox, right=2.8cm of tool] (compiler) {Rust\\Compiler};
  \node[procbox, right=2.8cm of compiler] (verifier) {Linux eBPF\\Verifier};

  %% --- what each layer guarantees ---
  \node[layerlab, above=3pt of tool] {Semantic Equivalence\\+ Source Safety};
  \node[layerlab, above=3pt of compiler] {Memory \&\\Type Safety};
  \node[layerlab, above=3pt of verifier] {Kernel\\Safety};

  %% --- trusted library feeding the compiler from below ---
  \node[libbox, below=1.6cm of compiler] (lib)
    {Trusted eBPF Rust\\Library (unsafe)};

  %% --- legacy C source entering \toolname ---
  \coordinate (srcstart) at ([xshift=-2.6cm]tool.west);
  \draw[arr] (srcstart) -- node[edgelab, above] {C eBPF\\Source} (tool.west);

  %% --- \toolname -> compiler (verified-equivalent safe Rust) ---
  \draw[arr] (tool.east) -- node[edgelab, above] {Equivalent \emph{Safe}\\Rust Source} (compiler.west);

  %% --- library -> compiler (Safe Interface) ---
  \draw[arr] (lib.north) -- node[edgelab, right] {Safe\\Interface} (compiler.south);

  %% --- compiler -> verifier (Safe eBPF Bytecode) ---
  \draw[arr] (compiler.east) -- node[edgelab, above] {\emph{Safe} eBPF\\Bytecode} (verifier.west);

  %% --- verifier -> output (Safer eBPF Bytecode) ---
  \coordinate (outend) at ([xshift=2.6cm]verifier.east);
  \draw[arr] (verifier.east) -- node[edgelab, above] {\emph{Safer} eBPF\\Bytecode} (outend);

\end{tikzpicture}%
}%
\caption{Layered verification with \toolname{} for safer eBPF programs (figure adapted from
\citet{gerhorst2025mitigating}).}
\label{fig:safety-layers}
\end{figure}

Safety decisions for eBPF must be layered since source and bytecode representations expose different obligations and neither subsumes the other. \toolname{} therefore adds a source-level layer in front of the kernel verifier (\Cref{fig:safety-layers}).  It translates legacy C into
safe Aya Rust, formally verifies the translation to be semantically
equivalent to the compiled original, and enforces source-level safety
obligations on the result.  The Rust compiler then enforces memory and type
safety, with residual \texttt{unsafe} code confined to the trusted Aya
library behind a safe interface, and the unmodified kernel verifier admits
the compiled bytecode at load time exactly as it does today.

Under our threat model, a trusted developer migrates
an eBPF program that processes untrusted inputs.  Since \toolname{} receives no
functional specification, it cannot verify that the program matches its
author's intent.  It instead enforces a non-exhaustive set of \emph{obligations}, conditions
that every correct program of a given hook type must satisfy no matter what
it computes.  A program must initialize every byte it publishes to a map or
event buffer, call each helper with the correct size of the buffer it operates on, check
error-valued helper returns before using them as data, read each map at its declared value type, and use only the context fields and action constants of
its own hook. 
Since these conditions hold under any intent, a violation is
a defect, and \toolname{} detects it without a specification. Correctness
beyond these obligations remains out of scope. Our goal is not completeness but a strict improvement over the status quo of loading a C program with the verifier alone.

The kernel's eBPF verifier does not enforce these obligations.  It checks memory
bounds, pointer validity, bounded execution, and permitted helper use on the
exact bytecode under the running kernel's
contracts~\cite{rex-atc25,prevail}, and some of its gaps are deliberate.  For
loaders with \texttt{CAP\_PERFMON} permissions, Linux enables the
\texttt{allow\_uninit\_stack} path even though a later read may expose
uninitialized data.\footnote{The
\texttt{allow\_uninit\_stack} path in \url{https://github.com/torvalds/linux/blob/master/kernel/bpf/verifier.c}.} The onus is on the developer to correctly initialize data.
% The tracing programs we study require this capability, so revoking it only
% prevents them from loading.  
The remaining obligations are erased by
compilation.  A struct field becomes a byte range, so an access can stay
inside the stack frame yet overrun the field it names.  An error-valued
\texttt{long} becomes an ordinary scalar, so a negative errno can flow into a
key or an index unchecked.  A map's value type and a hook's context and
action types become plain integers, so an offset or constant can be
numerically valid yet name a different field or the opposite action.  Only a
check that runs on source code, before this information is lost, can enforce the
obligations. A missing check of this kind does not mean the verifier is broken. Source-level checks complement the verifier rather than replace it, and the resulting bytecode needs no new runtime and no new admission path.
% Checks at the source-level complement the verifier rather than replacing it and the resulting bytecode needs no new runtime admission path. 

A second and independent reason to layer checks is that the verifier is
itself a large and complex trusted component with the most reported CVEs of
any component in the eBPF subsystem~\cite{sok-ebpf-oakland25,verifier-untenable-hotos23,verifier-state-embedding-osdi24,agni-verifier-bugs-cav23,10.1145/3731569.3764797}, and once it admits a program no further
memory-safety checks occur, making it a single point of failure. Furthermore, \citet{10.1145/3731569.3764797} argue that the safety enforcement of the verifier itself is inconsistent since its safety properties are often informally described in natural language on mailing lists and code comments. \toolname{} removes this single point
of failure without altering the deployed pipeline.  Every migrated program is
checked for memory safety at the source and must still pass the verifier at load time, so an unsafe program reaches the kernel only if two
independently constructed checks, working on different representations of the
same program, both miss it.  
% The source layer bounds the impact of a verifier
% defect, and the verifier bounds the impact of a translation error.  
This follows
the direction of \citet{sok-ebpf-oakland25}, which recommends enforcing memory
safety of eBPF at the source in a language such as Rust rather than relying solely on
the verifier. 

% We could check the obligations on the C source directly, but migrating to
% Rust and Aya makes them permanent.  Aya encodes the obligations in
% the program's types.  Contexts, map values, and helper results are typed,
% error returns must be inspected before use, and ownership enforces paired protocols such as ring-buffer reserve and submit.  
% \toolname{}'s safety
% policy additionally restricts the unsafe surface by rejecting raw pointers, \texttt{MaybeUninit}, and
% unnecessary \texttt{unsafe}.  
% A conforming program therefore cannot express
% an obligation violation, and \texttt{rustc} re-checks this on every later
% edit, whereas a C analyzer must be re-run and maintained outside the
% toolchain forever. Because migration can introduce errors of its own, \toolname{}
% formally verifies semantic equivalence between the compiled C and Rust
% programs.  This check also surfaces latent defects
% in the original C as concrete counterexamples. \Cref{sec:verifier-gaps}
% presents the bugs \toolname{} found this way in real-world eBPF tools.

\subsection{Verifier-Accepted Bugs in eBPF Programs}
\label{sec:verifier-gaps}

We present a non-exhaustive set of nine bug classes that evade the eBPF verifier. We validate all bugs on kernel 6.8.0-106-generic (Ubuntu) with clang 21.1.8 and bpftool v7.4.0 and confirm that the verifier accepts every program.  This subsection
discusses the five classes that cover the nine real-world bugs that \toolname{} with
Opus 4.6 identified and fixed. Eight of these nine bugs have been acknowledged and fixed by the developers.\footnote{The only bug that is not yet fixed is the one in BMC. Our PR is under review, but the project has not been active since 2021.} \Cref{sec:appendix-additional-patterns}
presents four additional classes, including verifier CVE patterns that violate memory safety. \Cref{sec:full_code_listings} presents the full code listings of the bugs.
% , and two further variants of a class introduced
% here.

\paragraph{Uninitialized stack state.}
\Cref{lst:bashreadline} shows \texttt{bashreadline} (full source in
\Cref{lst:bashreadline-full}), which declares an 84-byte
event struct without initializing it.  The handler assigns the process identifier and
uses \texttt{bpf\_probe\_read\_user\_str} to write a NUL-terminated prefix of the
string field, but it emits the complete event struct (line 10).  On short bash commands (e.g., \texttt{ls}), every event
we observed contained kernel-pointer-shaped values in the unwritten suffix. An unprivileged user can control how much data is leaked by varying the length of the command they run.
Across two independent runs, recurring kernel-text addresses produced the same
KASLR slide of \texttt{0x0e600000} and disclosed the live kernel-text base.  The verifier admits this flow through the
\texttt{allow\_uninit\_stack} policy used for tracing programs with
\texttt{CAP\_PERFMON}. 
The resulting leak is outside its kernel-safety contract and belongs to the broader class studied by
BPFflow~\cite{10.1145/3748355.3748374}.  
% and the verifier soundness bug identified by SpecCheck~\cite{10.1145/3731569.3764797}
\toolname{} prevents it because
\texttt{rustc} rejects missing or possibly uninitialized fields in the translated Aya Rust code, and \toolname{}'s safety policy rejects
\texttt{MaybeUninit}, raw-pointer, and \texttt{unsafe} escape paths.  
%\gtan{In "the safety policy rejects ...": have we explained what the safety policy is somewhere? I haven't read the intro since it's not ready.} \vishnu{yes, this will be mentioned in the intro}
% Accepted repairs, on the other hand, zero-initialize the complete event.

\begin{lstlisting}[style=cstyle,
caption={\texttt{bashreadline} contains an uninitialized event (BUG~1) and an
unchecked helper return (BUG~2).},
label=lst:bashreadline]
SEC("uretprobe/readline")
int BPF_URETPROBE(printret, const void *ret)
{
    struct str_t data;  /* BUG 1: str_t is never zero-initialized */
    /* ... */
    data.pid = bpf_get_current_pid_tgid() >> 32;
    bpf_probe_read_user_str(&data.str,
        sizeof(data.str), ret); /* BUG 2: return value ignored */
    /* emits all 84 bytes, incl. the uninitialized suffix */
    bpf_perf_event_output(ctx, &events,
        BPF_F_CURRENT_CPU, &data, sizeof(data));
    return 0;
}
\end{lstlisting}

The same class appears in \texttt{offcputime}
(\Cref{lst:offcputime-full}).  It writes only a prefix of
\texttt{val.comm}, assigns \texttt{delta}, and inserts the entire
\texttt{val} into a map.  We observed residual stack bytes after the NUL in the
stored value.  The Rust translation must initialize the complete value before
the map update, matching the accepted zero-initialization patch, and the
equivalence checker verifies that all other behavior is unchanged.

\paragraph{Unchecked helper return.}
The code in \Cref{lst:bashreadline} also discards the result of
\texttt{bpf\_probe\_read\_user\_str}.  A read failure may result in an invalid or unreadable user pointer (\texttt{\&data.str}),
leaving some or all of the 80-byte string unwritten; however the event is still
emitted (line 10) when it should have early returned on failure. 
The verifier checks whether the helper may access the buffer, but it
does not require the program to act on the helper result before using that
buffer.  In contrast, in Aya the result is of an option type \texttt{Result<T,E>}, and \toolname{} only
accepts translations that propagate the result or handle both success and failure.
% Rust's initialization rules independently ensure that no
% undefined bytes can be propagated. 
Public issue reports show that these helper failures do occur in production~\cite{tetragon-issue-3728,bcc-issue-3175,bcc-issue-622,bcc-issue-2245}. Across the 115 programs in our dataset, \toolname{} identified and fixed 137 unchecked helper returns.
% Tetragon issue~\#3728
% reports inaccurate arguments after unchecked \texttt{probe\_read} failures, and
% BCC issues~\#3175, \#622, and \#2245 report \texttt{-EFAULT} from
% \texttt{bpf\_probe\_read} at runtime
% ~\cite{tetragon-issue-3728,bcc-issue-3175,bcc-issue-622,bcc-issue-2245}.

\paragraph{Pointer-size mismatch.}
Many BPF helpers accept a raw pointer and a size for the pointer's underlying object.  The
compiler and verifier may establish that the selected byte interval is
accessible without establishing that the size matches the object named
by the pointer.  In each case below, the provided size does not match the object's true size: a larger provided size allows reads or writes into adjacent memory, and a smaller size may leave the object partially transferred.

\texttt{filetop} (\Cref{lst:filetop-overread}) passes the 4096-byte (\texttt{size}) capacity of its
destination as the size of a raw kernel read.  The source is a
\texttt{qstr} name (\texttt{dname.name}) that is usually much shorter, but the
program ignores \texttt{dname.len}, and the raw helper does not stop at a NUL byte.
It therefore copies memory following the filename into the map value.  We
reproduced the over-read and found kernel-text pointers that disclosed the KASLR
slide. This was the most egregious bug \toolname~found since it leaked
$\approx 4$\,KB of adjacent kernel memory---including kernel and module
pointers---into userspace for each filename.  The verifier can validate the
destination byte range but cannot recover the source string's logical extent.

\begin{lstlisting}[style=cstyle,
caption={\texttt{filetop} uses the destination capacity as the readable source
extent.},
label=lst:filetop-overread]
static void get_file_path(struct file *file,
                          char *buf, size_t size)
{
    struct qstr dname = BPF_CORE_READ(file, f_path.dentry, d_name);
    bpf_probe_read_kernel(buf, size, dname.name);
    /* BUG: raw read copies size bytes instead of dname.len bytes, ignoring the NUL in a short dname.name */
}

/* caller, in the tracepoint handler: */
get_file_path(file, valuep->filename,
              sizeof(valuep->filename)); /* size = 4096 capacity */
\end{lstlisting}

% The accepted patch uses \texttt{bpf\_probe\_read\_kernel\_str}.
\toolname{} prevents the leak through Aya's NUL-stopping
\texttt{bpf\_probe\_read\_kernel\_str\_bytes} wrapper and rejects a raw string
read with an independent byte count.  Its equivalence check exposes the
difference between the raw and string-aware operations.  We found the same
fixed-size read in \texttt{gethostlatency} (\Cref{lst:gethostlatency-full}),
which copies 80~bytes from a user string, and \texttt{slabratetop}
(\Cref{lst:slabratetop-full}), which copies 32~bytes from a kernel string.
Both read past the NUL and export the trailing bytes through a map value.

\texttt{biostacks} and
\texttt{biosnoop}\footnote{The bug in \texttt{biosnoop} was independently discovered by a third party. \toolname{} caught and fixed the same bug in the version of the dataset we collected.} instead pass a size that is too small: they apply \texttt{sizeof} to the address (e.g., \texttt{sizeof(\&data)} instead of \texttt{sizeof(data)})
of a 16-byte command array when calling \texttt{bpf\_get\_current\_comm}
(\Cref{lst:biostacks-full,lst:biosnoop-full}).  On a 64-bit system, the expression
evaluates to the eight-byte pointer size, so both programs silently truncate
command names.  The write remains within bounds and therefore satisfies the
verifier.

% \begin{lstlisting}[style=cstyle,
% caption={Two \texttt{sizeof(pointer)} errors truncate command names.},
% label=lst:comm-size-bugs]
% /* biostacks */
% bpf_get_current_comm(
%     &i_rqinfop->rqinfo.comm,
%     /* BUG: sizeof(&comm) is the 8-byte pointer size, not 16 */
%     sizeof(&i_rqinfop->rqinfo.comm));

% /* biosnoop */
% bpf_get_current_comm(
%     /* BUG: sizeof(&comm) is 8, so only half of comm is written */
%     &piddata.comm, sizeof(&piddata.comm));
% \end{lstlisting}

Aya's \texttt{bpf\_get\_current\_comm} returns a complete fixed-size array and
takes no caller-supplied length, so the faulty expression is impossible to
replicate.  \toolname{}'s safety policy rejects raw \texttt{unsafe} calls that
recreate the C pointer-and-length interface.
\Cref{sec:appendix-additional-patterns} presents two further instances, in which
the size spans an enclosing object rather than the field or prefix the pointer
names.

\paragraph{Action-code namespace confusion.}
BMC~\cite{bmc-cache} (\texttt{bmc\_kern.c}) attaches programs to Express Data Path (XDP) on ingress and traffic control
(TC) on egress.  The two kinds of hooks use different action namespaces with overlapping
integers.  Action \texttt{XDP\_PASS} is represented as 2 under XDP, while the same value means
\texttt{TC\_ACT\_SHOT} under TC and drops the frame.  The TC pass action,
\texttt{TC\_ACT\_OK}, is represented as 0.
\texttt{bmc\_tx\_filter\_main} and \texttt{bmc\_update\_cache\_main} (\Cref{lst:bmc-full}) both return
\texttt{XDP\_PASS} on early exits even though they are TC programs.
% \Cref{lst:bmc-action} shows the short-frame path; the two relevant TC entry
% functions appear in full in \Cref{lst:bmc-full}.

% \begin{lstlisting}[style=cstyle,
% caption={BMC returns an XDP action from a TC program, so the frame is dropped
% instead of passed.},
% label=lst:bmc-action]
% SEC("bmc_tx_filter") /* attached to the TC (sched_cls) hook */
% int bmc_tx_filter_main(struct __sk_buff *skb)
% {
%     /* ... */
%     if (ip + 1 > data_end)
%         return XDP_PASS;  /* BUG: 2 == TC_ACT_SHOT drops the frame */
%     /* ... */
%     return TC_ACT_OK;
% }
% \end{lstlisting}

The verifier accepts any integer return from a TC program and therefore reports
no error.  The kernel's test-run facility confirmed that the short-frame path
produces \texttt{TC\_ACT\_SHOT}.  The corrected programs return
\texttt{TC\_ACT\_OK}.  \toolname{} binds action constants to appropriate kinds of hooks (XDP or TC),
so a TC classifier may return only \texttt{TC\_ACT\_*} values and an XDP
program only \texttt{XDP\_*} values.  
% This differs from a context-type mismatch,
% where the hook interprets its input structure under the wrong type.

\paragraph{Generic C logic bugs.}
The verifier also accepts ordinary C defects that do not violate its
kernel-safety contract.  Katran~\cite{katran}, Meta's production Layer-4 load
balancer, contains a representative case on its LRU-miss statistics path
(\Cref{lst:katran-full}).  The code intends to compare two protocol fields but
uses \texttt{=} instead of \texttt{==}.  The expression is true whenever the
assigned protocol is nonzero and also overwrites the protocol in the map value (\texttt{lru\_miss\_stat\_vip->proto}).
A C compiler emits at most a \texttt{-Wparentheses} warning, and the verifier
accepts the map update.
% \begin{lstlisting}[style=cstyle,
% caption={An assignment in Katran mutates a map value where the code intends to
% compare it.},
% label=lst:katran-assign]
% /* balancer.bpf.c: lru_miss_stats path */
% bool proto_match =
%     lru_miss_stat_vip->proto = vip->proto; /* BUG: '=' instead of '==' */
% \end{lstlisting}

Rust assignments have unit type, so a faithful translation into a Boolean
binding fails to type check: the assignment yields the unit value \texttt{()}
where a \texttt{bool} is expected.  Mutating a value reached through a shared,
immutable map reference is likewise rejected as a write through a
non-mutable borrow.
More broadly,
\texttt{rustc}'s stronger typing, definite initialization, and explicit
conversions reject general C mistakes such as uninitialized reads and implicit
signed or unsigned conversions before the verifier sees any bytecode. 
% Such logic bugs can have unintended security consequences (e.g., unintended map updates) since eBPF programs are often deployed in security and networking monitoring applications.

\section{Methodology}
\label{sec:methodology}
We present \toolname, an automated pipeline that translates C eBPF programs to Rust and
formally verifies semantic equivalence.

%\gtan{Writing wise, I prefer a top-down approach, which means we should put sec 3.4 (translation pipeline) earlier. Then we can say the most challenging part is equivalence checking, which we elaborate next. Also, are there not other issues in the methodology we need to discuss? I thought the bug finding part might be worth discussion too.} \vishnu{1. reordering done 2. Do you mean a phase in the framework that automatically identifies bugs? We do not have that but our safety check static analysis engine + general translation to Rust/aya implicitly removes bugs}

%% ================================================================
\subsection{\toolname~Translation Pipeline}
\label{sec:pipeline-overview}

%\gtan{In this discussion, we should link back to the challenges, mentioning which challenge(s) a component aims to solve and discussing briefly how they are solved.} \vishnu{done}

\begin{figure}[t]
\centering
\resizebox{\columnwidth}{!}{%
\begin{tikzpicture}[
    >=Stealth,
    node distance=0.6cm,
    % styles
    inputbox/.style={
      draw, rounded corners=3pt, fill=black!6,
      font=\small, inner sep=5pt, minimum width=3.8cm,
      minimum height=0.7cm, align=center, line width=0.6pt,
    },
    llmbox/.style={
      draw, rounded corners=3pt, fill=orange!10,
      font=\small, inner sep=5pt, minimum width=3.8cm,
      minimum height=0.7cm, align=center, line width=0.6pt,
    },
    toolbox/.style={
      draw, rounded corners=3pt, fill=blue!8,
      font=\small, inner sep=5pt, minimum width=3.8cm,
      minimum height=0.7cm, align=center, line width=0.6pt,
    },
    decidebox/.style={
      draw, diamond, aspect=2.2, fill=yellow!12,
      font=\small, inner sep=2pt, align=center, line width=0.6pt,
    },
    resultbox/.style={
      draw, thick, rounded corners=3pt, fill=green!12,
      font=\small\bfseries, inner sep=5pt,
      minimum width=2.4cm, minimum height=0.7cm, align=center,
      line width=1.2pt,
    },
    stagelab/.style={
      font=\sffamily\scriptsize\bfseries, text=blue!70!black,
    },
    arr/.style={->, thick, color=black!50},
    looparr/.style={->, thick, color=red!60!black, densely dashed},
  ]

  %% --- Stage 1 ---
  \node[inputbox] (csrc) {C libbpf (.c)};
  \node[llmbox, below=0.5cm of csrc] (llm) {LLM Translation};
  \node[stagelab, left=0.2cm of llm] {Stage 1};
  \draw[arr] (csrc) -- (llm);

  %% --- Rust source (central node for repair loops) ---
  \node[inputbox, below=0.5cm of llm] (rsrc) {Rust Aya (.rs)};
  \draw[arr] (llm) -- (rsrc);

  %% --- Stage 2 ---
  \node[toolbox, below=0.5cm of rsrc] (compverify)
    {Compile $\rightarrow$ Kernel Verify};
  \node[stagelab, left=0.2cm of compverify] {Stage 2};
  \draw[arr] (rsrc) -- (compverify);

  %% --- Stage 3 ---
  \node[toolbox, below=0.5cm of compverify] (safety)
    {Safety Check};
  \node[stagelab, left=0.2cm of safety] {Stage 3};
  \draw[arr] (compverify) -- (safety);

  %% --- Compile decision ---
  \node[decidebox, below=0.6cm of safety] (compok) {pass?};
  \draw[arr] (safety) -- (compok);

  %% --- Compile fail loop ---
  \node[llmbox, right=1.6cm of safety, minimum width=2.9cm]
    (compfix) {\small LLM + error/safety feedback};
  \draw[looparr] (compok) -| node[below, pos=0.25, font=\scriptsize] {fail}
    (compfix);
  \coordinate (compfixturn) at ([xshift=0.45cm]rsrc.east);
  \draw[looparr, rounded corners=4pt]
    (compfix.north) |- (compfixturn) -- (rsrc.east);

  %% --- Stage 4 ---
  \node[inputbox, below=0.8cm of compok, minimum width=2.8cm] (cbin)
    {Rust eBPF Binary (.o)};
  \node[inputbox, right=0.8cm of cbin, minimum width=2.8cm] (rbin)
    {C eBPF Binary (.o)};
  \draw[arr] (compok) -- node[left, font=\scriptsize] {pass} (cbin);

  \node[toolbox, below=0.5cm of $(cbin.south)!0.5!(rbin.south)$,
    minimum width=3.8cm] (symbex)
    {Symbolic Execution (angr)};
  \node[stagelab, left=0.2cm of symbex] {Stage 4};
  \draw[arr] (cbin) -- (symbex);
  \draw[arr] (rbin) -- (symbex);

  \node[toolbox, below=0.5cm of symbex, minimum width=3.8cm] (z3)
    {Z3 Equivalence Check};
  \node[stagelab, left=0.2cm of z3] {Stage 5};
  \draw[arr] (symbex) -- (z3);

  %% --- Equivalence decision ---
  \node[decidebox, below=0.6cm of z3] (equivok) {result?};
  \draw[arr] (z3) -- (equivok);

  %% --- UNSAT: verified ---
  \node[resultbox, below=0.8cm of equivok] (verified)
    {\checkmark{} VERIFIED};
  \draw[arr] (equivok) -- node[left, font=\scriptsize] {UNSAT}
    (verified);

  %% --- SAT: Stage 6 LLM bug triage ---
  \node[llmbox, right=1.5cm of equivok, minimum width=2.7cm] (triage)
    {LLM Bug Triage};
  \node[stagelab, above=0.12cm of triage] {Stage 6};
  \draw[arr] (equivok.east) -- node[above, font=\scriptsize] {SAT}
    (triage.west);

  %% --- triage decision: candidate C bug vs. translation regression ---
  \node[decidebox, below=0.55cm of triage] (triageok) {LLM verdict?};
  \draw[arr] (triage) -- (triageok);

  %% --- LLM bug claim -> human review -> terminal BUG FOUND ---
  \node[resultbox, fill=red!15, below=0.8cm of triageok] (bugfound)
    {BUG FOUND};
  \draw[arr] (triageok) -- node[left, font=\scriptsize, align=right]
    {BUG claim} (bugfound);

  %% --- non-bug verdict -> counter-example repair loop back to Rust source ---
  \node[llmbox, right=1.3cm of triageok, minimum width=2.6cm] (cexfix)
    {\small LLM + counter-example};
  \draw[looparr] (triageok.east) -- node[above, font=\scriptsize] {not a bug}
    (cexfix.west);
  \coordinate (turnR) at ([xshift=0.6cm]cexfix.east);
  \coordinate (turnTop) at (turnR |- rsrc.east);
  \draw[looparr, rounded corners=4pt]
    (cexfix.east) -- (turnR)
    -- (turnTop)
    -- (rsrc.east);

\end{tikzpicture}%
}%
\caption{\toolname 's six-stage pipeline.
  \textcolor{orange!70!black}{\textbf{Orange}} boxes are LLM-driven steps,
  \textcolor{blue!60!black}{\textbf{blue}} boxes are deterministic tool steps.}
\label{fig:pipeline}
\end{figure}
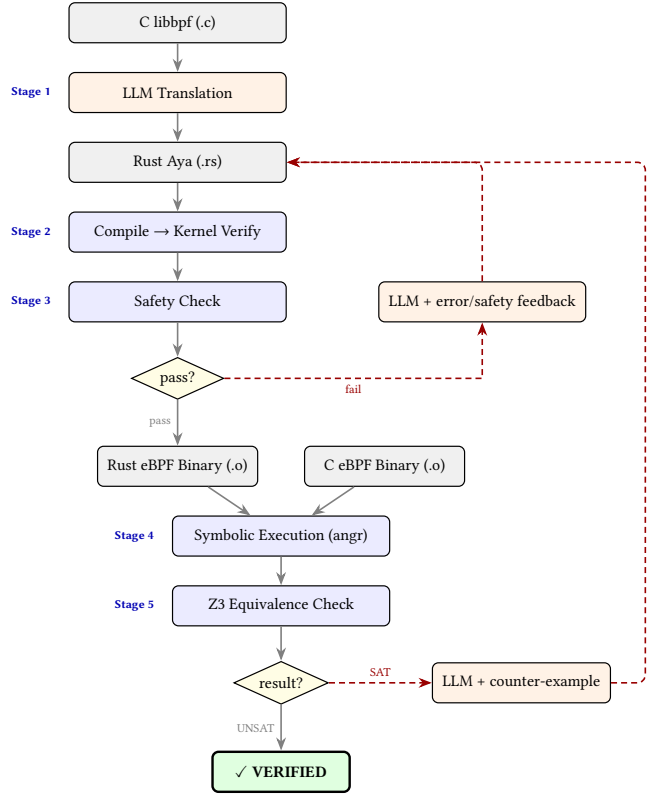

\Cref{fig:pipeline} presents the six stages of \toolname.
\toolname~can be instantiated in two ways: a \emph{deterministic} approach
that explicitly follows each stage sequentially via a scripted pipeline,
and an \emph{agentic} approach where an LLM agent implicitly follows the same
stages via tool-mediated reasoning. More details about the two approaches are provided in \Cref{sec:approaches}. We first summarize the end-to-end workflow and then detail the safety analysis, symbolic execution, and equivalence checking machinery that underpin Stages~3, 4, and 5.

\textbf{Stage~1: LLM Translation.}
The pipeline takes as input a C eBPF source file, the
entry point symbol name, and a list of map names with their types.
An LLM translates the C source to Rust using a prompt that includes Aya API mappings,
hook macro tables, map access patterns, and a complete working example. The prompts for the LLM translations are provided in \Cref{sec:llm_prompts}.

\textbf{Stage~2: Compile and Kernel Verify.}
The Rust source is compiled to eBPF bytecode with a set of compiler-supported lint guards.
These guards deny multiple unsafe operations within one block, undocumented
unsafe blocks, unnecessary unsafe, and unused \texttt{must\_use} results. 
For example, \texttt{\#![deny(unused\_unsafe)]} is added at the top of the produced Rust code
to deny unnecessary uses of unsafe blocks.
These guards force the translation toward small, scoped unsafe regions and prevent silently ignored API outcomes. Compiler errors are fed back to the LLM in an inner retry loop.
On successful compilation, the binary is loaded into the kernel
verifier, which ensures the translated program satisfies eBPF safety requirements
(bounded loops, memory safety, restricted helpers).  Stage~2 rejects
translations that are not accepted by the Rust toolchain or by the eBPF verifier.
This stage handles the part of the LLM-unsafety challenge (\Cref{sec:intro}) that Aya's typed API and the kernel verifier already encode (e.g.,\ hook/context mismatches and map-schema mismatches surface as \texttt{rustc} errors or verifier rejections).

\textbf{Stage~3: Safety Check via Static Analysis.}
This stage determines whether the translation used the safest available Aya
abstraction and idiomatic Rust.  The compiler can reject type errors and flag
overly broad or unnecessary \texttt{unsafe} blocks, but it cannot determine
whether a translation bypasses an available typed Aya interface with a
lower-level construct that still compiles and passes the kernel verifier.  We
therefore create a source-level safety engine that strips comments and applies
banned-pattern checks, regex bans, and file-level invariants to the Rust source.
The engine requires Aya's typed helper wrappers and map APIs wherever they are
available. 
% For example, it rejects direct generated helper calls that bypass the Aya interface, hand-written \texttt{unsafe extern "C"} trampolines, raw-pointer mutations outside the approved patterns, volatile or
% inline-assembly bypasses that recreate C's untyped interface, and requires every failable helper
% \texttt{Result} to be propagated with \texttt{?} or handled through explicit
% \texttt{Ok}/\texttt{Err} branches.
% It rejects direct generated-helper calls, helper-ID
% \texttt{transmute}, hand-written \texttt{unsafe extern "C"} trampolines,
% raw-pointer mutations outside the approved patterns, and volatile or
% inline-assembly bypasses that recreate C's untyped interface.  It also requires
% every failable helper
% \texttt{Result} to be propagated with \texttt{?} or handled through explicit
% \texttt{Ok}/\texttt{Err} branches.  
Together with the lint guards in Stage~2,
these rules reject unrestricted or unnecessary \texttt{unsafe} code while
retaining only the narrowly scoped operations required by approved Aya APIs.
% Any violation rejects the translation and produces structured feedback for
% repair.
% The full policy is listed in the appendix
% (\Cref{sec:safety_engine_full}).
% This stage deterministically rules out verifier-accepted translations that still bypass Aya's safer interfaces; safety regressions are fed back to the LLM with concrete safe alternatives.
In this way, Stage~3 closes the residual portion of the LLM-unsafety challenge
(\Cref{sec:intro}): even a translation that compiles and passes the kernel
verifier is rejected if it bypasses the typed Aya API boundary.
% \saeid{check this last sentence.}
% \saeid{the clause after `but' is not clear, as both sentences before and after `but' say passing. That is why I am confused.} \vishnu{is it clear now?} \saeid{Yes.}

\textbf{Stage~4: Symbolic Execution.}
Both the original C eBPF binary and the compiled Rust binary are loaded into angr using the symbolic execution backend described in
\Cref{sec:symbolic-execution}. 
Our symbolic execution explores all paths, producing structured formulae that capture return values and map side effects of the two input programs.
Operating on eBPF bytecode rather than on Rust or C source addresses the source-level-mismatch challenge (\Cref{sec:intro}), since ownership moves, integer promotion, and trait resolution do not survive compilation.
% Note that it is possible to explore all paths since eBPF programs have bounded loops.

\textbf{Stage~5: Equivalence Checking.}
The equivalence checker (\Cref{sec:equiv-checking}) submits the generated
formulae to Z3 and compares return values, map state, and output events over all
modeled inputs. An \texttt{UNSAT} result establishes conditional semantic
equivalence (\cref{def:equivalence}) and yields a \textit{Verified} outcome. A \texttt{SAT} result produces
a concrete counter example and proceeds to Stage~6. It is not sent directly to
the repair loop because the difference may expose a bug in the original C
program. Stage~5 addresses the challenge of semantic equivalence between the C and Rust source (\Cref{sec:intro}). 

\textbf{Stage~6: Bug Triage.}
The LLM receives the C source, the Rust translation, and the structured
counter example. It classifies the difference as a bug in the C program or a
translation mismatch. A non-bug verdict returns automatically to the repair loop,
where the counter example guides the next Rust candidate through Stages~2--5.
These routine counter examples do not stop the pipeline. Only a
\textit{Bug Found} verdict terminates the pipeline since the divergence is deemed unfixable due to a bug in the C code. The developer can then inspect the bug to either accept or reject it. This addresses the challenge of enforcing literal translations that might port over bugs (\Cref{sec:intro}). We perform experiments to validate the true and false positive rates of the bug triage stage in Section~\ref{sec:synthetic-eval}.

% The C formula is generated once and cached
% across all repair iterations.  
% The most challenging part of this workflow is
% the symbolic equivalence machinery in Stages~4 and~5, which we detail next.

\subsection{Static Analysis Safety Engine}
\label{sec:safety_engine}

Compilation and kernel verification do not imply idiomatic, safe Aya.  Rust that
type-checks and passes the verifier can still recreate C's untyped interface
through \texttt{unsafe} escape hatches, and we observe that LLMs reach for these
by default.  Every model we evaluated produced
verifier-passing programs that bypass Aya's typed surface.  Some reconstruct the
raw helper ABI outright, transmuting a helper id into a function pointer
(e.g., \texttt{transmute::<usize, unsafe extern "C" fn() -> u64>(5)})
or emitting hand-written \texttt{unsafe extern "C"} trampolines.  Others reproduce the leaks of \Cref{sec:verifier-gaps}
directly. LLM translations of \texttt{slabratetop} and
\texttt{gethostlatency} each pair a fixed-size string read with a discarded
\texttt{Result} (\texttt{bpf\_probe\_read\_kernel\_buf(..).ok()} and
\texttt{let \_ = bpf\_probe\_read\_user\_buf(..)}), the same over-read and
unchecked helper return as the C originals.  Across 152 translations, ten reintroduce the raw helper ABI or a fixed-size string read,
twenty-one discard a failable helper's \texttt{Result}, and nearly half read a
loader-set global through raw \texttt{read\_volatile} rather than Aya's typed
\texttt{Global<T>::load()} accessor.  Compilation and kernel verification therefore do
not certify a faithful, safe translation.

The inability of LLMs to generate idiomatic code necessitates a deterministic check and feedback loop to enforce safety properties. Stage~3 supplies the missing check.  The safety engine is a pattern-based
analyzer that strips comments and applies string, regex, and cross-pattern rules
to the Rust source. A hit rejects the candidate and returns a structured
violation---rule, message, and the problematic code region---that guides repair.  Each rule
encodes how a particular Aya interface is meant to be used. 
For example, in an eBPF program, a configuration value such as \texttt{target\_pid} is supplied from user space before the program runs, and Aya reads it through a typed \texttt{Global<T>::load()} accessor. However, LLMs instead declare a plain \texttt{static TARGET\_PID: i32} and read it with \texttt{core::ptr::read\_volatile(\&TARGET\_PID)}.  
That form compiles and verifies, but it bypasses the mechanism Aya uses to receive that user-space configuration and drops the checked type, so the program can silently run against the compile-time default rather than the configured value. Our safety engine therefore rejects a \texttt{read\_volatile} of any immutable \texttt{static} (or never-written \texttt{static mut}) and requires the \texttt{Global<T>} form. The remaining rules follow the same principle across the interfaces the
LLMs abused. Translations must call Aya's typed helper and map wrappers
rather than \texttt{mem::transmute} of helper ids, hand-written
\texttt{unsafe extern "C"} bindings, or \texttt{aya\_ebpf::helpers::generated::*},
all of which restore C-level ABI risk on every call. 
% We designed the safety engine by analyzing the common mistakes LLMs make when generating Aya Rust code. 
The complete rule set appears in \Cref{sec:safety_engine_full}.

% {\color{red}The remaining rules follow the same principle across the interfaces the
% baselines abused.  Translations must call Aya's typed helper and map wrappers
% rather than \texttt{mem::transmute} of helper ids, hand-written
% \texttt{unsafe extern "C"} bindings, or \texttt{aya\_ebpf::helpers::generated::*},
% all of which restore C-level ABI risk on every call; must handle every failable
% helper \texttt{Result} by branching on \texttt{Ok}/\texttt{Err} or propagating
% with \texttt{?} rather than discarding it with \texttt{let \_} or \texttt{.ok()},
% which hides the unchecked-helper-return bug of \Cref{sec:verifier-gaps}; must
% read strings with the NUL-stopping \texttt{\_str\_bytes} helpers rather than
% fixed-size \texttt{\_buf} reads that over-read past the terminator (the
% \texttt{filetop} leak); and must mutate map values through a
% \texttt{get}--copy--\texttt{insert} sequence or an atomic read-modify-write
% rather than non-atomic raw-pointer field writes through \texttt{get\_ptr\_mut},
% which silently re-create C's data-race semantics.  A candidate that passes the
% kernel verifier but violates any rule is still rejected; the complete rule set
% appears in \Cref{sec:safety_engine_full}.}

%% ================================================================
\subsection{Symbolic Execution}
\label{sec:symbolic-execution}
\label{sec:angr-backend}

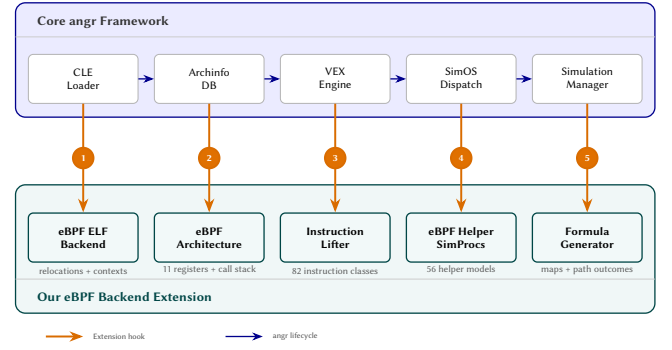
\begin{figure}[t]
\centering
\resizebox{\linewidth}{!}{%
\begin{tikzpicture}[
    >=Stealth,
    font=\sffamily,
    comp/.style={
      draw=black!30, rounded corners=3pt, fill=white,
      inner sep=3pt, align=center,
      minimum width=2.35cm, minimum height=1.05cm,
      font=\sffamily\footnotesize,
    },
    ext/.style={
      draw=teal!55!black, rounded corners=3pt, fill=teal!5, thick,
      inner sep=3pt, align=center,
      minimum width=2.35cm, minimum height=1.05cm,
      font=\sffamily\footnotesize\bfseries,
    },
    artifact/.style={
      draw=orange!75!black, rounded corners=5pt, fill=orange!8, thick,
      minimum width=1.6cm, minimum height=1.3cm,
      align=center, font=\sffamily\tiny\bfseries,
    },
    lanehdr/.style={font=\sffamily\small\bfseries, text=black!80},
    lanehdrext/.style={font=\sffamily\small\bfseries, text=teal!55!black},
    bandrule/.style={draw=black!20},
    hookdown/.style={->, very thick, draw=orange!85!black},
    coreflow/.style={->, thick, draw=blue!55!black},
    extflow/.style={->, thick, draw=teal!60!black},
    artflow/.style={->, thick, draw=orange!75!black},
    hooknum/.style={
      draw, circle, fill=orange!88!black, text=white,
      minimum size=4.4mm, inner sep=0pt,
      font=\sffamily\scriptsize\bfseries,
    },
    legend/.style={font=\sffamily\tiny, text=black!70},
  ]

  %% --- lane backgrounds with header bands ---
  %% Top lane: header at top, bandrule below header, boxes below bandrule
  \path[draw=blue!55!black, fill=blue!8, thick, rounded corners=6pt]
    (1.2,0) rectangle (14.9,2.45);
  \draw[bandrule] (1.2,1.72) -- (14.9,1.72);
  \node[lanehdr, anchor=west] at (1.55,2.03) {Core angr Framework};

  %% Bottom lane: boxes at top, bandrule below boxes, header at bottom
  \path[draw=teal!55!black, fill=teal!6, thick, rounded corners=6pt]
    (1.2,-4.20) rectangle (14.9,-1.45);
  \draw[bandrule] (1.2,-3.47) -- (14.9,-3.47);
  \node[lanehdrext, anchor=west] at (1.55,-3.84)
    {Our eBPF Backend Extension};

  %% --- core angr components (top lane) ---
  \node[comp] (cle)  at (2.65,0.82)  {CLE\\Loader};
  \node[comp] (arch) at (5.35,0.82)  {Archinfo\\DB};
  \node[comp] (vex)  at (8.05,0.82)  {VEX\\Engine};
  \node[comp] (sim)  at (10.75,0.82) {SimOS\\Dispatch};
  \node[comp] (mgr)  at (13.45,0.82) {Simulation\\Manager};

  %% --- extension components (bottom lane) ---
  \node[ext] (elf)   at (2.65,-2.58)  {eBPF ELF\\Backend};
  \node[ext] (archx) at (5.35,-2.58)  {eBPF\\Architecture};
  \node[ext] (lift)  at (8.05,-2.58)  {Instruction\\Lifter};
  \node[ext] (help)  at (10.75,-2.58) {eBPF Helper\\SimProcs};
  \node[ext] (form)  at (13.45,-2.58) {Formula\\Generator};

  %% --- detail subtitles under extension boxes ---
  \node[font=\sffamily\scriptsize, text=black!65] at (2.65,-3.28)
    {relocations + contexts};
  \node[font=\sffamily\scriptsize, text=black!65] at (5.35,-3.28)
    {11 registers + call stack};
  \node[font=\sffamily\scriptsize, text=black!65] at (8.05,-3.28)
    {82 instruction classes};
  \node[font=\sffamily\scriptsize, text=black!65] at (10.75,-3.28)
    {56 helper models};
  \node[font=\sffamily\scriptsize, text=black!65] at (13.45,-3.28)
    {maps + path outcomes};

  %% --- horizontal progression arrows (angr lifecycle, top lane) ---
  \draw[coreflow] (cle.east)  -- (arch.west);
  \draw[coreflow] (arch.east) -- (vex.west);
  \draw[coreflow] (vex.east)  -- (sim.west);
  \draw[coreflow] (sim.east)  -- (mgr.west);

  %% --- vertical hook arrows with numbered circles ---
  \draw[hookdown] (cle.south)  -- node[hooknum, pos=0.50] {1} (elf.north);
  \draw[hookdown] (arch.south) -- node[hooknum, pos=0.50] {2} (archx.north);
  \draw[hookdown] (vex.south)  -- node[hooknum, pos=0.50] {3} (lift.north);
  \draw[hookdown] (sim.south)  -- node[hooknum, pos=0.50] {4} (help.north);
  \draw[hookdown] (mgr.south)  -- node[hooknum, pos=0.50] {5} (form.north);

  %% --- legend ---
  \draw[hookdown] (1.85,-4.70) -- (2.65,-4.70);
  \node[legend, anchor=west] at (2.75,-4.70) {Extension hook};
  \draw[coreflow] (5.70,-4.70) -- (6.50,-4.70);
  \node[legend, anchor=west] at (6.60,-4.70) {angr lifecycle};

\end{tikzpicture}%
}%
\caption{Our eBPF support in angr.  Five numbered hooks pair each
angr component (top) with its eBPF extension (bottom). Horizontal arrows show the angr lifecycle from bytecode to formula generation.}
\label{fig:angr-ebpf-overview}
\end{figure}

One stage of \toolname{} performs symbolic execution on eBPF bytecode. No existing symbolic execution framework supported realistic eBPF programs at the bytecode level,\footnote{The existing angr-platforms eBPF backend supports bare-bones ALU/jump/load-store lifting with 2 helper stubs, with no map modeling, no atomic instructions, no BPF-to-BPF call support, and no relocation handling.} so we built a full eBPF backend for angr~\cite{angr}.  \Cref{fig:angr-ebpf-overview} shows how our eBPF support extends angr across five layers of its execution stack:
(1) CLE delegates object loading to an eBPF ELF backend,
(2) \texttt{archinfo} resolves \texttt{bpf} binaries to our
eBPF architecture definition,
(3) the VEX engine delegates instruction lifting to an eBPF lifter,
(4) SimOS dispatch invokes eBPF helper and runtime models, and
(5) the simulation manager hands terminated states to a formula generator.  
% Map support is cross-cutting rather
% than isolated to a single hook: loader relocations resolve map descriptors,
% helper models encode map operations, and the resulting map snapshots are
% preserved in the generated formulas.  Together these extensions enable
% end-to-end symbolic execution of
% compiled eBPF ELF objects produced by both Clang (C/libbpf) and
% \texttt{rustc}/LLVM (Rust/Aya).  
We provide a concrete end-to-end example of symbolic execution of eBPF bytecode in \Cref{sec:appendix-e2e-example}.

\paragraph{eBPF ELF backend.}
We extend angr's loader with an eBPF ELF backend that resolves the
eBPF-specific relocations needed to make a compiled program symbolically
executable: map references become handles consumed by the runtime model,
BPF-to-BPF call targets become reachable addresses, and remaining external
references (read-only constants, mutable globals, kernel functions) are bound
to memory regions or stubs.  The loader auto-detects the program
type (kprobe, XDP, etc.) from ELF section names, and the symbolic program context is sized appropriately for each program type.

\paragraph{Architecture definition.}
We register a 64-bit eBPF architecture with angr so every component treats the binary as native eBPF.  The
architecture declares the eBPF general-purpose registers along with the auxiliary state the lifter needs to track instruction flow and BPF-to-BPF call depth.

\paragraph{Instruction lifter.}
We implement a VEX IR lifter that decodes the eBPF instruction format and
lifts it into VEX IR for angr's symbolic engine.  The lifter covers the eBPF
instruction set in both 32-bit and 64-bit variants: arithmetic and bitwise
ALU operations, conditional and unconditional control flow (including helper
and BPF-to-BPF calls), load/store at all standard widths, and atomic memory
operations.

\paragraph{eBPF helper stubs.}
We extend angr's SimOS with an environment that initializes the
symbolic execution state according to the eBPF calling convention and
registers a model for every BPF helper a program may call.  Helpers that read
kernel state return symbolic bitvectors under a stable naming convention
shared between the C and Rust binaries, so the equivalence checker can unify reads across the two programs and
ensure both are evaluated under identical kernel state.

\paragraph{Formula generator.}
We extend angr's simulation manager with a formula generator that drives a
fully symbolic initial state, explores all feasible paths, and emits, for each
terminated path, the structured outputs the equivalence checker consumes: a
path predicate, the return value in R0, and the final state of each map.
Mutable ELF globals are tracked alongside maps and included in the same
per-path summary, so any divergence between the C and Rust binaries in either
map state or globals surface during equivalence checking.

% Previous Figure 2 retained for comparison while the alternative layout
% below is under consideration.

\subsection{Equivalence Checking}
\label{sec:equiv-checking}
\label{sec:map-modeling}
\label{sec:cross-binary-equiv}
% In case of mismatches, it guides the LLM in repairing the translation using concrete counterexamples. 
The equivalence checking stage in \toolname~asserts whether the Rust translation is formally equivalent to the C program. We formally define the equivalence checking problem for eBPF programs and then describe the process of checking the equivalence of maps. We provide additional details of the implementation in \Cref{sec:additional_equiv_check}.
% (e.g, \texttt{xpd\_md}, \texttt{sk\_buff}, etc.)
\begin{definition}[eBPF Program]
\label{def:ebpf-program}
An eBPF program $\mathcal{P}$ is a function
\begin{equation}
  \mathcal{P} : \mathcal{C} \times \mathcal{M} \longrightarrow \mathcal{R} \times \mathcal{M}' \times \mathcal{E}
\end{equation}
where $\mathcal{C}$ is the symbolic program context,
$\mathcal{M} = \{m_1, \ldots, m_k\}$ is the initial map state,
$\mathcal{R} \in \mathbb{B}^{64}$ is the return value (register R0), $\mathcal{M}'$ is the final map state, and $\mathcal{E}$ is the finite ordered sequence of emitted events through perf-event or ring-buffer output sinks.
\end{definition}
% \noindent This fixes the surface that equivalence must agree on: the C and Rust translations are compared as functions taking the same context and initial map state and producing the same return value and final map state.

% \begin{definition}[Symbolic Execution Paths]
% \label{def:paths}
% Symbolic execution of program $\mathcal{P}$ yields a set of paths
% $\Pi = \{\pi_1, \ldots, \pi_n\}$ where each path
% $\pi_i = (\phi_i, r_i, \mathcal{M}'_i)$ consists of:
% \begin{itemize}
%   \item $\phi_i$: the \emph{path predicate}, a conjunction of branch conditions
%     constraining the symbolic inputs
%   \item $r_i \in \mathbb{B}^{64}$: the symbolic return value along this path, which is the content of register R0
%   \item $\mathcal{M}'_i$: the map state at the end of this path, represented as
%     a sequence of \emph{map entry snapshots} per map
% \end{itemize}
% where the path predicates are mutually exclusive and collectively exhaustive i.e.,
% $\bigvee_i \phi_i = \mathsf{true}$ and $\phi_i \wedge \phi_j = \mathsf{false}$ for $i \neq j$.
% \end{definition}

\begin{definition}[Symbolic Execution Paths]
\label{def:paths}
Symbolic execution of program $\mathcal{P}$ yields a set of paths
$\Pi = \{\pi_1, \ldots, \pi_n\}$ where each path
$\pi_i = (\phi_i, r_i, \mathcal{M}'_i, \mathcal{E}_i)$ consists of $\phi_i$ the \emph{path predicate}, $r_i \in \mathbb{B}^{64}$ the symbolic return value along this path, $\mathcal{M}'_i$ the map state at the end of this path, and $\mathcal{E}_i$ the ordered sequence of events emitted along the path,
% \begin{itemize}
%   \item $\phi_i$: the \emph{path predicate}, a conjunction of branch conditions
%     constraining the symbolic inputs
%   \item $r_i \in \mathbb{B}^{64}$: the symbolic return value along this path, which is the content of register R0
%   \item $\mathcal{M}'_i$: the map state at the end of this path, represented as
%     a sequence of \emph{map entry snapshots} per map
% \end{itemize}
where the path predicates are mutually exclusive and collectively exhaustive i.e.,
$\bigvee_i \phi_i = \mathsf{true}$ and $\phi_i \wedge \phi_j = \mathsf{false}$ for $i \neq j$.
\end{definition}

% \begin{definition}[Program Output]
% \label{def:output}
% The output of program $\mathcal{P}$ on input $\vec{x} \in \mathcal{C} \times \mathcal{M}$ is encoded as a predicate $\mathit{out}_{\mathcal{P}}(\vec{x})$ over the return value $\mathcal{R}$, built in three layers:

% \begin{enumerate}
%     \item \emph{Per-path Binding.} Along each path $\pi_i$, the expression $\mathcal{R} = r_i$ asserts that the program's return value matches the symbolic register R0 value generated along $\pi_i$.
%     \item \emph{Path Guard.} The binding is conjoined with the path predicate,  $\phi_i(\vec{x}) \wedge \mathcal{R} = r_i$, so that the equality $\mathcal{R} = r_i$ is valid only on inputs that traverse $\pi_i$.
%     \item \emph{Path Enumeration.} Disjunction of the guarded bindings yields the full output predicate and covers the entire input space:
%     \begin{equation}
%       \label{eq:output-disjunction}
%       \mathit{out}_{\mathcal{P}}(\vec{x}) \;\triangleq\;
%       \bigvee_{i=1}^{n}\; \bigl(\phi_i(\vec{x}) \;\wedge\; \mathcal{R} = r_i\bigr).
%     \end{equation}
% \end{enumerate}
% \end{definition}

\begin{definition}[Program Output]
\label{def:output}
The output of program $\mathcal{P}$ on input $\vec{x} \in \mathcal{C} \times \mathcal{M}$ is the predicate over return value $\mathcal{R}$ and emitted-event sequence $\mathcal{E}$
\begin{equation}
  \label{eq:output-disjunction}
  \mathit{out}_{\mathcal{P}}(\vec{x}) \;\triangleq\;
  \bigvee_{i=1}^{n}\; \bigl(\phi_i(\vec{x}) \;\wedge\; \mathcal{R} = r_i \;\wedge\; \mathcal{E} = \mathcal{E}_i\bigr),
\end{equation}
where each disjunct binds the return value $\mathcal{R}$ to the symbolic R0 value $r_i$ produced along path $\pi_i$ and the emitted-event sequence $\mathcal{E}$ to $\mathcal{E}_i$, guarded by the path predicate $\phi_i(\vec{x})$ so the binding holds only on inputs traversing $\pi_i$. The disjunction over all $n$ paths covers the entire input space.
\end{definition}

% The output of program $\mathcal{P}$ on input
% $\vec{x} \in \mathcal{C} \times \mathcal{M}$ binds the program return-value $\mathcal{R}$ to the path return value via a path-indexed disjunction over all paths:
% \begin{equation}
%   \label{eq:output-disjunction}
%   \mathit{out}_{\mathcal{P}}(\vec{x}) \;\equiv\;
%   \bigvee_{i=1}^{n}\; \bigl(\phi_i(\vec{x}) \;\wedge\; \mathcal{R} = r_i\bigr).
% \end{equation} 

\begin{definition}[Map Output]
\label{def:map-output}
$\mathit{map}_{\mathcal{P}}^m(\vec{x}, k_q) \triangleq \bigl(P_m(k_q),\, V_m(k_q)\bigr)$
denotes the final \emph{(presence, value)} pair of map $m$ at query key
$k_q \in \mathcal{K}$ after executing program $\mathcal{P}$ on input $\vec{x}$,
where each component is encoded as an \emph{If-Then-Else} (ITE) chain.
\end{definition}
% Tuple
% equality of $\mathit{map}_{\mathcal{P}}^m$ in \Cref{def:equivalence} reduces to
% component-wise equality of $V_m$ and $P_m$, so equivalence requires
% agreement on both the value and presence at every key.
\noindent Informally, given a query key $k_q$, the $\mathit{map}_{\mathcal{P}}^m$ tuple answers the following questions: \emph{``would the lookup succeed?''} ($P_m$) and \emph{``if the lookup succeeds, what value would I read back?''} ($V_m$).  
% Both are encoded as ITE chains that traverse the program's writes in newest-first order, mirroring the kernel's last-write-wins semantics.

\paragraph{ITE-chain map encoding.}
\label{sec:ite-map-encoding}
A natural approach to reason about maps is to use Z3's Theory of Arrays. However, angr's constraint backend (Claripy) represents all symbolic values as
fixed-width bitvectors (Theory of Bitvectors) that cannot be directly combined with Z3
Arrays.\footnote{\url{https://github.com/angr/claripy/issues/171}}
Instead, we post-process the symbolic execution output into ITE chains
over bitvectors.  The encoding is two-level: an inner chain captures the
ordered write trace on a single path, and an outer chain selects the
trace whose path predicate $\phi_j$ is satisfied.
On a single path $\pi_j$, the entries
$\langle k_1, v_1, e_1 \rangle, \ldots, \langle k_n, v_n, e_n \rangle$
are sorted by temporal write sequence ($e_i$ is 1 after
\texttt{update\_elem}, 0 after \texttt{delete\_elem}).
The per-path value chain walks the trace newest-first and returns the first write whose key matches and whose post-state still exists,
falling through to the shared initial map $v_{\mathit{init}}$:
\begin{multline}
\label{eq:hash-ite-path}
  V_m^{\pi_j}(k_q) \triangleq \textsc{Ite}\Big( k_q=k_n \wedge e_n=1,\; v_n, \\
  \textsc{Ite}\big( k_q=k_{n-1} \wedge e_{n-1}=1,\; v_{n-1}, \; \dots,\; v_{\mathit{init}} \big) \Big).
\end{multline}
The presence chain $P_m^{\pi_j}(k_q)$ has the same structure but branches on $k_q = k_i$ alone (without conjoining $e_i = 1$) and returns $e_i$ directly, so a delete clears presence at $k_i$ while still letting the value chain fall through to the prior write at the same key.  This encoding yields last-write-wins under symbolic key aliasing and is equivalent to the read-over-write axioms of the Theory of Arrays~\cite{mccarthy1993towards}. 
Across all paths, we create per-program ITE chains for value ($V_m(k_q)$) and presence ($P_m(k_q)$) that are ITE chains over individual path chains combined with the respective path predicates:

\begin{multline}
\label{eq:hash-ite-program}
  % V_m(k_q) = \textsc{Ite}\Big( k_q=k_n \wedge e_n=1,\; v_n, \\
  % \textsc{Ite}\big( k_q=k_{n-1} \wedge e_{n-1}=1,\; v_{n-1}, \dots, v_{\mathit{init}} \big) \Big).
  V_m(k_q) \triangleq \textsc{Ite}\Big( \phi_n(\vec{x}),\; V_m^{\pi_n}(k_q),\\
  \textsc{Ite}\big( \phi_{n-1}(\vec{x}),\; V_m^{\pi_{n-1}}(k_q) ,\; \dots,\;  V_{init} \big) \Big).
\end{multline}

where $V_{init}$ is the shared initial value for the program. We provide an example of the ITE chain construction in \Cref{sec:worked-example}.
% The full per-program selector $V_m(k_q)$, the parallel presence chains
% $P_m^{\pi_j}$ and $P_m$, and a worked instantiation are given in
% \Cref{sec:worked-example}.  

Several eBPF programs mutate map values through pointers returned by
\texttt{bpf\_map\_lookup\_elem} rather than via \texttt{bpf\_map\_update\_elem}
(e.g.\ \texttt{\_\_sync\_fetch\_and\_add} on a hash-map slot).
To account for such pointer updates, we model map values as memory-backed regions in angr. All
pointer-level updates are recorded in the same write sequence as
helper-mediated updates and flow through the same ITE chain. 
% A
% translation that preserves the observable write trace---regardless of
% whether it goes through pointers or helpers---remains equivalent.

\begin{definition}[Conditional Semantic Equivalence]
\label{def:equivalence}
Let $\Phi_{\mathit{safe}} : \mathcal{C} \times \mathcal{M} \to \{\mathsf{true}, \mathsf{false}\}$ be a \emph{safety condition} on inputs. Programs $\mathcal{P}_C$ (C original) and $\mathcal{P}_R$ (Rust translation) are \emph{equivalent modulo $\Phi_{\mathit{safe}}$} iff for all inputs $\vec{x} \in \mathcal{C} \times \mathcal{M}$ and all query keys $k_q \in \mathcal{K}$:
\begin{equation}
\label{eq:equiv}
\begin{aligned}
  \Phi_{\mathit{safe}}(\vec{x}) \Longrightarrow{}
    &\; \mathit{out}_{\mathcal{P}_C}(\vec{x}) = \mathit{out}_{\mathcal{P}_R}(\vec{x}) \\
    &\;\wedge\; \forall m \in \mathcal{M}:\;
       \mathit{map}_{\mathcal{P}_C}^m(\vec{x}, k_q)
       = \mathit{map}_{\mathcal{P}_R}^m(\vec{x}, k_q).
\end{aligned}
\end{equation}
% \textcolor{red}{By Definition~\ref{def:output}, equality of $\mathit{out}$ requires equality of both the return value and the ordered emitted-event sequence.}
\end{definition}

% , each of which is a deliberate scoping choice that lets safety-improving Rust translations coexist with bytecode-level equivalence to legacy C
$\Phi_{\mathit{safe}}$ is the conjunction of two safety conditions:
\begin{inparaenum}[(i)]
  \item \emph{Helper-success ($\Phi_{\mathit{safe}}^{(i)})$.} Equivalence is required only on inputs that drive every helper call to success. On the success paths, the typed \texttt{Err} arm is unreachable during symbolic execution, so a Rust translation that short-circuits on \texttt{Err} (e.g.,\ \texttt{Err(\_) => return}) produces the same state as a C source that ignores the helper-failure return on every input that $\Phi_{\mathit{safe}}^{(i)}$ admits.
  \item \emph{Zero-initialized memory ($\Phi_{\mathit{safe}}^{(ii)})$.} Equivalence is checked under a memory model in which an uninitialized read always returns zero. This is analogous to restricting the symbolic input state to executions without nonzero unwritten gaps.
  % \toolname~\emph{enforces} this assumption on the Rust side: Aya's typed output buffers, the Rust compiler, and the safety policy's mandate zero-fill and explicit-padding rules, guaranteeing that every byte of a map value or emitted record is definitely written before use. 
  Under this model, a Rust translation that zero-fills a buffer before population (closing the \texttt{bashreadline}-style KASLR leak) is equivalent to a C source that leaves those bytes uninitialized.
  % because the only bytes that could differ are exactly the ones the memory model fixes to zero.
\end{inparaenum}

For bug discovery, the checker can toggle either safety
condition. Disabling helper-success modeling allows helpers to fail, while
disabling zero-read modeling leaves unwritten C bytes symbolic. In the latter
mode, those bytes can differ from Rust's zero-initialized bytes and produce the
\texttt{SAT} counter-example. Our notion of conditional semantic equivalence allows us to claim that a Rust program generated by \toolname~preserves the behavior of the legacy C program on \emph{safe paths} while diverging on \emph{unsafe paths}. Typically, the unchecked helpers and uninitialized state do not affect the safe paths in the program. However, if a bug in the C program affects a safe path (e.g., the \texttt{biostacks} sizeof bug), then the Rust translation will diverge on safe paths. In such cases, Stage 6 of the translation pipeline catches the divergence and reports it as a bug. 
The only way to claim equivalence to the C program for bugs that affect the safe paths is to first fix the bug in the C source and then run the equivalence checker against the Rust program.

% Neither safety condition excludes ordinary
% helper-success executions over initialized state. A Rust correction that changes
% behavior on such a safe path therefore cannot be equivalent to the C source and
% also produces a \texttt{SAT} counter-example. The counter-example enters LLM
% bug triage. Only a \textit{Bug} verdict stops translation for human inspection,
% while every other verdict returns to the repair loop.

Verification of equivalence reduces to checking unsatisfiability.  We negate
\Cref{eq:equiv} and ask Z3 whether there exists an input $\vec{x}$ and a
query key $k_q$ such that:
\begin{multline}
\label{eq:mismatch}
  \mathit{out}_{\mathcal{P}_C}(\vec{x}) \neq \mathit{out}_{\mathcal{P}_R}(\vec{x})
  \;\vee\;\\
  \exists m \in \mathcal{M}:\;
    \mathit{map}_{\mathcal{P}_C}^m(\vec{x}, k_q) \neq \mathit{map}_{\mathcal{P}_R}^m(\vec{x}, k_q).
\end{multline}

If the formula is \texttt{UNSAT}, the programs are equivalent.  If \texttt{SAT},
the satisfying assignment is a concrete counter-example. Unless otherwise specified, ``equivalent'' and ``verified'' refer to conditional equivalence in Definition~\ref{def:equivalence}.

\paragraph{Counterexample feedback.}
When Z3 returns \texttt{SAT}, the satisfying assignment provides concrete values
for the shared program context, initial map contents, and helper returns. The
checker evaluates both programs under those values and classifies the observed
difference as a sign extension, truncation, missing write, or another
supported category. It then presents the LLM with both outputs and the program
locations associated with the difference. The LLM first performs the bug-triage
decision from Stage~6. A non-bug verdict feeds the same evidence into targeted
repair.
% , forming a loop analogous to counter-example-guided inductive synthesis
% (CEGIS)~\cite{solar2013program,alur2013syntax}.

% Given two structured formulae---one from the C program, one from the Rust
% translation---the equivalence checker determines whether they are semantically
% identical for all inputs using Z3~\cite{z3}.  The checker first unifies
% symbolic variables from the two independent symbolic execution runs into a
% shared namespace, constructs divergence conditions for return values and map
% states, and queries Z3 for satisfiability of \Cref{eq:mismatch}.

%%% Local Variables:
%%% mode: LaTeX
%%% TeX-master: "main"
%%% End:

\section{Evaluation}
\label{sec:evaluation}

%% ================================================================
\subsection{Experimental Setup}
\label{sec:setup}

\paragraph{Hardware.}
All experiments were conducted on an x86\_64 server equipped with 24 physical cores and 335 GB of RAM. The system runs Ubuntu 22.04.3 LTS with Linux Kernel 6.8.0. The environment is hosted on a KVM-based virtual machine with hardware acceleration.

\paragraph{Software.}
We use Python~3.12, angr~9.2 for symbolic execution, Z3~4.13 as the
SMT solver, and the Aya\footnote{Upstream Commit ID: \href{https://github.com/aya-rs/aya/commit/748755e955a2bf05260d17c70506eb8f563edcb9}{748755e955a2bf05260d17c70506eb8f563edcb9}} eBPF framework with rustc 1.93.1. The C libbpf programs were compiled with clang 14.0.0, bpftool v7.4.0, and libbpf v1.4. In our experiments with command-line agents, we use claude code 2.1.112, codex 0.116.0, and gemini-cli 0.36.0. The C libbpf programs were compiled at the \texttt{-O2} optimization level. The Rust programs were compiled with the \texttt{--release} flag.  We open-source our codebase and make all artifacts public.\footnote{\url{https://github.com/vdasu/heimdall}}
% \footnote{Anonymous Link: \url{https://anonymous.4open.science/r/heimdall-6D3D/}}
% For the LLM-in-the-loop pipeline we use Claude Sonnet~4.6; for the
% agentic approaches we use Claude Opus~4.6 via Claude Code.

%% ================================================================
\subsection{Translation Approaches}
\label{sec:approaches}

We experiment with the following \emph{fully autonomous} translation approaches:

\begin{enumerate}
\item \textbf{Baseline.} For the baseline, we consider code agents with state-of-the-art LLMs inside a coding harness e.g., Claude Code. The agent is asked to translate the C libbpf to Rust Aya, along with the source .c file. It is unrestricted and runs with full-tool reasoning, i.e., file browsing, creating bash/.py scripts, etc. The agent iterates until it produces a compilable Aya Rust program that it believes is equivalent to the C program. Our baseline setup mimics the setup of software engineering benchmarks \cite{merrill2026terminalbenchbenchmarkingagentshard,jimenez2024swebenchlanguagemodelsresolve,miserendino2025swelancerfrontierllmsearn}, and our experiments serve as a benchmark of agentic capability in C to Rust translation for eBPF.
  
  % with full tool access---file browsing,
  % multi-step reasoning, Aya source inspection---but \emph{no}
  % equivalence checking feedback.  The agent iterates until the
  % translation compiles and passes the kernel verifier, then stops. TerminalBench\cite{merrill2026terminalbenchbenchmarkingagentshard}, SWE-bench\cite{jimenez2024swebenchlanguagemodelsresolve}, SWE-lancer\cite{miserendino2025swelancerfrontierllmsearn}

\item \textbf{\toolname~(Deterministic).} In this setting, we instantiate \toolname~(Figure~\ref{fig:pipeline}) in a scripted pipeline that \emph{explicitly} follows the 6-stage approach. The LLM is invoked as a stateless translator: on the first attempt it receives the C source and is asked to produce a Rust Aya translation; on subsequent attempts it additionally receives the previous Rust code along with structured feedback from the failing stage (compiler errors, kernel verifier diagnostics, safety-policy violations, or Z3 counter-examples). No tool-mediated reasoning is permitted. %The LLM cannot browse the filesystem, run shell commands, or otherwise interact with the environment. 
Transitions between stages are driven by an automaton-style controller in our pipeline rather than by the model itself, so the LLM never decides what to do next. It only produces the next candidate translation given the current stage's feedback.

\item \textbf{\toolname~(Agentic).}  In this setting, we instantiate \toolname~(Figure~\ref{fig:pipeline}) as an agentic harness layered on top of the same code agents used in the baseline, so the 6-stage approach is followed \emph{implicitly} rather than enforced by an external controller. The agent retains full freedom to perform tool-mediated reasoning---reading files, inspecting C and Rust bytecode, browsing the Aya source tree, spawning subagents, and writing helper scripts---while being directed to follow the stages of our approach to produce an equivalent and idiomatic Rust Aya translation. The agent has access to the same structured signals as the deterministic pipeline (compiler output, kernel verifier diagnostics, safety analyzer reports, and equivalence-checker counterexamples), but it can also act on them with arbitrary tool calls.\end{enumerate}

% All three approaches are \emph{fully autonomous} with zero human intervention. The agentic approaches are run in unrestricted mode (e.g., \texttt{--dangerously-skip-permissions} in claude-code) within a sandbox, and the agents self-report the program's status once translation is complete. 
% Later, we validate the agent's findings by compiling, running the safety static analysis engine, kernel verifying, and equivalence checking.

\subsection{Dataset}
\label{sec:dataset}

% \Cref{tab:dataset} summarizes our evaluation dataset drawn from ten
% sources spanning four major eBPF program categories: kernel tracing
% (kprobes, tracepoints, fentry), user-space tracing (uprobes, USDT),
% networking (XDP, TC, socket filters), and security (LSM).

\begin{table}[t]
\centering
\caption{Dataset of eBPF programs. ``Total'' is the number of programs in each source, ``Valid'' is the subset supported by the Rust/Aya target, and ``Entries'' is the number of entry points.}
\label{tab:dataset}
\resizebox{\columnwidth}{!}{
\begin{tabular}{l|r|rrrr}
\toprule
\textbf{Source} & \textbf{Total} & \textbf{Valid} & \textbf{Entries} & \textbf{Avg LoC} & \textbf{Avg Maps} \\
\midrule
libbpf-tools~\cite{libbpf-tools}     & 57 & 57 & 365 & 175 & 2.2 \\
libbpf-bootstrap~\cite{libbpf-tools} & 15 & 13 &  21 &  45 & 0.5 \\
KEN samples~\cite{10.1145/3672197.3673434}       & 26 & 25 &  38 &  40 & 1.0 \\
DAE~\cite{dae}                 &  1 &  1 &   6 & 206 & 2.0 \\
Fluvia~\cite{fluvia}               &  1 &  1 &   1 & 162 & 1.0 \\
Hercules~\cite{hercules}                 &  1 &  1 &   1 & 193 & 3.0 \\
Suricata~\cite{suricata}  &  6 &  5 &   5 & 216 & 2.6 \\
CRAB~\cite{crab}              &  1 &  1 &   1 & 370 & 4.0 \\
XDP-FW~\cite{hxdp}             &  3 &  3 &   3 & 111 & 1.3 \\
BMC~\cite{bmc-cache}  &  1 &  1 &   7 & 526 & 6.0 \\
cache\_ext~\cite{cacheext}         &  7 &  7 &  42 & 325 & 3.0 \\
\midrule
\textbf{Total}                        & \textbf{119} & \textbf{115} & \textbf{490} & --- & --- \\
\bottomrule
\end{tabular}
}
\end{table}
% Table 2 updated on 2026-04-20 against the checked-in dataset artifacts.
% 2026-06-26: cgroup_array maps are now supported, so the 13 cgroup-filtering
% libbpf-tools programs move from excluded to valid: libbpf-tools Valid 44->57,
% Entries 322->365 (+43, counted as 'g F' symbols in the C .o, the same method
% that reproduces the original 322); Total Valid 102->115, Entries 447->490.
% libbpf-tools Avg LoC/Maps stay 175/2.2 (consistent 57-program recompute = 175.1/2.2).
% We keep libbpf-bootstrap/task_iter in the unsupported set: the vendored Aya
% tree has user-space loader support for iterator programs (aya/src/programs/iter.rs),
% but our Rust eBPF translation target still lacks a corresponding aya-ebpf
% iterator macro/context path, so SEC("iter/task") is not currently translatable
% in this pipeline. Program totals/valid counts, entry totals, and average map
% counts were recomputed over the supported subsets from the current source/object
% files; the LoC values were left unchanged here because the draft mixed older
% manual counting conventions across rows.

\Cref{tab:dataset} summarizes our evaluation dataset, which is diverse in both source and eBPF subsystem coverage.  It includes
production tracing tools, tutorial programs, benchmark samples, network
functions, research systems, and network-IDS filters, spanning kprobes,
tracepoints, fentry/fexit, uprobes, LSM, XDP, TC, and socket-filter programs. Of the 119 collected programs, 4 are excluded because Aya does not support
USDT arguments or legacy \texttt{BPF\_LD\_ABS}/\texttt{BPF\_LD\_IND}
socket-filter instructions: two libbpf-bootstrap programs require an iterator
or USDT-argument support, one KEN sample uses \texttt{bpf\_usdt\_arg}, and one
Suricata program depends on legacy socket-filter instructions. 
This
leaves 115 valid programs for translation and verification. We refer to each \texttt{.c} file as a program and each \texttt{SEC()}-annotated function within it as an entry point. The data sources are of varying complexity, measured using average LoC and number of eBPF maps (shown in the last two columns of \Cref{tab:dataset}).
% \saeid{Is there any way we can show in Table 2 that we are going to have 51 programs for translation?} \vishnu{we actually have 102 in our full-scale translation. 51 is only for benchmarking}\saeid{got it!}
%% ================================================================

\subsection{Benchmarks}

To benchmark the three translation approaches, we create 11 experimental settings. For the baseline, we use Claude Code (sonnet-4.6 and opus-4.6), Codex (gpt-5.4), and Gemini-CLI (gemini-3-flash-preview). For \toolname~(Deterministic), we instantiate the pipeline with opus-4.6, sonnet-4.6, and gpt-5.4.\footnote{We only use three LLMs here due to resource constraints and limited API credits.} Lastly, we instantiate \toolname~(Agentic) with the same agents as the baseline. 
% \gtan{Can we also say a few words about why that particular setting costs a lot more credits than other settings?} \vishnu{actually, this reasoning is weird: anthropic, openai, etc. heavily subsidize their agentic frameworks to encourage new signups (i spent over \$500 last month but only paid a flat \$100. With API credit, there is no subsidy (so far i spent \$450). if i explain this, it makes the USD estimates we provide shaky since we didnt actually spend that much per program}
We evaluate these settings against 44 libbpf-tools programs and 7 randomly picked programs (DAE, Fluvia, Hercules, CRAB, and XDP-FW). 
This gives us 561 translations across all settings. Of these, the \texttt{ksnoop} program from libbpf-tools is the only translation that cannot be verified since it exceeds symbolic execution limits. We consider this a de facto non-equivalent result since it is a limitation of our verification framework. We first run \toolname~(Agentic) with opus-4.6, fix all the bugs it caught, and then run the other models on the fixed C programs.

\begin{table*}[t]
\centering
\small
\setlength{\tabcolsep}{4pt}
\caption{Evaluation results of the three translation approaches across 11 experiment settings and 51 eBPF programs. \toolname~(Agentic) with opus-4.6 is the most successful with 98\% formally verified and equivalent translations.}
\label{tab:benchmarks}
\resizebox{\textwidth}{!}{%
\begin{tabular}{l|l|cccccccccc}
\toprule
\textbf{Translation Approach}
& \textbf{Model}
& \textbf{Compile}
& \textbf{Kernel Verified (KV)}
& \textbf{Safety}
& \textbf{Equivalent}
& \textbf{KV $\cap$ Safety $\cap$ Equiv.}
& \textbf{Time (min)}
& \textbf{Bytecode Size Overhead}
& \textbf{Unsafe Ops}
& \textbf{Tokens In/Out}
& \textbf{Cost} \\
\midrule
\multirow{4}{*}{Baseline}
  & sonnet-4.6             & \textbf{51/51 (100\%)} & 39/51 (76\%) & 21/51 (41\%) & 10/51 (20\%) & 5/51 (10\%) & 6.7 & 2.62$\times$ & 18.0 & 594K~/~24K & \$0.78 \\
  & opus-4.6               & \textbf{51/51 (100\%)} & 49/51 (96\%) & 18/51 (35\%) & 11/51 (22\%) & 6/51 (12\%) & 4.6 & 2.41$\times$ & 19.7 & 800K~/~16K & \$1.23 \\
  & gpt-5.4                & \textbf{51/51 (100\%)} & 47/51 (92\%) & 23/51 (45\%) & 6/51 (12\%) & 3/51 (6\%) & 3.6 & 2.46$\times$ & 6.3 & 861K~/~\textbf{9K} & \$0.48 \\
  & gemini-3-flash-preview & \textbf{51/51 (100\%)} & 36/51 (71\%) & 19/51 (37\%) & 8/51 (16\%) & 2/51 (4\%) & \textbf{2.7} & 2.33$\times$ & 26.4 & 597K~/~11K & \textbf{\$0.13} \\
\midrule
\multirow{3}{*}{\toolname~(Deterministic)}
  & sonnet-4.6             & \textbf{51/51 (100\%)} & 49/51 (96\%) & 43/51 (84\%) & 23/51 (45\%) & 23/51 (45\%) & 16.5 & 2.62$\times$ & 30.3 & 164K~/~66K & \$1.42 \\
  & opus-4.6               & \textbf{51/51 (100\%)} & 47/51 (92\%) & 43/51 (84\%) & 30/51 (59\%) & 28/51 (55\%) & 18.7 & \textbf{2.12$\times$} & 29.5 & 286K~/~79K & \$3.26 \\
  & gpt-5.4                & 49/51 (96\%) & 49/51 (96\%) & 49/51 (96\%) & 32/51 (63\%) & 32/51 (63\%) & 41.7 & 2.63$\times$ & \textbf{6.2} & \textbf{53K}~/~167K & \$2.64 \\
\midrule
\multirow{4}{*}{\toolname~(Agentic)}
  & sonnet-4.6             & \textbf{51/51 (100\%)} & \textbf{51/51 (100\%)} & \textbf{50/51 (98\%)} & 49/51 (96\%) & 49/51 (96\%) & 25.8 & 2.77$\times$ & 17.9 & 3906K~/~89K & \$3.22 \\
  & opus-4.6               & \textbf{51/51 (100\%)} & \textbf{51/51 (100\%)} & \textbf{50/51 (98\%)} & \textbf{50/51 (98\%)} & \textbf{50/51 (98\%)} & 23.4 & 2.63$\times$ & 19.7 & 2636K~/~62K & \$3.94 \\
  & gpt-5.4                & \textbf{51/51 (100\%)} & \textbf{51/51 (100\%)} & 44/51 (86\%) & 36/51 (71\%) & 35/51 (69\%) & 8.1 & 2.56$\times$ & 9.5 & 3018K~/~17K & \$1.29 \\
  & gemini-3-flash-preview & \textbf{51/51 (100\%)} & \textbf{51/51 (100\%)} & 45/51 (88\%) & 44/51 (86\%) & 40/51 (78\%) & 11.1 & 2.68$\times$ & 26.7 & 3782K~/~25K & \$0.62 \\
\bottomrule
\end{tabular}%
}
\end{table*}

Table~\ref{tab:benchmarks} highlights the results for the 561 translations in
our benchmark study. \emph{Compile}, \emph{Kernel Verified (KV)}, \emph{Safety}, \emph{Equivalent (Equiv.)}, and the \emph{KV~$\cap$~Safety~$\cap$~Equiv.} columns report counts and percentages over the 51-program comparison set, while \emph{Time}, \emph{Bytecode Size Overhead}, \emph{Unsafe Ops}, \emph{Tokens In/Out}, and \emph{Cost} are per-program averages.
We measure \emph{Bytecode Size Overhead} as the ratio of Rust to C verifier-loaded \texttt{.text}/\texttt{prog\_*} bytecode sections and exclude debug and symbol sections. 
\emph{Unsafe Ops} is the per-program count of unsafe \emph{operations} (raw pointer dereferences, function/method calls inside an unsafe context, inline-assembly invocations, and mutable-static reads), counted via tree-sitter-rust on the produced Rust source.
% GT: removed details.
%this is formatting-invariant and tracks auditor burden better than line- or block-counts. \gtan{"formatting-invariant"? Not sure I understand.} \vishnu{unsafe block spanning multiple lines is treated the same as unsafe block (long one) in one line if both have identical ops}

The first trend is that compilation is not a meaningful proxy for translation success. Every setting produces 51/51 compiled Rust artifacts except \toolname~(Deterministic) with gpt-5.4 (49/51), but the baseline agents only fully verify 2--6/51 (4--12\%) translations.
\toolname{}'s staged repair loop closes most of this gap on the same underlying models. \toolname~(Deterministic) reaches 23--32/51 (45--63\%) for all three levels of checking  (\emph{KV~$\cap$~Safety~$\cap$~Equiv}), and \toolname{} (Agentic) reaches 35--50/51 (69--98\%). 
% --- a $\sim$90 percentage-point absolute improvement over the strongest baseline. 
Additionally, kernel verification is a poor signal of equivalence. Baseline KV pass rates of 71--96\% drop to 4--12\% \emph{KV~$\cap$~Safety~$\cap$~Equiv.} once safety policy and symbolic equivalence are layered in, confirming that the three downstream checks catch disjoint failure classes (520/561 pass KV, 405/561 pass safety, 299/561 pass equivalence, but only 273/561 pass all three).

Since every setting but \toolname~(Deterministic) gpt-5.4 (49 compiled) compiles all 51 programs. As a result, only 2 of the 288 translations that fail at least one check fail for syntactic reasons. 
The remaining 286 fail in eBPF-specific semantics or safety policy. Of these 288, 126 (43.8\%) are safety violations, 121 (42.0\%) are equivalence failures, 39 (13.5\%) are KV rejections, and 2 (0.7\%) are compile failures.
We charge each failing translation to its earliest failing stage in the order compile $\rightarrow$ KV $\rightarrow$ safety $\rightarrow$ equivalence, so these four sub-counts are disjoint and sum to 288. In contrast, the per-stage columns in Table~\ref{tab:benchmarks} count every stage a translation fails, so a translation that fails both KV and a safety check appears in both columns but is charged only to KV in the 2/39/126/121 split.
Within safety violations, 73.0\% of failing translations declare read-only configuration globals (e.g., \texttt{target\_pid}, \texttt{targ\_tgid}) as \texttt{static mut} and read them via \texttt{read\_volatile} --- Aya emits \texttt{static mut} as a writable BPF map rather than read-only \texttt{.rodata}, exposing what the C source intends as immutable configuration to userspace writes.
% The remaining classes are discarded fallible helper return values from \texttt{bpf\_probe\_read\_*} and \texttt{bpf\_get\_current\_comm} (25.4\%), untyped or unzeroed ringbuf entries that replicate the C-style partial-initialization leak (11.4\%), unsafe raw-pointer writes into \texttt{HashMap} entries (11.4\%), and a long tail of \texttt{mem::transmute} escape hatches and direct calls to generated helper bindings (each $\sim$4\%); these classes are not mutually exclusive and a single translation can hit several (mean 1.7 violations per failing program, max 12).
Within equivalence failures, the top two categories are both map-related and account for 63.7\% of the total: 36.4\% are wrong map values for the same key (the LLM emits an incorrect update or skips one), and 27.3\% are state-representation mismatches between \texttt{.data} globals and BPF-map slots (e.g., the C version uses a global and the translation uses a map, or vice-versa).
% The remaining $\sim$23\% of equivalence failures split across entry-symbol mismatches (10.1\%), section/program-type mismatches (3.7\%), angr/cle lifter limits (4.6\%), and \texttt{ksnoop}-style symbex blow-ups or framework crashes (4.6\% combined).
Of the KV failures, 26/39 ($\approx$ 67\%) are a heterogeneous mix of helper-call and pointer-tracking errors. The remaining KV failures are split across pointer-bounds violations, unreleased reference-counted resources, invalid memory accesses, and out-of-range branches.
%each accounting for 3--4/39 ($\sim$8--10\%).

% The two \toolname{} configurations cover the same five-stage pipeline but trade off audit-ability against repair power. The deterministic variant is easy to audit since every state transition is scripted by an external controller, so each repair has a fixed prompt and a fixed input set. The agentic variant additionally inspects C/Rust bytecode, browses Aya source, and spawns subagents to ground each fix --- behavior that goes beyond simply following the pipeline (\Cref{sec:fullscale} reports per-program tool-call and subagent statistics). This extra capability lets opus-4.6 agentic pass KV and safety on all 51 programs and reach 50/51 (98\%) equivalence; the only miss is \texttt{ksnoop}, whose symbolic state space exceeds our checker's limits.

\toolname~(Agentic) excels at the expense of higher running time and token consumption.
Opus-4.6 agentic averages 23.4~min and \$3.94 per program vs.\ 18.7~min and \$3.26 for opus-4.6 deterministic. Gemini-3-flash-preview agentic finishes in 11.1~min at \$0.62 per program.
Agentic input-token use is roughly an order of magnitude higher than deterministic (2.6--3.9M vs.\ 53--286K) because the agent reads files, inspects bytecode, and invokes subagents to ground each fix. We observe \toolname~(Agentic) with gpt-5.4 has a much lower runtime and output token consumption since the agent tends to give up early without attempting to fix the errors.

The \emph{Unsafe Ops} column tracks the number of unsafe operations in the translations. The operations counted here are unavoidable sanctioned unsafe (e.g., certain helper invocations, ctx-pointer reads) rather than safety violations.
The 3$\times$ gap between gpt-5.4 (6.2--9.5 ops) and the other models (17.9--30.3) is driven by translation choices. For example, gpt-5.4 abstracts the unsafe \texttt{bpf\_probe\_read\_kernel} function into a safe wrapper function whose body holds the only \texttt{unsafe} block. Every kernel read at a call-site is a plain safe-Rust call. The other models inline \texttt{unsafe { bpf\_probe\_read\_kernel(\ldots) }} at each read instead. So if a program performs ten kernel reads, gpt-5.4 contributes \emph{one} unsafe op (the wrapper body) while the other models contribute \emph{ten}, even though the runtime behavior is the same.

\subsection{Full-scale Evaluation}
\label{sec:fullscale}

% 6 unverified programs: ksnoop (symbex intractable), research__bmc_kern,
% research__cache_ext_lhd, research__cache_ext_mglru, suricata__xdp_filter,
% suricata__xdp_lb (the latter five hit the agent's iteration budget).
% Reproduction:
%   python3 paper_scripts/analyze_agent_sessions.py
%   python3 paper_scripts/classify_bash_commands.py
%   python3 paper_scripts/generate_paper_figures.py
% Resource aggregates (time/cost/tokens/BC overhead/unsafe ops) and all
% plots/subagent statistics are over the 109 fully-verified translations
% (the 6 unverified programs are excluded); see
% agent_analysis/per_program.json for the raw data.

We evaluate \toolname~(Agentic) with opus-4.6, the most successful approach in the benchmarks, across all 115 valid programs in our dataset. Of these, \toolname~produces \textbf{109/115 (94.8\%)} fully verified translations. Of the remaining six, our system fails to certify the translations. Three are partially verified (\texttt{bmc\_kern}, \texttt{cache\_ext\_lhd}, \texttt{cache\_ext\_mglru}) with some entry points verified and the others timing out due to symbolic execution time and memory limits. The other three (\texttt{ksnoop}, \texttt{suricata\_xdp\_filter}, \texttt{suricata\_xdp\_lb}) are unverified since all entry points exceed solver limits. \Cref{sec:sample-translations} presents three translations, and the remaining are in our codebase.
% We extend the comparison-set evaluation to all 102~valid Aya-supported
% programs in our dataset (\Cref{tab:dataset}).  Of these, \toolname{}'s
% agentic pipeline with opus-4.6 fully verifies \textbf{96/102 (94.1\%)}:
% compile, kernel verifier, strict safety policy, and symbex equivalence
% all pass.  The remaining six are \texttt{ksnoop}, whose Rust
% translation compiles and passes KV/safety but whose symbex equivalence
% is intractable due to path explosion, plus five
% (\texttt{suricata\_xdp\_filter}, \texttt{research\_bmc\_kern},
% \texttt{research\_cache\_ext\_lhd}, \texttt{research\_cache\_ext\_mglru},
% \texttt{suricata\_xdp\_lb}) that compile and pass KV/safety but time
% out during symbex equivalence on at least one entry.

Across all 109~verified translations, the agent averages 17.7~min/program and consumes 2.16M input / 47K output tokens on average. This yields an average translation cost of \$3.04 per program and \$332 in total across all. The average is lower than the \$3.94 cost from Table~\ref{tab:benchmarks} since the 51-program subset in the benchmarks is considerably harder to translate. The bytecode size overhead of the produced Rust \texttt{.o} averages 2.91$\times$ over the C counterpart, and each translation contains 15.5~unsafe operations on average.

% produced Rust \texttt{.o} averages 2.89$\times$ the verifier-loaded bytecode of the C source.
% Aggregating across all 102~attempted translations, the agent averages
% \textbf{27.3~min/program} (median 11.7~min; total 46.4~CPU-hours),
% \textbf{38~turns} per program (median 34), \textbf{\$2.94~in API cost}
% (median \$2.09; total \$297 for the full corpus), and processes
% \textbf{2.0M input / 47K output tokens} per program on average
% (corpus totals 202.6M / 4.71M).  The produced Rust \texttt{.o} averages
% \textbf{2.89$\times$} the verifier-loaded bytecode of the C source.
% Static unsafe-surface analysis (tree-sitter-rust) shows the typical
% program contains \textbf{18.3~unsafe operations} (median 11) spread
% across \textbf{16.0 unsafe blocks}, with only \textbf{19.9\%} of blocks
% containing more than one unsafe op --- consistent with the
% agentic-opus row of \Cref{tab:benchmarks} on the 51-program subset.
% The six unverified programs sit in the right tail of all four
% distributions; restricted to just the 96 fully verified, the means
% shift down to \textbf{16.5~min}, \textbf{\$2.73}, and
% \textbf{13.9~unsafe ops/prog} (with \textbf{2.84$\times$} bytecode
% overhead).

\begin{figure}[t]
\centering
\includegraphics[width=0.8\columnwidth]{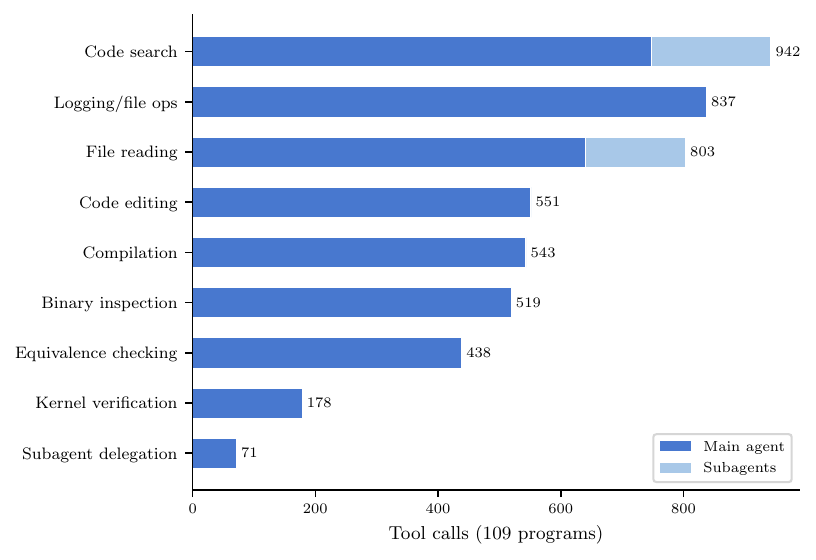}
\caption{Tool-call distribution across the 109-program verified
  evaluation, separated into main-agent and subagent contributions.}
\label{fig:tool-distribution}
\end{figure}

\begin{figure}[t]
\centering
\includegraphics[width=0.85\columnwidth]{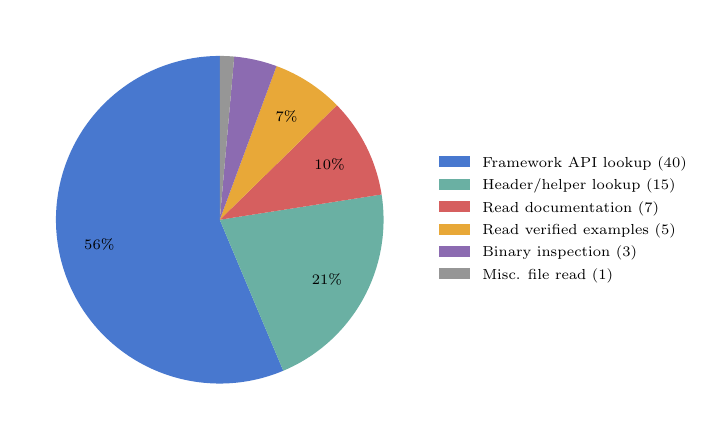}
\caption{Purpose distribution of 71~subagent invocations across
  52~programs in the verified set.}
\label{fig:subagent-purposes}
\end{figure}

% \subsection{Agent's Tool Calls}
%\label{sec:agents-toolcalls}
% \paragraph{Agent's Tool Calls.}
To assess the effectiveness of \toolname~(Agentic), we investigate the operations the agent performs.
\Cref{fig:tool-distribution} groups the agent's tool calls by purpose, separating main-agent and subagent contributions. The two largest categories are Code search with 942~calls (locating where a symbol or pattern lives across many files, via shell \texttt{grep}/\texttt{find} or the agent's native search primitives) and file reading with 803~calls (loading the contents of a known file: C source under translation, the Aya framework source tree, and prior verified translations consumed as in-context examples). Together, these two exploration activities account for roughly 37\,\% of all 4{,}706 tool calls. The agent locates and ingests far more than it generates (code editing: 551 calls). The next-largest categories cover the four deterministic gates each translation must clear: compilation runs (543), equivalence checking (438), binary inspection (519) via \texttt{llvm-objdump}, and kernel verification (178). %\gtan{What is binary inspection?} \vishnu{agent invokes llvm-objdump to see the eBPF intsructions in case of errors.}
Another advantage of the agentic approach is subagent delegation. A subagent is invoked 71 times while translating 52/109 programs. The subagent makes 358 of the 4{,}706 combined tool calls (7.6\%). \Cref{fig:subagent-purposes} breaks the subagent invocations down by purpose: 40~resolve Aya framework-API questions, 15 resolve C-side BPF headers (\texttt{bits.bpf.h}, \texttt{maps.bpf.h}, helper-macro definitions), and the remainder read documentation, prior verified translations, or inspect binaries.
\toolname~(Agentic) thus exploits tool-mediated reasoning with test-time scaling --- searching widely, ingesting and learning in-context from verified outputs, and delegating focused lookups to subagents --- along with the six-stage pipeline to generate formally verified and idiomatic translations. 
% The agent invokes a subagent on \textbf{48/96 (50\%)} of programs;
% 67~subagent invocations made 421~tool calls in total
% (10.7\% of all 3,928~tool calls), almost all of which are framework
% API lookups or reads of previously-verified translations
% (\Cref{fig:subagent-purposes}).
% \Cref{fig:agent-iterations,fig:tool-distribution,fig:subagent-purposes}
% show per-program iteration effort, tool-call category breakdown, and
% subagent-purpose distribution over the 96-program verified set.

% \begin{figure}[t]
% \centering
% \includegraphics[width=\columnwidth]{figures/agent_iterations.pdf}
% \caption{Per-program iteration effort over the 96-program verified
%   set (top~25 by total iterations).  Most programs converge within a
%   few attempts; outliers drive a long tail of compile- and
%   equivalence-fix loops.}
% \label{fig:agent-iterations}
% \end{figure}

\subsection{Runtime Validation}
\label{sec:runtime-validation}

To validate that the formally equivalent programs actually behave the same as their C counterparts, we built common userspace loaders
for 10 verified \toolname~(Agentic) opus~4.6 translations and ran each of C and Rust under the
same controlled workload (file opens, signals, forking children, loopback
connections). We repeat each program $N{=}100$ times ($1{,}000$ paired trials
total). We restrict our comparison to our trigger-induced events only since certain fields always change across runs; for example, timestamps, PIDs, and ephemeral ports. We read
per-program ns/invocation from \texttt{kernel.bpf\_stats\_enabled}. All
$1{,}000/1{,}000$ trials match exactly with identical filtered record counts. Since the equivalence side is uniform, we
report runtime overhead in \Cref{tab:runtime-overhead} as the geometric mean of the per-trial Rust/C ratio
with a Student's $t$ $95\%$ confidence interval computed on log-ratios. Seven of ten per-program CIs lie strictly above
$1\times$, \texttt{tcplife}'s CI overlaps $1\times$ ($[0.96, 1.24]$), and
\texttt{tcpconnlat}'s CI lies strictly below ($[0.39, 0.48]$). This sub-$1\times$
ratio is a genuine code-generation artifact we confirmed in the bytecode: the C object
zero-initializes the per-event struct on every invocation before the
\texttt{TCP\_SYN\_SENT} guard, whereas \texttt{rustc}/LLVM moves that
initialization after the guard. Because \texttt{tcp\_rcv\_state\_process} fires on
every state transition but only proceeds for \texttt{TCP\_SYN\_SENT}, most
invocations exit early and skip the additional work entirely.
% Our runtime validation experiments also validate the safety gaps \toolname{} closes. In
% \texttt{opensnoop}, the C side leaves the \texttt{ts} field unwritten and
% emits stack residue, while the Rust translation emits $\texttt{ts}{=}0$. We ignore this field on both sides since the divergence is intentional
% and confirms the safety pass enforces zeroing on a production program.

\begin{table}[t]
\centering
\small
\setlength{\tabcolsep}{6pt}
\caption{Runtime overhead and pass rate of Rust translations over $N{=}100$ paired trials across 10 programs.}
\label{tab:runtime-overhead}
\begin{tabular}{cccc}
\toprule
\multirow{2}{*}{\textbf{Program}} & \multirow{2}{*}{\textbf{Pass}}
  & \multicolumn{2}{c}{\textbf{Runtime Overhead}} \\
\cmidrule(lr){3-4}
                    &                    & \textbf{mean}         & \textbf{95\% CI} \\
\midrule
\texttt{bootstrap}  & 100/100            & $1.47\times$          & $[1.33, 1.62]$ \\
\texttt{execsnoop}  & 100/100            & $1.33\times$          & $[1.17, 1.51]$ \\
\texttt{exitsnoop}  & 100/100            & $1.60\times$          & $[1.36, 1.87]$ \\
\texttt{opensnoop}  & 100/100            & $1.86\times$          & $[1.44, 2.42]$ \\
\texttt{sigsnoop}   & 100/100            & $1.24\times$          & $[1.17, 1.30]$ \\
\texttt{solisten}   & 100/100            & $1.14\times$          & $[1.01, 1.29]$ \\
\texttt{statsnoop}  & 100/100            & $1.89\times$          & $[1.55, 2.31]$ \\
\texttt{syncsnoop}  & 100/100            & $1.17\times$          & $[1.08, 1.27]$ \\
\texttt{tcpconnlat} & 100/100            & $0.44\times$          & $[0.39, 0.48]$ \\
\texttt{tcplife}    & 100/100            & $1.09\times$          & $[0.96, 1.24]$ \\
\bottomrule
\end{tabular}
\end{table}

\section{Related Work}
\label{sec:related}

% \textit{TODO: Related work section to be completed.  Will cover:
% \begin{itemize}
%   \item eBPF verification and analysis (PREVAIL, Jitterbug, Agni)
%   \item Translation validation (CompCert, Alive2)
%   \item LLM-assisted code translation
%   \item Symbolic execution frameworks (angr, KLEE, S2E)
%   \item Rust adoption in systems programming
%   \item eBPF safety and security (SoK)
% \end{itemize}
% }

\subsection{eBPF Analysis}

A substantial body of work has tackled the soundness and analysis of
the eBPF substrate itself, rather than its migration to safer
languages.  \citet{sok-ebpf-oakland25} survey complementary paths
toward eBPF memory safety such as verifier hardening, runtime checks,
language-based approaches and identify language-based migration to Rust/Aya as a
promising but underexplored direction that \toolname{} concretizes
with formal per-program equivalence.
PREVAIL~\cite{prevail} replaces the in-kernel verifier with an
abstract-interpretation-based static analyzer. Complementary work
formally models the verifier to find soundness bugs or improve its
precision via proof-guided abstraction
refinement~\cite{agni-verifier-bugs-cav23,10.1145/3731569.3764796},
or fuzzes it with differential, state-embedding, and
specification-based oracles~\cite{bpfchecker-ccs24,verifier-state-embedding-osdi24,10.1145/3731569.3764797}.
Other axes add runtime
checks~\cite{10.1145/3694715.3695950,309490} that confine or augment
extension execution beyond what the verifier proves, hardware
isolation via Intel MPK~\cite{299591}, or proof-carrying code via an
annotation-guided toolchain~\cite{306007}. Jitterbug~\cite{jitterbug}
verifies the eBPF JIT compilers themselves.

For language-based security and symbolic execution specifically,
Rex~\cite{rex-atc25} sidesteps the in-kernel verifier entirely by
running Rust eBPF programs directly with type-system and
runtime-check-based safety, but does not address the legacy migration
problem. eBPF-SE~\cite{ebpf_se} extends KLEE~\cite{klee} at the C
source level (requiring C-side annotations and not supporting Rust),
and Serval~\cite{10.1145/3341301.3359641} lifts an interpreter into a
Rosette-based BPF verifier used to check JIT-compiler correctness on
individual BPF instructions. Both target verifier- or JIT-side
properties. Neither models BPF maps, the broader helper set, ELF
relocations, or BPF-to-BPF subprogram calls. On the synthesis side,
K2~\cite{10.1145/3672197.3673434} generates safe and efficient
packet-processing eBPF programs from high-level specifications, and
KEN/KGent~\cite{ken_ebpf} drives natural-language-to-eBPF code
generation with an LLM agent. None of these systems translate
between the C and Rust eBPF dialects or check semantic equivalence
between two eBPF objects.

% Recent paper that discussed some related works:
% \url{https://pure.royalholloway.ac.uk/ws/portalfiles/portal/71513373/llm4code26.pdf}
% use that to make sure we are up to date.
% eBPF papers reading list: https://pchaigno.github.io/bpf/2025/01/07/research-papers-bpf.html

\subsection{C-to-Rust Translation}

\paragraph{Rule-based approaches.}
C2Rust~\cite{c2rust} produces a faithful AST-level transliteration
from C that is functionally correct but pervasively unsafe.  A series
of static-analysis post-processors --- Laertes~\cite{10.1145/3485498},
Crown~\cite{zhang2023ownershipguidedcrust},
Concrat~\cite{hong2023concratautomaticctorustlock},
Forcrat~\cite{hong2025forcratautomaticioapi}, and
PR2~\cite{gao2025pr2peepholerawpointer} --- apply ownership inference,
pointer lifting, or API-specific rewrites (locks, I/O) on top of
C2Rust output to reduce unsafe usage, but retain unsafe in the general
case and provide no semantic equivalence guarantee.

\paragraph{LLM-based and hybrid approaches.}
A growing line of work translates C to safer Rust with LLMs, either
end-to-end~\cite{eniser2024translatingrealworldcodellms,
shiraishi2024contextawarecodesegmentationctorust,
wang2025evoc2rustskeletonguidedframeworkprojectlevel,
farrukh2025safetransllmassistedtranspilationc,
cai2025rustmapprojectscalectorustmigration,
hong2024tagtagtranslatingcs}
or by combining LLMs with static analysis or differential testing on
top of a C2Rust skeleton~\cite{nitin2025c2saferrusttransformingcprojects,
zhou2025llmdrivenmultisteptranslationc,
shetty2024syzygydualcodetestc}.  ReCodeAgent~\cite{ibrahimzada2026recodeagentmultiagentworkflowlanguageagnostic} proposes a multi-agent coding workflow for
language-agnostic translation.  Most closely related on the
verification axis is VERT~\cite{yang2024vertverifiedequivalentrust},
which lifts the C source through Wasm into an oracle Rust program and checks bounded equivalence between the oracle and LLM-produced Rust.
However, VERT's Wasm path does not exist for eBPF (which compiles to a
distinct bytecode with its own helpers, maps, and verifier
constraints), and its bounded model checking does not scale to
programs with complex data structures.

With the exception of VERT, the systems above establish equivalence
through differential fuzzing or generated test cases rather than a
formal proof, so silent semantic divergences that aren't surfaced by
the test workload escape detection.  More importantly, none of them target
eBPF: their toolchains assume libc, std, and a hosted runtime, none of which are available to in-kernel eBPF programs.  \toolname{} closes both gaps --- formal symbolic equivalence as the success criterion, and a translation pipeline that produces eBPF-valid Aya Rust --- within a single autonomous loop.

\section{Limitations}
\label{sec:discussion}

Symbolic execution and equivalence checking are only as faithful as their modeling. Our angr backend caps path exploration at $n{=}50{,}000$, models 56~helper stubs, eBPF maps, and limits ringbuf reservations to 512~bytes for tractable equivalence checking. 
% A program that exercises behavior beyond these caps may be incorrectly admitted or rejected. 
Furthermore, we rely on monolithic symbolic execution over the full eBPF bytecode to verify equivalence. This times out on larger programs with 6/115 translations remaining unverified. 
% We hypothesize this is due to path explosion in programs with deeply nested branching, large constant tables, or long write traces against tracked map state. 
Additionally, our static safety policy is a hand-curated set of pattern-based rules. While it strictly enforces every rule it covers, it cannot anticipate idiomaticity issues we have not yet encountered. Another limitation of our approach is that we detect dropped atomicity via a static bytecode scan that compares atomic-opcode totals between the C and Rust binaries, which is a coarse heuristic rather than a per-site atomicity proof. Finally, layered verification narrows the trusted computing base but does not empty it: the Rust toolchain and Aya are themselves software with their own bugs that can be exploited~\cite{gerhorst2025mitigating}, so like the kernel verifier they are not infallible, and we do not claim to eliminate every defect. Our claim is more modest; pairing \texttt{rustc} and Aya with the kernel verifier gives two independent checks instead of one, making both the migrated programs and the eBPF loading process more robust than relying on either layer alone.

\section{Conclusion}
\label{sec:conclusion}

We presented \toolname, the first system for formally verified automated migration of eBPF programs from C to Rust. It combines LLM-driven translation with bytecode-level symbolic equivalence checking and a static safe-Aya analyzer, producing translations that are formally equivalent to the original C program across all inputs and free of the residual unsafe-Rust patterns. Across 115~real-world eBPF programs, \toolname~generates 109 (94.8\,\%) verified translations and catches nine bugs. Future work includes optimizing for performance metrics (bytecode size, runtime) into the feedback loop, compositional verification at the function level to scale to larger programs, and targeting alternative eBPF frontends like Go, Python, and Java.

% For future work, one may explore strategies to incorporate performance optimization techniques like bytecode size and execution time into the feedback loop. Additionally, compositional reasoning over individual functions or basic blocks would enable scaling up to much larger and more complicated eBPF programs. Lastly, since our eBPF symbolic execution and equivalence engine is frontend-agnostic, one may explore translation strategies across other languages like Go, Python, and Java. 

%% ----------------------------------------------------------------
%%  Bibliography
%% ----------------------------------------------------------------
\bibliographystyle{abbrvnat}
\bibliography{references}

@Inbook{mccarthy1993towards,
author="McCarthy, J.",
editor="Colburn, Timothy R.
and Fetzer, James H.
and Rankin, Terry L.",
title="Towards a Mathematical Science of Computation",
bookTitle="Program Verification: Fundamental Issues in Computer Science",
year="1993",
publisher="Springer Netherlands",
address="Dordrecht",
pages="35--56",
isbn="978-94-011-1793-7",
doi="10.1007/978-94-011-1793-7_2",
url="https://doi.org/10.1007/978-94-011-1793-7_2"
}

@INPROCEEDINGS {sok-ebpf-oakland25,
author = { Huang, Kaiming and Payer, Mathias and Qian, Zhiyun and Sampson, Jack and Tan, Gang and Jaeger, Trent },
booktitle = { 2025 IEEE Symposium on Security and Privacy (SP) },
title = {{ SoK: Challenges and Paths Toward Memory Safety for eBPF }},
year = {2025},
volume = {},
ISSN = {},
pages = {848-866},
doi = {10.1109/SP61157.2025.00134},
url = {https://doi.ieeecomputersociety.org/10.1109/SP61157.2025.00134},
publisher = {IEEE Computer Society},
address = {Los Alamitos, CA, USA},
month =May}

@misc{aya,
  author       = {{Aya Contributors}},
  title        = {Aya: An {eBPF} Library for {Rust}},
  year         = {2024},
  howpublished = {\url{https://github.com/aya-rs/aya}},
}

@inproceedings{angr,
  author={Shoshitaishvili, Yan and Wang, Ruoyu and Salls, Christopher and Stephens, Nick and Polino, Mario and Dutcher, Andrew and Grosen, John and Feng, Siji and Hauser, Christophe and Kruegel, Christopher and Vigna, Giovanni},
  booktitle={2016 IEEE Symposium on Security and Privacy (SP)}, 
  title={SOK: (State of) The Art of War: Offensive Techniques in Binary Analysis}, 
  year={2016},
  volume={},
  number={},
  pages={138-157},
  doi={10.1109/SP.2016.17}
}

@misc{libbpf-tools,
  author       = {{BCC Contributors}},
  title        = {libbpf-tools: {BPF} {CO-RE} Tools},
  year         = {2024},
  howpublished = {\url{https://github.com/iovisor/bcc/tree/master/libbpf-tools}},
}

@inproceedings{xdp,
  author = {H\o{}iland-J\o{}rgensen, Toke and Brouer, Jesper Dangaard and Borkmann, Daniel and Fastabend, John and Herbert, Tom and Ahern, David and Miller, David},
title = {The eXpress data path: fast programmable packet processing in the operating system kernel},
year = {2018},
isbn = {9781450360807},
publisher = {Association for Computing Machinery},
address = {New York, NY, USA},
url = {https://doi.org/10.1145/3281411.3281443},
doi = {10.1145/3281411.3281443},
booktitle = {Proceedings of the 14th International Conference on Emerging Networking EXperiments and Technologies},
pages = {54–66},
numpages = {13},
location = {Heraklion, Greece},
series = {CoNEXT '18}
}

@inproceedings{prevail,
author = {Gershuni, Elazar and Amit, Nadav and Gurfinkel, Arie and Narodytska, Nina and Navas, Jorge A. and Rinetzky, Noam and Ryzhyk, Leonid and Sagiv, Mooly},
title = {Simple and precise static analysis of untrusted Linux kernel extensions},
year = {2019},
isbn = {9781450367127},
publisher = {Association for Computing Machinery},
address = {New York, NY, USA},
url = {https://doi.org/10.1145/3314221.3314590},
doi = {10.1145/3314221.3314590},
booktitle = {Proceedings of the 40th ACM SIGPLAN Conference on Programming Language Design and Implementation},
pages = {1069–1084},
numpages = {16},
location = {Phoenix, AZ, USA},
series = {PLDI 2019}
}

@misc{bcc,
  author       = {{IO Visor Project}},
  title        = {{BCC}: Tools for {BPF}-based {Linux} {IO} Analysis, Networking,
                  Monitoring, and More},
  year         = {2024},
  howpublished = {\url{https://github.com/iovisor/bcc}},
}

@inproceedings{parca-agent,
  author       = {{Polar Signals}},
  title        = {Parca Agent: {eBPF}-based Always-On Continuous Profiler},
  year         = {2024},
  howpublished = {\url{https://github.com/parca-dev/parca-agent}},
}

@misc{ebpfio,
  author       = {{eBPF.io}},
  title        = {{eBPF}},
  howpublished = {\url{https://ebpf.io/}},
}

@misc{gregg2019bpf,
  author       = {Brendan Gregg},
  title        = {{BPF}: A New Type of Software},
  year         = {2019},
  howpublished = {\url{https://www.brendangregg.com/blog/2019-12-02/bpf-a-new-type-of-software.html}},
}

@misc{llm-translation-formal,
  title        = {{LLM}-Based Code Translation Needs Formal Compositional
                  Reasoning},
  author       = {Divyam Anshumaan and Tej Chajed and Varun Chandrasekaran
                  and Alvin Cheung and Sarthak Choudhary and Adwait Godbole
                  and Somesh Jha and Nils Palumbo and Elizabeth Polgreen
                  and Sanjit A. Seshia and Tianyang Zhou},
  year         = {2025},
  url          = {https://openreview.net/forum?id=wGj8LU2EOf},
}

@inproceedings{rex-atc25,
  author = {Jia, Jinghao and Qin, Ruowen and Craun, Milo and Lukiyanov, Egor and Bansal, Ayush and Phan, Minh and Le, Michael V. and Franke, Hubertus and Jamjoom, Hani and Xu, Tianyin and Williams, Dan},
title = {Rex: closing the language-verifier gap with safe and usable kernel extensions},
year = {2025},
isbn = {978-1-939133-48-9},
publisher = {USENIX Association},
address = {USA},
booktitle = {Proceedings of the 2025 USENIX Conference on Usenix Annual Technical Conference},
articleno = {20},
numpages = {18},
location = {Boston, MA, USA},
series = {USENIX ATC '25}
}

@inproceedings{verifier-untenable-hotos23,
author = {Jia, Jinghao and Sahu, Raj and Oswald, Adam and Williams, Dan and Le, Michael V. and Xu, Tianyin},
title = {Kernel extension verification is untenable},
year = {2023},
isbn = {9798400701955},
publisher = {Association for Computing Machinery},
address = {New York, NY, USA},
url = {https://doi.org/10.1145/3593856.3595892},
doi = {10.1145/3593856.3595892},
booktitle = {Proceedings of the 19th Workshop on Hot Topics in Operating Systems},
pages = {150–157},
numpages = {8},
location = {Providence, RI, USA},
series = {HotOS '23}
}

@inproceedings{agni-verifier-bugs-cav23,
 author = {Vishwanathan, Harishankar and Shachnai, Matan and Narayana, Srinivas and Nagarakatte, Santosh},
title = {Verifying the Verifier: eBPF Range Analysis Verification},
year = {2023},
isbn = {978-3-031-37708-2},
publisher = {Springer-Verlag},
address = {Berlin, Heidelberg},
url = {https://doi.org/10.1007/978-3-031-37709-9_12},
doi = {10.1007/978-3-031-37709-9_12},
booktitle = {Computer Aided Verification: 35th International Conference, CAV 2023, Paris, France, July 17–22, 2023, Proceedings, Part III},
pages = {226–251},
numpages = {26},
location = {Paris, France}
}

@inproceedings{verifier-state-embedding-osdi24,
  author    = {Sun, Hao and Su, Zhendong},
  title     = {Validating the {eBPF} Verifier via State Embedding},
  booktitle = {Proceedings of the 18th USENIX Symposium on Operating
               Systems Design and Implementation (OSDI)},
  year      = {2024},
  pages     = {615--628},
  publisher = {USENIX Association},
}

@inproceedings{bpfchecker-ccs24,
  author = {Peng, Chaoyuan and Jiang, Muhui and Wu, Lei and Zhou, Yajin},
title = {Toss a Fault to BpfChecker: Revealing Implementation Flaws for eBPF runtimes with Differential Fuzzing},
year = {2024},
isbn = {9798400706363},
publisher = {Association for Computing Machinery},
address = {New York, NY, USA},
url = {https://doi.org/10.1145/3658644.3690237},
doi = {10.1145/3658644.3690237},
booktitle = {Proceedings of the 2024 on ACM SIGSAC Conference on Computer and Communications Security},
pages = {3928–3942},
numpages = {15},
location = {Salt Lake City, UT, USA},
series = {CCS '24}
}

@misc{shetty2024syzygydualcodetestc,
      title={Syzygy: Dual Code-Test C to (safe) Rust Translation using LLMs and Dynamic Analysis}, 
      author={Manish Shetty and Naman Jain and Adwait Godbole and Sanjit A. Seshia and Koushik Sen},
      year={2024},
      eprint={2412.14234},
      archivePrefix={arXiv},
      primaryClass={cs.SE},
      url={https://arxiv.org/abs/2412.14234}, 
}

@misc{zhou2025llmdrivenmultisteptranslationc,
      title={SACTOR: LLM-Driven Correct and Idiomatic C to Rust Translation with Static Analysis and FFI-Based Verification}, 
      author={Tianyang Zhou and Ziyi Zhang and Haowen Lin and Somesh Jha and Mihai Christodorescu and Kirill Levchenko and Varun Chandrasekaran},
      year={2025},
      eprint={2503.12511},
      archivePrefix={arXiv},
      primaryClass={cs.SE},
      url={https://arxiv.org/abs/2503.12511},
}

@misc{nitin2025c2saferrusttransformingcprojects,
      title={C2SaferRust: Transforming C Projects into Safer Rust with NeuroSymbolic Techniques}, 
      author={Vikram Nitin and Rahul Krishna and Luiz Lemos do Valle and Baishakhi Ray},
      year={2025},
      eprint={2501.14257},
      archivePrefix={arXiv},
      primaryClass={cs.SE},
      url={https://arxiv.org/abs/2501.14257}, 
}

@misc{shiraishi2024contextawarecodesegmentationctorust,
      title     = {SmartC2Rust: Iterative, Feedback-Driven C-to-Rust Translation via Large Language Models for Safety and Equivalence},
  author    = {Momoko Shiraishi and Yinzhi Cao and Takahiro Shinagawa},
  booktitle = {Proceedings of the 48th IEEE/ACM International Conference on Software Engineering (ICSE 2026)},
  month     = Apr,
  year      = 2026
}

@misc{eniser2024translatingrealworldcodellms,
      title={Towards Translating Real-World Code with LLMs: A Study of Translating to Rust}, 
      author={Hasan Ferit Eniser and Hanliang Zhang and Cristina David and Meng Wang and Maria Christakis and Brandon Paulsen and Joey Dodds and Daniel Kroening},
      year={2024},
      eprint={2405.11514},
      archivePrefix={arXiv},
      primaryClass={cs.SE},
      url={https://arxiv.org/abs/2405.11514}, 
}

@misc{yang2024vertverifiedequivalentrust,
      title={VERT: Verified Equivalent Rust Transpilation with Large Language Models as Few-Shot Learners}, 
      author={Aidan Z. H. Yang and Yoshiki Takashima and Brandon Paulsen and Josiah Dodds and Daniel Kroening},
      year={2024},
      eprint={2404.18852},
      archivePrefix={arXiv},
      primaryClass={cs.PL},
      url={https://arxiv.org/abs/2404.18852}, 
}

@misc{zhang2023ownershipguidedcrust,
      title={Ownership guided C to Rust translation}, 
      author={Hanliang Zhang and Cristina David and Yijun Yu and Meng Wang},
      year={2023},
      eprint={2303.10515},
      archivePrefix={arXiv},
      primaryClass={cs.PL},
      url={https://arxiv.org/abs/2303.10515}, 
}

@article{10.1145/3485498,
author = {Emre, Mehmet and Schroeder, Ryan and Dewey, Kyle and Hardekopf, Ben},
title = {Translating C to safer Rust},
year = {2021},
issue_date = {October 2021},
publisher = {Association for Computing Machinery},
address = {New York, NY, USA},
volume = {5},
number = {OOPSLA},
url = {https://doi.org/10.1145/3485498},
doi = {10.1145/3485498},
journal = {Proc. ACM Program. Lang.},
month = oct,
articleno = {121},
numpages = {29}
}

@misc{hong2023concratautomaticctorustlock,
      title={Concrat: An Automatic C-to-Rust Lock API Translator for Concurrent Programs}, 
      author={Jaemin Hong and Sukyoung Ryu},
      year={2023},
      eprint={2301.10943},
      archivePrefix={arXiv},
      primaryClass={cs.SE},
      url={https://arxiv.org/abs/2301.10943}, 
}

@misc{gao2025pr2peepholerawpointer,
      title={Raw Pointer Rewriting with LLMs for Translating C to Safer Rust}, 
      author={Yifei Gao and Chengpeng Wang and Pengxiang Huang and Xuwei Liu and Mingwei Zheng and Xiangyu Zhang},
      year={2026},
      eprint={2505.04852},
      archivePrefix={arXiv},
      primaryClass={cs.SE},
      url={https://arxiv.org/abs/2505.04852}, 
}

@misc{farrukh2025safetransllmassistedtranspilationc,
      title={SafeTrans: LLM-assisted Transpilation from C to Rust}, 
      author={Muhammad Farrukh and Baris Coskun and Tapti Palit and Michalis Polychronakis},
      year={2026},
      eprint={2505.10708},
      archivePrefix={arXiv},
      primaryClass={cs.CR},
      url={https://arxiv.org/abs/2505.10708}, 
}

@misc{wang2025evoc2rustskeletonguidedframeworkprojectlevel,
      title={EVOC2RUST: A Skeleton-guided Framework for Project-Level C-to-Rust Translation}, 
      author={Chaofan Wang and Tingrui Yu and Jie Wang and Dong Chen and Wenrui Zhang and Yuling Shi and Xiaodong Gu and Beijun Shen},
      year={2025},
      eprint={2508.04295},
      archivePrefix={arXiv},
      primaryClass={cs.SE},
      url={https://arxiv.org/abs/2508.04295}, 
}

@misc{hong2025forcratautomaticioapi,
      title={Forcrat: Automatic I/O API Translation from C to Rust via Origin and Capability Analysis}, 
      author={Jaemin Hong and Sukyoung Ryu},
      year={2025},
      eprint={2506.01427},
      archivePrefix={arXiv},
      primaryClass={cs.SE},
      url={https://arxiv.org/abs/2506.01427}, 
}

@inproceedings{10.1145/3672197.3673434,
author = {Zheng, Yusheng and Yang, Yiwei and Chen, Maolin and Quinn, Andrew},
title = {Kgent: Kernel Extensions Large Language Model Agent},
year = {2024},
isbn = {9798400707124},
publisher = {Association for Computing Machinery},
address = {New York, NY, USA},
url = {https://doi.org/10.1145/3672197.3673434},
doi = {10.1145/3672197.3673434},
booktitle = {Proceedings of the ACM SIGCOMM 2024 Workshop on EBPF and Kernel Extensions},
pages = {30–36},
numpages = {7},
location = {Sydney, NSW, Australia},
series = {eBPF '24}
}

@misc{c2rust,
	author = {{Galois}},
        day = {14},
	month = {08},
	title = {{C2Rust}},
	year = {2018},
	url = {https://galois.com/blog/2018/08/c2rust/}
}

@inproceedings{bmc-cache,
     author = {Yoann Ghigoff and Julien Sopena and Kahina Lazri and Antoine Blin and Gilles Muller},
    title = {{BMC}: Accelerating Memcached using Safe In-kernel Caching and Pre-stack Processing},
    booktitle = {18th USENIX Symposium on Networked Systems Design and Implementation (NSDI 21)},
    year = {2021},
    isbn = {978-1-939133-21-2},
    pages = {487--501},
    url = {https://www.usenix.org/conference/nsdi21/presentation/ghigoff},
    publisher = {USENIX Association},
    month = apr
}

@misc{bcc-issue-3175,
  title        = {\texttt{bpf\_probe\_read\_user} returns error ($-14$) and \texttt{opensnoop} emits empty filenames},
  howpublished = {\url{https://github.com/iovisor/bcc/issues/3175}},
  year         = {2020},
}

@misc{bcc-issue-622,
  title        = {Can't read in struct fields even with \texttt{bpf\_probe\_read}: \texttt{-EFAULT}},
  howpublished = {\url{https://github.com/iovisor/bcc/issues/622}},
  year         = {2016},
}

@misc{bcc-issue-2245,
  title        = {\texttt{bpf\_probe\_read()} returned \texttt{-14}},
  howpublished = {\url{https://github.com/iovisor/bcc/issues/2245}},
  year         = {2019},
}

@misc{tetragon-issue-3728,
  title        = {Tetragon does not raise an event when the resolved value is null},
  howpublished = {\url{https://github.com/cilium/tetragon/issues/3728}},
  year         = {2025},
}

@misc{katran,
  author       = {{Meta Incubator}},
  title        = {Katran: A High Performance Layer 4 Load Balancer},
  howpublished = {\url{https://github.com/facebookincubator/katran}},
  year         = {2018},
}

@inproceedings{crab,
  author    = {Kogias, Marios and Iyer, Rishabh and Bugnion, Edouard},
  title     = {Bypassing the Load Balancer Without Regrets},
  year      = {2020},
  isbn      = {9781450381376},
  publisher = {Association for Computing Machinery},
  address   = {New York, NY, USA},
  url       = {https://doi.org/10.1145/3419111.3421304},
  doi       = {10.1145/3419111.3421304},
  booktitle = {Proceedings of the 11th ACM Symposium on Cloud Computing},
  pages     = {193--207},
  numpages  = {15},
  location  = {Virtual Event, USA},
  series    = {SoCC '20}
}

@inproceedings{hxdp,
  author    = {Spaziani Brunella, Marco and Belocchi, Giacomo and Bonola, Marco
               and Pontarelli, Salvatore and Siracusano, Giuseppe and Bianchi, Giuseppe
               and Cammarano, Aniello and Palumbo, Alessandro and Petrucci, Luca
               and Bifulco, Roberto},
  title     = {{hXDP}: Efficient Software Packet Processing on {FPGA} {NICs}},
  booktitle = {14th USENIX Symposium on Operating Systems Design and
               Implementation (OSDI 20)},
  year      = {2020},
  isbn      = {978-1-939133-19-9},
  pages     = {973--990},
  url       = {https://www.usenix.org/conference/osdi20/presentation/brunella},
  publisher = {USENIX Association},
  month     = nov
}

@misc{fluvia,
  author       = {{NTT Communications}},
  title        = {Fluvia: {IPFIX} Exporter Using {XDP}},
  howpublished = {\url{https://github.com/nttcom/fluvia}},
  year         = {2023},
}

@misc{hercules,
  author       = {{Network Security Group, ETH Z\"{u}rich}},
  title        = {Hercules: High-Speed Bulk Data Transfer Using {XDP}},
  howpublished = {\url{https://github.com/netsec-ethz/hercules}},
  year         = {2023},
}

@misc{dae,
  author       = {{daeuniverse}},
  title        = {{dae}: A Linux High-Performance Transparent Proxy Based on {eBPF}},
  howpublished = {\url{https://github.com/daeuniverse/dae}},
  year         = {2023},
}

@misc{suricata,
  author       = {{Open Information Security Foundation}},
  title        = {Suricata: Open Source {IDS}/{IPS}/{NSM} Engine},
  howpublished = {\url{https://suricata.io/}},
  year         = {2024},
}

@inproceedings{cacheext,
author = {Zussman, Tal and Zarkadas, Ioannis and Carin, Jeremy and Cheng, Andrew and Franke, Hubertus and Pfefferle, Jonas and Cidon, Asaf},
title = {cache\_ext: Customizing the Page Cache with eBPF},
year = {2025},
isbn = {9798400718700},
publisher = {Association for Computing Machinery},
address = {New York, NY, USA},
url = {https://doi.org/10.1145/3731569.3764820},
doi = {10.1145/3731569.3764820},
booktitle = {Proceedings of the ACM SIGOPS 31st Symposium on Operating Systems Principles},
pages = {462–478},
numpages = {17},
location = {Lotte Hotel World, Seoul, Republic of Korea},
series = {SOSP '25}
}

@inproceedings{ken_ebpf,
author = {Zheng, Yusheng and Yang, Yiwei and Chen, Maolin and Quinn, Andrew},
title = {Kgent: Kernel Extensions Large Language Model Agent},
year = {2024},
isbn = {9798400707124},
publisher = {Association for Computing Machinery},
address = {New York, NY, USA},
url = {https://doi.org/10.1145/3672197.3673434},
doi = {10.1145/3672197.3673434},
booktitle = {Proceedings of the ACM SIGCOMM 2024 Workshop on EBPF and Kernel Extensions},
pages = {30–36},
numpages = {7},
location = {Sydney, NSW, Australia},
series = {eBPF '24}
}

@misc{merrill2026terminalbenchbenchmarkingagentshard,
      title={Terminal-Bench: Benchmarking Agents on Hard, Realistic Tasks in Command Line Interfaces}, 
      author={Mike A. Merrill and Alexander G. Shaw and Nicholas Carlini and Boxuan Li and Harsh Raj and Ivan Bercovich and Lin Shi and Jeong Yeon Shin and Thomas Walshe and E. Kelly Buchanan and Junhong Shen and Guanghao Ye and Haowei Lin and Jason Poulos and Maoyu Wang and Marianna Nezhurina and Jenia Jitsev and Di Lu and Orfeas Menis Mastromichalakis and Zhiwei Xu and Zizhao Chen and Yue Liu and Robert Zhang and Leon Liangyu Chen and Anurag Kashyap and Jan-Lucas Uslu and Jeffrey Li and Jianbo Wu and Minghao Yan and Song Bian and Vedang Sharma and Ke Sun and Steven Dillmann and Akshay Anand and Andrew Lanpouthakoun and Bardia Koopah and Changran Hu and Etash Guha and Gabriel H. S. Dreiman and Jiacheng Zhu and Karl Krauth and Li Zhong and Niklas Muennighoff and Robert Amanfu and Shangyin Tan and Shreyas Pimpalgaonkar and Tushar Aggarwal and Xiangning Lin and Xin Lan and Xuandong Zhao and Yiqing Liang and Yuanli Wang and Zilong Wang and Changzhi Zhou and David Heineman and Hange Liu and Harsh Trivedi and John Yang and Junhong Lin and Manish Shetty and Michael Yang and Nabil Omi and Negin Raoof and Shanda Li and Terry Yue Zhuo and Wuwei Lin and Yiwei Dai and Yuxin Wang and Wenhao Chai and Shang Zhou and Dariush Wahdany and Ziyu She and Jiaming Hu and Zhikang Dong and Yuxuan Zhu and Sasha Cui and Ahson Saiyed and Arinbjörn Kolbeinsson and Jesse Hu and Christopher Michael Rytting and Ryan Marten and Yixin Wang and Alex Dimakis and Andy Konwinski and Ludwig Schmidt},
      year={2026},
      eprint={2601.11868},
      archivePrefix={arXiv},
      primaryClass={cs.SE},
      url={https://arxiv.org/abs/2601.11868}, 
}

@misc{jimenez2024swebenchlanguagemodelsresolve,
      title={SWE-bench: Can Language Models Resolve Real-World GitHub Issues?}, 
      author={Carlos E. Jimenez and John Yang and Alexander Wettig and Shunyu Yao and Kexin Pei and Ofir Press and Karthik Narasimhan},
      year={2024},
      eprint={2310.06770},
      archivePrefix={arXiv},
      primaryClass={cs.CL},
      url={https://arxiv.org/abs/2310.06770}, 
}

@misc{miserendino2025swelancerfrontierllmsearn,
      title={SWE-Lancer: Can Frontier LLMs Earn \$1 Million from Real-World Freelance Software Engineering?}, 
      author={Samuel Miserendino and Michele Wang and Tejal Patwardhan and Johannes Heidecke},
      year={2025},
      eprint={2502.12115},
      archivePrefix={arXiv},
      primaryClass={cs.LG},
      url={https://arxiv.org/abs/2502.12115}, 
}

@inproceedings {jitterbug,
author = {Luke Nelson and Jacob Van Geffen and Emina Torlak and Xi Wang},
title = {Specification and verification in the field: Applying formal methods to {BPF} just-in-time compilers in the Linux kernel},
booktitle = {14th USENIX Symposium on Operating Systems Design and Implementation (OSDI 20)},
year = {2020},
isbn = {978-1-939133-19-9},
pages = {41--61},
url = {https://www.usenix.org/conference/osdi20/presentation/nelson},
publisher = {USENIX Association},
month = nov
}

@misc{ibrahimzada2026recodeagentmultiagentworkflowlanguageagnostic,
      title={ReCodeAgent: A Multi-Agent Workflow for Language-agnostic Translation and Validation of Large-scale Repositories}, 
      author={Ali Reza Ibrahimzada and Brandon Paulsen and Daniel Kroening and Reyhaneh Jabbarvand},
      year={2026},
      eprint={2604.07341},
      archivePrefix={arXiv},
      primaryClass={cs.SE},
      url={https://arxiv.org/abs/2604.07341}, 
}

@misc{cai2025rustmapprojectscalectorustmigration,
      title={RustMap: Towards Project-Scale C-to-Rust Migration via Program Analysis and LLM}, 
      author={Xuemeng Cai and Jiakun Liu and Xiping Huang and Yijun Yu and Haitao Wu and Chunmiao Li and Bo Wang and Imam Nur Bani Yusuf and Lingxiao Jiang},
      year={2025},
      eprint={2503.17741},
      archivePrefix={arXiv},
      primaryClass={cs.SE},
      url={https://arxiv.org/abs/2503.17741}, 
}

@misc{hong2024tagtagtranslatingcs,
      title={To Tag, or Not to Tag: Translating C's Unions to Rust's Tagged Unions}, 
      author={Jaemin Hong and Sukyoung Ryu},
      year={2024},
      eprint={2408.11418},
      archivePrefix={arXiv},
      primaryClass={cs.SE},
      url={https://arxiv.org/abs/2408.11418}, 
}

@inproceedings {ebpf_se,
author = {Rishabh Iyer and Katerina Argyraki and George Candea},
title = {Performance Interfaces for Network Functions},
booktitle = {19th USENIX Symposium on Networked Systems Design and Implementation (NSDI 22)},
year = {2022},
isbn = {978-1-939133-27-4},
address = {Renton, WA},
pages = {567--584},
url = {https://www.usenix.org/conference/nsdi22/presentation/iyer},
publisher = {USENIX Association},
month = apr
}

@inproceedings{klee,
author = {Cadar, Cristian and Dunbar, Daniel and Engler, Dawson},
title = {KLEE: unassisted and automatic generation of high-coverage tests for complex systems programs},
year = {2008},
publisher = {USENIX Association},
address = {USA},
booktitle = {Proceedings of the 8th USENIX Conference on Operating Systems Design and Implementation},
pages = {209–224},
numpages = {16},
location = {San Diego, California},
series = {OSDI'08}
}

@inproceedings{10.1145/3694715.3695950,
author = {Dwivedi, Kumar Kartikeya and Iyer, Rishabh and Kashyap, Sanidhya},
title = {Fast, Flexible, and Practical Kernel Extensions},
year = {2024},
isbn = {9798400712517},
publisher = {Association for Computing Machinery},
address = {New York, NY, USA},
url = {https://doi.org/10.1145/3694715.3695950},
doi = {10.1145/3694715.3695950},
booktitle = {Proceedings of the ACM SIGOPS 30th Symposium on Operating Systems Principles},
pages = {249–264},
numpages = {16},
location = {Austin, TX, USA},
series = {SOSP '24}
}

@inproceedings {306007,
author = {Xiwei Wu and Yueyang Feng and Tianyi Huang and Xiaoyang Lu and Shengkai Lin and Lihan Xie and Shizhen Zhao and Qinxiang Cao},
title = {{VEP}: A Two-stage Verification Toolchain for Full {eBPF} Programmability},
booktitle = {22nd USENIX Symposium on Networked Systems Design and Implementation (NSDI 25)},
year = {2025},
isbn = {978-1-939133-46-5},
address = {Philadelphia, PA},
pages = {277--299},
url = {https://www.usenix.org/conference/nsdi25/presentation/wu-xiwei},
publisher = {USENIX Association},
month = apr
}

@inproceedings{10.1145/3731569.3764796,
author = {Sun, Hao and Su, Zhendong},
title = {Prove It to the Kernel: Precise Extension Analysis via Proof-Guided Abstraction Refinement},
year = {2025},
isbn = {9798400718700},
publisher = {Association for Computing Machinery},
address = {New York, NY, USA},
url = {https://doi.org/10.1145/3731569.3764796},
doi = {10.1145/3731569.3764796},
booktitle = {Proceedings of the ACM SIGOPS 31st Symposium on Operating Systems Principles},
pages = {736–751},
numpages = {16},
location = {Lotte Hotel World, Seoul, Republic of Korea},
series = {SOSP '25}
}

@inproceedings{10.1145/3731569.3764797,
author = {Lyu, Tao and Dwivedi, Kumar Kartikeya and Bourgeat, Thomas and Payer, Mathias and Xu, Meng and Kashyap, Sanidhya},
title = {eBPF Misbehavior Detection: Fuzzing with a Specification-Based Oracle},
year = {2025},
isbn = {9798400718700},
publisher = {Association for Computing Machinery},
address = {New York, NY, USA},
url = {https://doi.org/10.1145/3731569.3764797},
doi = {10.1145/3731569.3764797},
booktitle = {Proceedings of the ACM SIGOPS 31st Symposium on Operating Systems Principles},
pages = {701–718},
numpages = {18},
location = {Lotte Hotel World, Seoul, Republic of Korea},
series = {SOSP '25}
}

@inproceedings {309490,
author = {Hao Sun and Zhendong Su},
title = {Approximation Enforced Execution of Untrusted Linux Kernel Extensions},
booktitle = {34th USENIX Security Symposium (USENIX Security 25)},
year = {2025},
isbn = {978-1-939133-52-6},
address = {Seattle, WA},
pages = {7467--7485},
url = {https://www.usenix.org/conference/usenixsecurity25/presentation/sun-hao},
publisher = {USENIX Association},
month = aug
}

@inproceedings{10.1145/3341301.3359641,
author = {Nelson, Luke and Bornholt, James and Gu, Ronghui and Baumann, Andrew and Torlak, Emina and Wang, Xi},
title = {Scaling symbolic evaluation for automated verification of systems code with Serval},
year = {2019},
isbn = {9781450368735},
publisher = {Association for Computing Machinery},
address = {New York, NY, USA},
url = {https://doi.org/10.1145/3341301.3359641},
doi = {10.1145/3341301.3359641},
booktitle = {Proceedings of the 27th ACM Symposium on Operating Systems Principles},
pages = {225–242},
numpages = {18},
location = {Huntsville, Ontario, Canada},
series = {SOSP '19}
}

@inproceedings {299591,
author = {Hongyi Lu and Shuai Wang and Yechang Wu and Wanning He and Fengwei Zhang},
title = {{MOAT}: Towards Safe {BPF} Kernel Extension},
booktitle = {33rd USENIX Security Symposium (USENIX Security 24)},
year = {2024},
isbn = {978-1-939133-44-1},
address = {Philadelphia, PA},
pages = {1153--1170},
url = {https://www.usenix.org/conference/usenixsecurity24/presentation/lu-hongyi},
publisher = {USENIX Association},
month = aug
}

@misc{gerhorst2025mitigating,
  author       = {Luis Gerhorst and Timo H\"{o}nig},
  title        = {Mitigating Bugs in the {Linux} {eBPF} Verifier using {Rust}- or {PREVAIL}-based Layered Verification},
  howpublished = {Talk at FOSDEM 2025},
  month        = feb,
  year         = {2025},
  note         = {Brussels, Belgium, February 1, 2025},
  url          = {https://archive.fosdem.org/2025/events/attachments/fosdem-2025-5095-mitigating-bugs-in-the-linux-ebpf-verifier-using-rust-or-prevail-based-layered-verification/slides/238243/25-01-31_MJ0MluR.pdf}
}

@inproceedings{10.1145/3748355.3748374,
author = {Dimobi, Chinecherem and Tiwari, Rahul and Ji, Zhengjie and Williams, Dan},
title = {BPFflow - Preventing information leaks from eBPF},
year = {2025},
isbn = {9798400720840},
publisher = {Association for Computing Machinery},
address = {New York, NY, USA},
url = {https://doi.org/10.1145/3748355.3748374},
doi = {10.1145/3748355.3748374},
booktitle = {Proceedings of the 3rd Workshop on EBPF and Kernel Extensions},
pages = {87–93},
numpages = {7},
location = {Coimbra, Portugal},
series = {eBPF '25}
}

%% ----------------------------------------------------------------
%%  Appendix
%% ----------------------------------------------------------------
\appendix
\section{Open Science}
\label{sec:open_science}

% We open-source and release all artifacts needed to evaluate the contributions of this paper at the following anonymous URL: \url{https://anonymous.4open.science/r/heimdall-6D3D/}

We open-source and release all artifacts needed to evaluate the contributions of this paper at the following URL: \url{https://github.com/vdasu/heimdall}

\begin{itemize}                                                                                        
    \item The \toolname{} translation pipeline.

    \item The eBPF backend for angr.

    \item The Z3 equivalence checker.

    \item The safe-Aya static analysis engine.

    \item The dataset of 115 valid Aya-supported eBPF programs and the 109 verified Rust translations.                                                
\end{itemize}  

\section{Ethical Considerations}
\label{sec:ethical_considerations}

Our framework translates C libbpf programs into Aya Rust to address safety and security concerns that the eBPF verifier misses in C programs. Additionally, our paper discovers information leaks in open-source tools. We notify the developers in line with responsible disclosure practices.

\section{Generative AI Usage}
\label{sec:gen_ai_usage}

Generative AI tools (Claude Code and Codex) were used to assist in the implementation of the approach, polish the writing of all sections of the paper (grammatical errors, clarity, etc.), and create the LaTeX code for listings, tables, and figures. All AI generated content has been validated by the author(s).

\section{Full Code Listings for Real World Bugs}
\label{sec:full_code_listings}

This appendix lists the source of every real-world program discussed in
\Cref{sec:verifier-gaps}.  Each listing is the original source with the bug site
annotated inline via a \texttt{/* BUG: ... */} comment.  For BMC, whose kernel
program spans several hundred lines, only the two relevant TC entry functions
are shown. For Katran, whose load balancer is far too large to reproduce, only
a short excerpt around the assignment that carries the bug is shown.  Unrelated
logic is commented with \texttt{/* ... */}.

\begin{lstlisting}[style=cstyle,
caption={\texttt{bashreadline.bpf.c} (uninitialized stack state and unchecked
helper return).  \texttt{struct str\_t} is declared without zero-initialization,
the return value of \texttt{bpf\_probe\_read\_user\_str} is ignored, and the
complete 84-byte struct is emitted.},
label=lst:bashreadline-full]
/* SPDX-License-Identifier: GPL-2.0 */
/* Copyright (c) 2021 Facebook */
#include "vmlinux.h"
#include <bpf/bpf_helpers.h>
#include <bpf/bpf_tracing.h>

#define TASK_COMM_LEN 16
#define MAX_LINE_SIZE 80

struct str_t {
    __u32 pid;
    char str[MAX_LINE_SIZE];
};

struct {
    __uint(type, BPF_MAP_TYPE_PERF_EVENT_ARRAY);
    __uint(key_size, sizeof(__u32));
    __uint(value_size, sizeof(__u32));
} events SEC(".maps");

SEC("uretprobe/readline")
int BPF_URETPROBE(printret, const void *ret) {
    struct str_t data;   /* BUG 1: str_t is never zero-initialized */
    char comm[TASK_COMM_LEN];
    u32 pid;

    if (!ret)
        return 0;

    bpf_get_current_comm(&comm, sizeof(comm));
    if (comm[0] != 'b' || comm[1] != 'a' || comm[2] != 's' || comm[3] != 'h' || comm[4] != 0 )
        return 0;

    pid = bpf_get_current_pid_tgid() >> 32;
    data.pid = pid;
    /* BUG 2: return value ignored; on failure data.str stays uninitialized */
    bpf_probe_read_user_str(&data.str, sizeof(data.str), ret);

    /* emits all 84 bytes, including the uninitialized suffix */
    bpf_perf_event_output(ctx, &events, BPF_F_CURRENT_CPU, &data, sizeof(data));

    return 0;
};

char LICENSE[] SEC("license") = "GPL";
\end{lstlisting}

\begin{lstlisting}[style=cstyle,
caption={\texttt{offcputime.bpf.c} (uninitialized stack state).  \texttt{val} is
declared on the stack without zero-initialization; only a NUL-terminated prefix
of \texttt{val.comm} is written, yet the whole struct is stored into the
\texttt{info} map, so the bytes after the NUL are stale stack residue.},
label=lst:offcputime-full]
// SPDX-License-Identifier: GPL-2.0
// Copyright (c) 2021 Wenbo Zhang
#include "vmlinux.h"
#include <bpf/bpf_helpers.h>
#include <bpf/bpf_core_read.h>
#include <bpf/bpf_tracing.h>
#include "core_fixes.bpf.h"

#define PF_KTHREAD      0x00200000  /* I am a kernel thread */
#define MAX_ENTRIES     10240
#define TASK_COMM_LEN   16
#define MAX_PID_NR      30
#define MAX_TID_NR      30

struct key_t {
    __u32 pid;
    __u32 tgid;
    int user_stack_id;
    int kern_stack_id;
};

struct val_t {
    __u64 delta;
    char comm[TASK_COMM_LEN];
};

const volatile bool kernel_threads_only = false;
const volatile bool user_threads_only = false;
const volatile __u64 max_block_us = -1;
const volatile __u64 min_block_us = 1;
const volatile bool filter_by_tgid = false;
const volatile bool filter_by_pid = false;
const volatile long state = -1;

struct internal_key {
    u64 start_ts;
    struct key_t key;
};

struct {
    __uint(type, BPF_MAP_TYPE_HASH);
    __type(key, u32);
    __type(value, struct internal_key);
    __uint(max_entries, MAX_ENTRIES);
} start SEC(".maps");

struct {
    __uint(type, BPF_MAP_TYPE_STACK_TRACE);
    __uint(key_size, sizeof(u32));
} stackmap SEC(".maps");

struct {
    __uint(type, BPF_MAP_TYPE_HASH);
    __type(key, struct key_t);
    __type(value, struct val_t);
    __uint(max_entries, MAX_ENTRIES);
} info SEC(".maps");

struct {
    __uint(type, BPF_MAP_TYPE_HASH);
    __type(key, u32);
    __type(value, u8);
    __uint(max_entries, MAX_PID_NR);
} tgids SEC(".maps");

struct {
    __uint(type, BPF_MAP_TYPE_HASH);
    __type(key, u32);
    __type(value, u8);
    __uint(max_entries, MAX_TID_NR);
} pids SEC(".maps");

static bool allow_record(struct task_struct *t)
{
    u32 tgid = BPF_CORE_READ(t, tgid);
    u32 pid = BPF_CORE_READ(t, pid);

    if (filter_by_tgid && !bpf_map_lookup_elem(&tgids, &tgid))
        return false;
    if (filter_by_pid && !bpf_map_lookup_elem(&pids, &pid))
        return false;
    if (user_threads_only && (BPF_CORE_READ(t, flags) & PF_KTHREAD))
        return false;
    else if (kernel_threads_only && !(BPF_CORE_READ(t, flags) & PF_KTHREAD))
        return false;
    if (state != -1 && get_task_state(t) != state)
        return false;
    return true;
}

static int handle_sched_switch(void *ctx, bool preempt, struct task_struct *prev, struct task_struct *next)
{
    struct internal_key *i_keyp, i_key;
    struct val_t *valp, val;   /* BUG: val is never zero-initialized */
    s64 delta;
    u32 pid;

    if (allow_record(prev)) {
        pid = BPF_CORE_READ(prev, pid);
        /* To distinguish idle threads of different cores */
        if (!pid)
            pid = bpf_get_smp_processor_id();
        i_key.key.pid = pid;
        i_key.key.tgid = BPF_CORE_READ(prev, tgid);
        i_key.start_ts = bpf_ktime_get_ns();

        if (BPF_CORE_READ(prev, flags) & PF_KTHREAD)
            i_key.key.user_stack_id = -1;
        else
            i_key.key.user_stack_id = bpf_get_stackid(ctx, &stackmap, BPF_F_USER_STACK);
        i_key.key.kern_stack_id = bpf_get_stackid(ctx, &stackmap, 0);
        bpf_map_update_elem(&start, &pid, &i_key, 0);
        /* writes only a NUL-terminated prefix of comm */
        bpf_probe_read_kernel_str(&val.comm, sizeof(prev->comm), BPF_CORE_READ(prev, comm));
        val.delta = 0;
        /* BUG: stores the full struct; comm's suffix past the NUL is stale stack */
        bpf_map_update_elem(&info, &i_key.key, &val, BPF_NOEXIST);
    }

    pid = BPF_CORE_READ(next, pid);
    i_keyp = bpf_map_lookup_elem(&start, &pid);
    if (!i_keyp)
        return 0;
    delta = (s64)(bpf_ktime_get_ns() - i_keyp->start_ts);
    if (delta < 0)
        goto cleanup;
    delta /= 1000U;
    if (delta < min_block_us || delta > max_block_us)
        goto cleanup;
    valp = bpf_map_lookup_elem(&info, &i_keyp->key);
    if (!valp)
        goto cleanup;
    __sync_fetch_and_add(&valp->delta, delta);

cleanup:
    bpf_map_delete_elem(&start, &pid);
    return 0;
}

SEC("tp_btf/sched_switch")
int BPF_PROG(sched_switch, bool preempt, struct task_struct *prev, struct task_struct *next)
{
    return handle_sched_switch(ctx, preempt, prev, next);
}

SEC("raw_tp/sched_switch")
int BPF_PROG(sched_switch_raw, bool preempt, struct task_struct *prev, struct task_struct *next)
{
    return handle_sched_switch(ctx, preempt, prev, next);
}

char LICENSE[] SEC("license") = "GPL";
\end{lstlisting}

\begin{lstlisting}[style=cstyle,
caption={\texttt{filetop.bpf.c} (pointer-size mismatch, oversized read).
\texttt{get\_file\_path} passes the 4096-byte destination capacity as the size
of a raw kernel read of a short filename, so the helper over-reads adjacent
kernel memory into the map value.},
label=lst:filetop-full]
/* SPDX-License-Identifier: (LGPL-2.1 OR BSD-2-Clause) */
/* Copyright (c) 2021 Hengqi Chen */
#include "vmlinux.h"
#include <bpf/bpf_helpers.h>
#include <bpf/bpf_core_read.h>
#include <bpf/bpf_tracing.h>

#define MAX_ENTRIES 10240
#define PATH_MAX    4096
#define TASK_COMM_LEN   16
#define S_IFMT      00170000
#define S_IFSOCK    0140000
#define S_IFLNK     0120000
#define S_IFREG     0100000
#define S_IFBLK     0060000
#define S_IFDIR     0040000
#define S_IFCHR     0020000
#define S_IFIFO     0010000
#define S_ISUID     0004000
#define S_ISGID     0002000
#define S_ISVTX     0001000

#define S_ISLNK(m)  (((m) & S_IFMT) == S_IFLNK)
#define S_ISREG(m)  (((m) & S_IFMT) == S_IFREG)
#define S_ISDIR(m)  (((m) & S_IFMT) == S_IFDIR)
#define S_ISCHR(m)  (((m) & S_IFMT) == S_IFCHR)
#define S_ISBLK(m)  (((m) & S_IFMT) == S_IFBLK)
#define S_ISFIFO(m) (((m) & S_IFMT) == S_IFIFO)
#define S_ISSOCK(m) (((m) & S_IFMT) == S_IFSOCK)

enum op {
    READ,
    WRITE,
};

struct file_id {
    __u64 inode;
    __u32 dev;
    __u32 rdev;
    __u32 pid;
    __u32 tid;
};

struct file_stat {
    __u64 reads;
    __u64 read_bytes;
    __u64 writes;
    __u64 write_bytes;
    __u32 pid;
    __u32 tid;
    char filename[PATH_MAX];
    char comm[TASK_COMM_LEN];
    char type;
};

const volatile pid_t target_pid = 0;
const volatile bool regular_file_only = true;
static struct file_stat zero_value = {};

struct {
    __uint(type, BPF_MAP_TYPE_HASH);
    __uint(max_entries, MAX_ENTRIES);
    __type(key, struct file_id);
    __type(value, struct file_stat);
} entries SEC(".maps");

static void get_file_path(struct file *file, char *buf, size_t size)
{
    struct qstr dname;

    dname = BPF_CORE_READ(file, f_path.dentry, d_name);
    /* BUG: raw read copies `size` bytes (the 4096-byte destination capacity),
       ignoring the NUL in a short dname.name, and over-reads adjacent memory */
    bpf_probe_read_kernel(buf, size, dname.name);
}

static int probe_entry(struct pt_regs *ctx, struct file *file, size_t count, enum op op)
{
    __u64 pid_tgid = bpf_get_current_pid_tgid();
    __u32 pid = pid_tgid >> 32;
    __u32 tid = (__u32)pid_tgid;
    int mode;
    struct file_id key = {};
    struct file_stat *valuep;

    if (target_pid && target_pid != pid)
        return 0;

    mode = BPF_CORE_READ(file, f_inode, i_mode);
    if (regular_file_only && !S_ISREG(mode))
        return 0;

    key.dev = BPF_CORE_READ(file, f_inode, i_sb, s_dev);
    key.rdev = BPF_CORE_READ(file, f_inode, i_rdev);
    key.inode = BPF_CORE_READ(file, f_inode, i_ino);
    key.pid = pid;
    key.tid = tid;
    valuep = bpf_map_lookup_elem(&entries, &key);
    if (!valuep) {
        bpf_map_update_elem(&entries, &key, &zero_value, BPF_ANY);
        valuep = bpf_map_lookup_elem(&entries, &key);
        if (!valuep)
            return 0;
        valuep->pid = pid;
        valuep->tid = tid;
        bpf_get_current_comm(&valuep->comm, sizeof(valuep->comm));
        /* size = sizeof(valuep->filename) = 4096 capacity */
        get_file_path(file, valuep->filename, sizeof(valuep->filename));
        if (S_ISREG(mode)) {
            valuep->type = 'R';
        } else if (S_ISSOCK(mode)) {
            valuep->type = 'S';
        } else {
            valuep->type = 'O';
        }
    }
    if (op == READ) {
        valuep->reads++;
        valuep->read_bytes += count;
    } else {    /* op == WRITE */
        valuep->writes++;
        valuep->write_bytes += count;
    }
    return 0;
};

SEC("kprobe/vfs_read")
int BPF_KPROBE(vfs_read_entry, struct file *file, char *buf, size_t count, loff_t *pos)
{
    return probe_entry(ctx, file, count, READ);
}

SEC("kprobe/vfs_write")
int BPF_KPROBE(vfs_write_entry, struct file *file, const char *buf, size_t count, loff_t *pos)
{
    return probe_entry(ctx, file, count, WRITE);
}

char LICENSE[] SEC("license") = "Dual BSD/GPL";
\end{lstlisting}

\begin{lstlisting}[style=cstyle,
caption={\texttt{gethostlatency.bpf.c} (pointer-size mismatch, oversized
read).  The fixed-size \texttt{bpf\_probe\_read\_user} copies the full 80-byte
\texttt{host} buffer from a user string instead of the string length, so bytes
past the NUL are captured and later emitted through the \texttt{events} perf
array.},
label=lst:gethostlatency-full]
/* SPDX-License-Identifier: GPL-2.0 */
/* Copyright (c) 2021 Hengqi Chen */
#include "vmlinux.h"
#include <bpf/bpf_helpers.h>
#include <bpf/bpf_core_read.h>
#include <bpf/bpf_tracing.h>

#define MAX_ENTRIES 10240
#define TASK_COMM_LEN   16
#define HOST_LEN    80

struct event {
    __u64 time;
    __u32 pid;
    char comm[TASK_COMM_LEN];
    char host[HOST_LEN];
};

const volatile pid_t target_pid = 0;

struct {
    __uint(type, BPF_MAP_TYPE_HASH);
    __uint(max_entries, MAX_ENTRIES);
    __type(key, __u32);
    __type(value, struct event);
} starts SEC(".maps");

struct {
    __uint(type, BPF_MAP_TYPE_PERF_EVENT_ARRAY);
    __uint(key_size, sizeof(__u32));
    __uint(value_size, sizeof(__u32));
} events SEC(".maps");

static int probe_entry(struct pt_regs *ctx)
{
    if (!PT_REGS_PARM1(ctx))
        return 0;

    __u64 pid_tgid = bpf_get_current_pid_tgid();
    __u32 pid = pid_tgid >> 32;
    __u32 tid = (__u32)pid_tgid;
    struct event event = {};

    if (target_pid && target_pid != pid)
        return 0;

    event.time = bpf_ktime_get_ns();
    event.pid = pid;
    bpf_get_current_comm(&event.comm, sizeof(event.comm));
    /* BUG: reads the full 80-byte host buffer, not the string length; the
       fixed-size read does not stop at the NUL and captures trailing bytes */
    bpf_probe_read_user(&event.host, sizeof(event.host), (void *)PT_REGS_PARM1(ctx));
    bpf_map_update_elem(&starts, &tid, &event, BPF_ANY);
    return 0;
}

static int probe_return(struct pt_regs *ctx)
{
    __u32 tid = (__u32)bpf_get_current_pid_tgid();
    struct event *eventp;

    eventp = bpf_map_lookup_elem(&starts, &tid);
    if (!eventp)
        return 0;

    /* update time from timestamp to delta */
    eventp->time = bpf_ktime_get_ns() - eventp->time;
    /* the trailing bytes of host captured above are emitted here */
    bpf_perf_event_output(ctx, &events, BPF_F_CURRENT_CPU, eventp, sizeof(*eventp));
    bpf_map_delete_elem(&starts, &tid);
    return 0;
}

SEC("uprobe")
int BPF_UPROBE(handle_entry)
{
    return probe_entry(ctx);
}

SEC("uretprobe")
int BPF_URETPROBE(handle_return)
{
    return probe_return(ctx);
}

char LICENSE[] SEC("license") = "GPL";
\end{lstlisting}

\begin{lstlisting}[style=cstyle,
caption={\texttt{slabratetop.bpf.c} (pointer-size mismatch, oversized read).
The fixed-size \texttt{bpf\_probe\_read\_kernel} copies the full 32-byte
\texttt{name} buffer from a kernel string instead of the string length, so bytes
past the NUL are stored in the \texttt{slab\_entries} map value.},
label=lst:slabratetop-full]
/* SPDX-License-Identifier: (LGPL-2.1 OR BSD-2-Clause) */
/* Copyright (c) 2022 Rong Tao */
#include "vmlinux.h"
#include <bpf/bpf_helpers.h>
#include <bpf/bpf_core_read.h>
#include <bpf/bpf_tracing.h>

#define MAX_ENTRIES 10240
#define CACHE_NAME_SIZE 32

struct slabrate_info {
    char name[CACHE_NAME_SIZE];
    __u64 count;
    __u64 size;
};

const volatile pid_t target_pid = 0;

static struct slabrate_info slab_zero_value = {};

struct {
    __uint(type, BPF_MAP_TYPE_HASH);
    __uint(max_entries, MAX_ENTRIES);
    __type(key, char *);
    __type(value, struct slabrate_info);
} slab_entries SEC(".maps");

static int probe_entry(struct kmem_cache *cachep)
{
    __u64 pid_tgid = bpf_get_current_pid_tgid();
    __u32 pid = pid_tgid >> 32;
    struct slabrate_info *valuep;
    const char *name = BPF_CORE_READ(cachep, name);

    if (target_pid && target_pid != pid)
        return 0;

    valuep = bpf_map_lookup_elem(&slab_entries, &name);
    if (!valuep) {
        bpf_map_update_elem(&slab_entries, &name, &slab_zero_value, BPF_ANY);
        valuep = bpf_map_lookup_elem(&slab_entries, &name);
        if (!valuep)
            return 0;
        /* BUG: reads the full 32-byte name buffer, not the string length; the
           fixed-size read does not stop at the NUL and stores trailing bytes */
        bpf_probe_read_kernel(&valuep->name, sizeof(valuep->name), name);
    }

    valuep->count++;
    valuep->size += BPF_CORE_READ(cachep, size);

    return 0;
}

SEC("kprobe/kmem_cache_alloc")
int BPF_KPROBE(kmem_cache_alloc, struct kmem_cache *cachep)
{
    return probe_entry(cachep);
}

SEC("kprobe/kmem_cache_alloc_noprof")
int BPF_KPROBE(kmem_cache_alloc_noprof, struct kmem_cache *cachep)
{
    return probe_entry(cachep);
}

char LICENSE[] SEC("license") = "Dual BSD/GPL";
\end{lstlisting}

\begin{lstlisting}[style=cstyle,
caption={\texttt{biostacks.bpf.c} (pointer-size mismatch, undersized length).
\texttt{sizeof(\&i\_rqinfop->rqinfo.comm)} is the 8-byte pointer size rather
than the 16-byte array size, so \texttt{bpf\_get\_current\_comm} writes only
half of \texttt{comm}.},
label=lst:biostacks-full]
// SPDX-License-Identifier: GPL-2.0
// Copyright (c) 2020 Wenbo Zhang
#include "vmlinux.h"
#include <bpf/bpf_helpers.h>
#include <bpf/bpf_core_read.h>
#include <bpf/bpf_tracing.h>
#include "bits.bpf.h"
#include "maps.bpf.h"
#include "core_fixes.bpf.h"

#define MAX_ENTRIES 10240

#define DISK_NAME_LEN   32
#define TASK_COMM_LEN   16
#define MAX_SLOTS   20
#define MAX_STACK   20

#define MINORBITS   20
#define MINORMASK   ((1U << MINORBITS) - 1)

#define MKDEV(ma, mi)   (((ma) << MINORBITS) | (mi))

struct rqinfo {
    __u32 pid;
    int kern_stack_size;
    __u64 kern_stack[MAX_STACK];
    char comm[TASK_COMM_LEN];
    __u32 dev;
};

struct hist {
    __u32 slots[MAX_SLOTS];
};

const volatile bool targ_ms = false;
const volatile bool filter_dev = false;
const volatile __u32 targ_dev = -1;

struct internal_rqinfo {
    u64 start_ts;
    struct rqinfo rqinfo;
};

struct {
    __uint(type, BPF_MAP_TYPE_HASH);
    __uint(max_entries, MAX_ENTRIES);
    __type(key, struct request *);
    __type(value, struct internal_rqinfo);
} rqinfos SEC(".maps");

struct {
    __uint(type, BPF_MAP_TYPE_HASH);
    __uint(max_entries, MAX_ENTRIES);
    __type(key, struct rqinfo);
    __type(value, struct hist);
} hists SEC(".maps");

static struct hist zero;

static __always_inline
int trace_start(void *ctx, struct request *rq, bool merge_bio)
{
    struct internal_rqinfo *i_rqinfop = NULL, i_rqinfo = {};
    struct gendisk *disk = get_disk(rq);
    u32 dev;

    dev = disk ? MKDEV(BPF_CORE_READ(disk, major),
            BPF_CORE_READ(disk, first_minor)) : 0;
    if (filter_dev && targ_dev != dev)
        return 0;

    if (merge_bio)
        i_rqinfop = bpf_map_lookup_elem(&rqinfos, &rq);
    if (!i_rqinfop)
        i_rqinfop = &i_rqinfo;

    i_rqinfop->start_ts = bpf_ktime_get_ns();
    i_rqinfop->rqinfo.pid = bpf_get_current_pid_tgid();
    i_rqinfop->rqinfo.kern_stack_size =
        bpf_get_stack(ctx, i_rqinfop->rqinfo.kern_stack,
            sizeof(i_rqinfop->rqinfo.kern_stack), 0);
    bpf_get_current_comm(&i_rqinfop->rqinfo.comm,
            /* BUG: sizeof(&comm) is the 8-byte pointer size, not 16 */
            sizeof(&i_rqinfop->rqinfo.comm));
    i_rqinfop->rqinfo.dev = dev;

    if (i_rqinfop == &i_rqinfo)
        bpf_map_update_elem(&rqinfos, &rq, i_rqinfop, 0);
    return 0;
}

static __always_inline
int trace_done(void *ctx, struct request *rq)
{
    u64 slot, ts = bpf_ktime_get_ns();
    struct internal_rqinfo *i_rqinfop;
    struct hist *histp;
    s64 delta;

    i_rqinfop = bpf_map_lookup_elem(&rqinfos, &rq);
    if (!i_rqinfop)
        return 0;
    delta = (s64)(ts - i_rqinfop->start_ts);
    if (delta < 0)
        goto cleanup;
    histp = bpf_map_lookup_or_try_init(&hists, &i_rqinfop->rqinfo, &zero);
    if (!histp)
        goto cleanup;
    if (targ_ms)
        delta /= 1000000U;
    else
        delta /= 1000U;
    slot = log2l(delta);
    if (slot >= MAX_SLOTS)
        slot = MAX_SLOTS - 1;
    __sync_fetch_and_add(&histp->slots[slot], 1);

cleanup:
    bpf_map_delete_elem(&rqinfos, &rq);
    return 0;
}

SEC("kprobe/blk_account_io_merge_bio")
int BPF_KPROBE(blk_account_io_merge_bio, struct request *rq)
{
    return trace_start(ctx, rq, true);
}

SEC("fentry/blk_account_io_start")
int BPF_PROG(blk_account_io_start, struct request *rq)
{
    return trace_start(ctx, rq, false);
}

SEC("fentry/blk_account_io_done")
int BPF_PROG(blk_account_io_done, struct request *rq)
{
    return trace_done(ctx, rq);
}

SEC("tp_btf/block_io_start")
int BPF_PROG(block_io_start, struct request *rq)
{
    return trace_start(ctx, rq, false);
}

SEC("tp_btf/block_io_done")
int BPF_PROG(block_io_done, struct request *rq)
{
    return trace_done(ctx, rq);
}

char LICENSE[] SEC("license") = "GPL";
\end{lstlisting}

\begin{lstlisting}[style=cstyle,
caption={\texttt{biosnoop.bpf.c} (pointer-size mismatch, undersized length).  As
in \texttt{biostacks}, \texttt{sizeof(\&piddata.comm)} yields the 8-byte pointer
size, so only half of \texttt{comm} is written before it is stored and later
emitted.},
label=lst:biosnoop-full]
// SPDX-License-Identifier: GPL-2.0
// Copyright (c) 2020 Wenbo Zhang
#include "vmlinux.h"
#include <bpf/bpf_helpers.h>
#include <bpf/bpf_core_read.h>
#include <bpf/bpf_tracing.h>
#include "core_fixes.bpf.h"

#define MAX_ENTRIES 10240
#define DISK_NAME_LEN   32
#define TASK_COMM_LEN   16
#define RWBS_LEN    8

#define MINORBITS   20
#define MINORMASK   ((1U << MINORBITS) - 1)

#define MKDEV(ma, mi)   (((ma) << MINORBITS) | (mi))

struct event {
    char comm[TASK_COMM_LEN];
    __u64 delta;
    __u64 qdelta;
    __u64 ts;
    __u64 sector;
    __u32 len;
    __u32 pid;
    __u32 cmd_flags;
    __u32 dev;
};

const volatile bool filter_cg = false;
const volatile bool targ_queued = false;
const volatile bool filter_dev = false;
const volatile __u32 targ_dev = 0;
const volatile __u64 min_ns = 0;

extern __u32 LINUX_KERNEL_VERSION __kconfig;

struct {
    __uint(type, BPF_MAP_TYPE_CGROUP_ARRAY);
    __type(key, u32);
    __type(value, u32);
    __uint(max_entries, 1);
} cgroup_map SEC(".maps");

struct piddata {
    char comm[TASK_COMM_LEN];
    u32 pid;
};

struct {
    __uint(type, BPF_MAP_TYPE_HASH);
    __uint(max_entries, MAX_ENTRIES);
    __type(key, struct request *);
    __type(value, struct piddata);
} infobyreq SEC(".maps");

struct stage {
    u64 insert;
    u64 issue;
    __u32 dev;
};

struct {
    __uint(type, BPF_MAP_TYPE_HASH);
    __uint(max_entries, MAX_ENTRIES);
    __type(key, struct request *);
    __type(value, struct stage);
} start SEC(".maps");

struct {
    __uint(type, BPF_MAP_TYPE_PERF_EVENT_ARRAY);
    __uint(key_size, sizeof(u32));
    __uint(value_size, sizeof(u32));
} events SEC(".maps");

static __always_inline
int trace_pid(struct request *rq)
{
    u64 id = bpf_get_current_pid_tgid();
    struct piddata piddata = {};

    piddata.pid = id >> 32;
    /* BUG: sizeof(&piddata.comm) is 8, so only half of comm is written */
    bpf_get_current_comm(&piddata.comm, sizeof(&piddata.comm));
    bpf_map_update_elem(&infobyreq, &rq, &piddata, 0);
    return 0;
}

SEC("fentry/blk_account_io_start")
int BPF_PROG(blk_account_io_start, struct request *rq)
{
    if (filter_cg && !bpf_current_task_under_cgroup(&cgroup_map, 0))
        return 0;

    return trace_pid(rq);
}

SEC("tp_btf/block_io_start")
int BPF_PROG(block_io_start, struct request *rq)
{
    if (filter_cg && !bpf_current_task_under_cgroup(&cgroup_map, 0))
        return 0;

    return trace_pid(rq);
}

SEC("kprobe/blk_account_io_merge_bio")
int BPF_KPROBE(blk_account_io_merge_bio, struct request *rq)
{
    if (filter_cg && !bpf_current_task_under_cgroup(&cgroup_map, 0))
        return 0;

    return trace_pid(rq);
}

static __always_inline
int trace_rq_start(struct request *rq, bool insert)
{
    struct stage *stagep, stage = {};
    u64 ts = bpf_ktime_get_ns();

    stagep = bpf_map_lookup_elem(&start, &rq);
    if (!stagep) {
        struct gendisk *disk = get_disk(rq);

        stage.dev = disk ? MKDEV(BPF_CORE_READ(disk, major),
                BPF_CORE_READ(disk, first_minor)) : 0;
        if (filter_dev && targ_dev != stage.dev)
            return 0;
        stagep = &stage;
    }
    if (insert)
        stagep->insert = ts;
    else
        stagep->issue = ts;
    if (stagep == &stage)
        bpf_map_update_elem(&start, &rq, stagep, 0);
    return 0;
}

SEC("tp_btf/block_rq_insert")
int BPF_PROG(block_rq_insert)
{
    if (filter_cg && !bpf_current_task_under_cgroup(&cgroup_map, 0))
        return 0;

    /* commit a54895fa changed the tracepoint argument list; see
       https://github.com/torvalds/linux/commit/a54895fa */
    if (LINUX_KERNEL_VERSION >= KERNEL_VERSION(5, 10, 137))
        return trace_rq_start((void *)ctx[0], true);
    else
        return trace_rq_start((void *)ctx[1], true);
}

SEC("tp_btf/block_rq_issue")
int BPF_PROG(block_rq_issue)
{
    if (filter_cg && !bpf_current_task_under_cgroup(&cgroup_map, 0))
        return 0;

    /* commit a54895fa changed the tracepoint argument list; see
       https://github.com/torvalds/linux/commit/a54895fa */
    if (LINUX_KERNEL_VERSION >= KERNEL_VERSION(5, 10, 137))
        return trace_rq_start((void *)ctx[0], false);
    else
        return trace_rq_start((void *)ctx[1], false);
}

SEC("tp_btf/block_rq_complete")
int BPF_PROG(block_rq_complete, struct request *rq, int error,
         unsigned int nr_bytes)
{
    if (filter_cg && !bpf_current_task_under_cgroup(&cgroup_map, 0))
        return 0;

    u64 ts = bpf_ktime_get_ns();
    struct piddata *piddatap;
    struct event event = {};
    struct stage *stagep;
    s64 delta;

    stagep = bpf_map_lookup_elem(&start, &rq);
    if (!stagep)
        return 0;
    delta = (s64)(ts - stagep->issue);
    if (delta < 0 || delta < min_ns)
        goto cleanup;
    piddatap = bpf_map_lookup_elem(&infobyreq, &rq);
    if (!piddatap) {
        event.comm[0] = '?';
    } else {
        __builtin_memcpy(&event.comm, piddatap->comm,
                sizeof(event.comm));
        event.pid = piddatap->pid;
    }
    event.delta = delta;
    if (targ_queued && BPF_CORE_READ(rq, q, elevator)) {
        if (!stagep->insert)
            event.qdelta = -1; /* missed or don't insert entry */
        else
            event.qdelta = stagep->issue - stagep->insert;
    }
    event.ts = ts;
    event.sector = BPF_CORE_READ(rq, __sector);
    event.len = BPF_CORE_READ(rq, __data_len);
    event.cmd_flags = BPF_CORE_READ(rq, cmd_flags);
    event.dev = stagep->dev;
    bpf_perf_event_output(ctx, &events, BPF_F_CURRENT_CPU, &event,
            sizeof(event));

cleanup:
    bpf_map_delete_elem(&start, &rq);
    bpf_map_delete_elem(&infobyreq, &rq);
    return 0;
}

char LICENSE[] SEC("license") = "GPL";
\end{lstlisting}

\begin{lstlisting}[style=cstyle,
caption={Relevant TC entry functions of \texttt{bmc\_kern.c} (action-code
namespace confusion).  Both functions are attached to the TC (\texttt{sched\_cls})
hook but return \texttt{XDP\_PASS} (numeric 2) on early exits, which TC
interprets as \texttt{TC\_ACT\_SHOT} and drops the frame; cache-update internals
are elided with \texttt{/* ... */}.},
label=lst:bmc-full]
SEC("bmc_tx_filter")   /* attached to the TC (sched_cls) hook */
int bmc_tx_filter_main(struct __sk_buff *skb)
{
    void *data_end = (void *)(long)skb->data_end;
    void *data     = (void *)(long)skb->data;
    struct ethhdr *eth = data;
    struct iphdr *ip = data + sizeof(*eth);
    struct udphdr *udp = data + sizeof(*eth) + sizeof(*ip);
    char *payload = data + sizeof(*eth) + sizeof(*ip) + sizeof(*udp)
                    + sizeof(struct memcached_udp_header);
    unsigned int zero = 0;

    /* ... drop oversized frames: return TC_ACT_OK ... */

    if (ip + 1 > data_end)
        return XDP_PASS;   /* BUG: 2 == TC_ACT_SHOT under TC, drops the frame */

    if (ip->protocol != IPPROTO_UDP)
        return TC_ACT_OK;

    if (udp + 1 > data_end)
        return TC_ACT_OK;

    __be16 sport = udp->source;
    if (sport == htons(11211) && payload+5+1 <= data_end
        && payload[0] == 'V' && payload[1] == 'A' && payload[2] == 'L'
        && payload[3] == 'U' && payload[4] == 'E' && payload[5] == ' ') {
        struct bmc_stats *stats = bpf_map_lookup_elem(&map_stats, &zero);
        if (!stats)
            return XDP_PASS;   /* BUG: should be a TC_ACT_* value */
        stats->get_resp_count++;
        bpf_tail_call(skb, &map_progs_tc, BMC_PROG_TC_UPDATE_CACHE);
    }

    return TC_ACT_OK;
}

SEC("bmc_update_cache")   /* attached to the TC (sched_cls) hook */
int bmc_update_cache_main(struct __sk_buff *skb)
{
    void *data_end = (void *)(long)skb->data_end;
    void *data = (void *)(long)skb->data;
    char *payload = data + sizeof(struct ethhdr) + sizeof(struct iphdr)
                    + sizeof(struct udphdr) + sizeof(struct memcached_udp_header);
    unsigned int zero = 0;
    u32 hash = FNV_OFFSET_BASIS_32;

    /* ... compute key hash, look up and update the cache entry ... */

    if (count == 2) { /* copy OK */
        entry->valid = 1;
        entry->hash = hash;
        bpf_spin_unlock(&entry->lock);
        struct bmc_stats *stats = bpf_map_lookup_elem(&map_stats, &zero);
        if (!stats)
            return XDP_PASS;   /* BUG: should be a TC_ACT_* value */
        stats->update_count++;
    } else {
        bpf_spin_unlock(&entry->lock);
    }

    return TC_ACT_OK;
}
\end{lstlisting}

\begin{lstlisting}[style=cstyle,
caption={\texttt{balancer.bpf.c} (Katran, generic C logic bug): excerpt of
\texttt{update\_vip\_lru\_miss\_stats()}.  \texttt{proto\_match} uses \texttt{=}
instead of \texttt{==}, so it assigns \texttt{vip->proto} into the map value
(\texttt{lru\_miss\_stat\_vip->proto}) instead of comparing it.},
label=lst:katran-full]
/* balancer.bpf.c: update_vip_lru_miss_stats() */
struct vip_definition* lru_miss_stat_vip =
    bpf_map_lookup_elem(&vip_miss_stats, &vip_miss_stats_key);
if (!lru_miss_stat_vip)
    return XDP_DROP;

bool address_match = /* ... compare stored VIP address vs vip ... */;
bool port_match  = lru_miss_stat_vip->port  == vip->port;
bool proto_match = lru_miss_stat_vip->proto = vip->proto; /* BUG: '=' not '==' */
bool vip_match   = address_match && port_match && proto_match;
if (vip_match) {
    /* ... increment this real's LRU-miss counter ... */
}
\end{lstlisting}

\section{Additional Verifier-Accepted Bug Patterns}
\label{sec:appendix-additional-patterns}
\label{sec:appendix-bug-listings}

\Cref{sec:verifier-gaps} presents only the findings in real-world eBPF programs.  This appendix
covers four additional classes and two further instances of the
pointer-size mismatch, using sample programs that exercise each
\toolname{} safety rule.  The final case adapts a pointer pattern from published
verifier vulnerabilities.  \Cref{tab:bug-classes} records the enforcement
mechanism for all nine classes discussed in the paper. All of these patterns have been validated on kernel 6.8.0-106-generic (Ubuntu) with clang 21.1.8 and bpftool v7.4.0 and confirm that the verifier accepts every program. 

\begin{table*}[t]
\centering
\small
\setlength{\tabcolsep}{5pt}
\caption{Source-level bug classes covered by \toolname{} and the obligation each
enforces.}
\label{tab:bug-classes}
\begin{tabular}{@{}p{.32\textwidth}p{.58\textwidth}@{}}
\toprule
\textbf{Bug class} & \textbf{Enforced obligation} \\
\midrule
Uninitialized stack state &
Complete initialization before use (\texttt{rustc}) \\
Unchecked helper return &
Explicit success or error branch (safety policy) \\
Pointer-size mismatch &
Typed or string-aware API eliminates an independent byte count \\
Action-code namespace &
Hook-bound action constants (safety policy) \\
Generic C logic bugs &
Stronger typing rejects the defect (\texttt{rustc}) \\
Hook/context mismatch &
Hook-specific Aya context type \\
Map schema confusion &
Generic map key and value types \\
Signed/error-domain confusion &
\texttt{Result} separates errors from values \\
Shifted ring-buffer pointer &
Owned entry submits its original base \\
\bottomrule
\end{tabular}
\end{table*}

\paragraph{Additional pointer-size mismatch cases.}

The production examples in \Cref{sec:verifier-gaps} pass a size larger than a
source string or smaller than a destination array.  Two further programs exhibit
the same mismatch in other outputs.  In the first
(\Cref{lst:output-size-leak-full}), the pointer names only the public prefix,
but the byte count names the enclosing object. Both objects are initialized
, and the selected stack interval is valid, so byte-level bounds checks do not
recover the developer's intended output schema.  A typed
\texttt{PerfEventArray<T>} output operation derives the size from a value of
type \texttt{T}.

\begin{lstlisting}[style=cstyle,
caption={The output pointer identifies a public prefix, while the size exposes
the enclosing private fields.},
label=lst:output-size-leak-full]
struct public_event {
    __u32 pid;
    __u32 reserved;
    __u64 timestamp;
};
struct full_event {
    struct public_event pub;
    __u64 task_ptr;
    __u64 ip_ptr;
};

SEC("kprobe/do_sys_openat2")
int output_size_leak(struct pt_regs *ctx)
{
    struct full_event evt = {};
    evt.pub.pid = bpf_get_current_pid_tgid() >> 32;
    evt.pub.timestamp = bpf_ktime_get_ns();
    evt.task_ptr = (__u64)bpf_get_current_task();
    evt.ip_ptr = PT_REGS_IP_CORE(ctx);

    /* BUG: prefix pointer, enclosing-object size. */
    bpf_perf_event_output(ctx, &events,
        BPF_F_CURRENT_CPU, &evt.pub, sizeof(evt));
    return 0;
}
\end{lstlisting}

In the second (\Cref{lst:helper-size-overwrite-full}), \texttt{filename} is 16
bytes but the helper receives the size of the complete 24-byte event, so a
successful read overwrites the adjacent \texttt{verdict} field.
Aya's slice-based read wrappers derive
the destination extent from \texttt{\&mut [u8; 16]}, and \toolname{} disallows
a raw helper call that reintroduces an independent length.

\begin{lstlisting}[style=cstyle,
caption={A valid enclosing-object extent overwrites the field adjacent to the
intended destination.},
label=lst:helper-size-overwrite-full]
struct event {
    char filename[16];
    __u64 verdict;
};

SEC("kprobe/security_file_open")
int overwrite_field(struct pt_regs *ctx)
{
    const void *src = (const void *)PT_REGS_PARM1(ctx);
    struct event evt = { .verdict = 1 };

    /* BUG: filename is 16 bytes; evt is 24 bytes. */
    bpf_probe_read_kernel(&evt.filename,
                          sizeof(evt), src);
    bpf_perf_event_output(ctx, &events,
        BPF_F_CURRENT_CPU, &evt, sizeof(evt));
    return 0;
}
\end{lstlisting}

\paragraph{Hook/context mismatch.}
C section names and C parameter types are not coupled.  The program in
\Cref{lst:wrong-ctx-full} is attached as XDP but interprets its context using
the TC \texttt{struct \_\_sk\_buff} layout.  An in-bounds load at a common
offset can pass verification while denoting a different field under
\texttt{struct xdp\_md}.  Aya's hook macro supplies an
\texttt{XdpContext}; \toolname{} rejects casts to an unrelated context type.

\begin{lstlisting}[style=cstyle,
caption={An XDP hook interprets its context using the TC layout.},
label=lst:wrong-ctx-full]
SEC("xdp")
int monitor_packets(struct __sk_buff *skb)
{
    /* Valid offset, but a different field under XDP. */
    bpf_printk("field=%u", skb->protocol);
    return XDP_PASS;
}
\end{lstlisting}

\paragraph{Map-schema confusion.}
The C helper returns \texttt{void *}; a same-width cast can reinterpret an
eight-byte map value without an out-of-bounds access.  In
\Cref{lst:map-schema-confusion-full}, the verifier can establish that the load
is within the value but cannot reconstruct the erased source schema.  Aya maps
carry \texttt{K} and \texttt{V} in their types, so the lookup result cannot be
silently treated as an unrelated value type.

\begin{lstlisting}[style=cstyle,
caption={A map value is read through an unrelated same-size source type.},
label=lst:map-schema-confusion-full]
struct conn  { __u32 src_ip, dst_ip; };
struct stats { __u64 bytes; };

struct {
    __uint(type, BPF_MAP_TYPE_ARRAY);
    __type(key, __u32);
    __type(value, struct conn);
    __uint(max_entries, 1);
} data SEC(".maps");

SEC("xdp")
int map_schema_bug(struct xdp_md *ctx)
{
    __u32 key = 0;
    struct stats *s = (struct stats *)
        bpf_map_lookup_elem(&data, &key);
    if (s)
        bpf_printk("bytes=%llu", s->bytes);
    return XDP_PASS;
}
\end{lstlisting}

\paragraph{Signed/error-domain confusion.}
Some helpers return either a nonnegative value or a negative errno.  Converting
that result directly to an unsigned map field, as in
\Cref{lst:sign-confusion-c-full}, turns failure into a large valid-looking
identifier.  The access is memory-safe and thus within the verifier's contract.
The allowed Aya wrapper exposes the operation as \texttt{Result}; \toolname{}
requires an explicit error arm before the successful identifier enters the map.

\begin{lstlisting}[style=cstyle,
caption={A negative stack-helper errno becomes an unsigned identifier.},
label=lst:sign-confusion-c-full]
struct key {
    __u32 pid;
    __u32 stack_id;
};

SEC("kprobe/finish_task_switch")
int unsigned_errno(struct pt_regs *ctx)
{
    struct key key = {};
    long id = bpf_get_stackid(ctx, &stackmap, 0);

    key.pid = bpf_get_current_pid_tgid() >> 32;
    key.stack_id = id;  /* BUG: negative errno wraps */
    bpf_map_update_elem(&counts, &key, &one, BPF_ANY);
    return 0;
}
\end{lstlisting}

\paragraph{Shifted ring-buffer pointer.}
\Cref{lst:ringbuf-shifted-pointer} models a plausible mistake in
which the developer reserves a complete record but submits a pointer to its
payload field.  Submitting a pointer other than the base returned by
\texttt{bpf\_ringbuf\_reserve} appears in both eBPF ring-buffer verifier CVEs
covered by the SoK survey~\cite{sok-ebpf-oakland25}.  CVE-2021-4204 exposed a
submission helper with no size argument for validating the pointer.
CVE-2023-2163 used unsound path pruning to admit an out-of-bounds pointer
derived from the reserved base. 

\begin{lstlisting}[style=cstyle,
caption={An accidental field pointer is submitted instead of the base returned
by the reserve helper.},
label=lst:ringbuf-shifted-pointer]
struct record {
    struct header hdr;
    struct payload body;
};

struct record *rec =
    bpf_ringbuf_reserve(&events, sizeof(*rec), 0);
if (!rec)
    return 0;
rec->hdr.pid = bpf_get_current_pid_tgid() >> 32;
fill_payload(&rec->body);

/* BUG: not the pointer returned by reserve. */
bpf_ringbuf_submit(&rec->body, 0);
\end{lstlisting}

The idiomatic Aya interface does not accept a submission pointer.  Reserving a
record produces an owned \texttt{RingBufEntry<Record>} and submitting consumes that
entry and internally uses the base pointer associated with its reservation.
Consequently, ordinary code cannot substitute a pointer to \texttt{body} at the
submission step.  \toolname{}'s safety policy rejects direct raw
\texttt{bpf\_ringbuf\_reserve}/\texttt{submit} calls, pointer arithmetic, and
unsafe conversions that would bypass the owned-entry interface.

This is defense in depth with a deliberately narrow claim.  \toolname{} does
not fix these CVEs or any other verifier defect, and vulnerable kernels
still require patches.  A malicious author can submit C, raw bytecode, or
unsafe Rust that \toolname{} rejects.  The guarantee is only that accepted
translations in \toolname{}'s policy-constrained Aya subset cannot express
this raw-pointer pattern through the owned entry API.

\section{End-to-End Symbolic Execution Example}
\label{sec:appendix-e2e-example}

\begin{figure}[t]
\centering
\begin{minipage}[t]{\linewidth}
\centering
\resizebox{\linewidth}{!}{%
\begin{tikzpicture}[
    >=Stealth,
    node distance=0.35cm and 0.6cm,
    % styles
    bytebox/.style={
      draw, rounded corners=2pt, fill=black!8,
      inner sep=4pt, minimum width=5.8cm, align=left,
    },
    stagebox/.style={
      draw, rounded corners=3pt, fill=blue!6,
      inner sep=5pt, align=left, minimum width=5.8cm,
    },
    resultbox/.style={
      draw, thick, rounded corners=3pt, fill=green!8,
      inner sep=5pt, align=left, minimum width=5.8cm,
    },
    stagelabel/.style={
      font=\sffamily\scriptsize\bfseries, text=blue!70!black,
    },
    arr/.style={->, thick, color=black!60},
  ]

  %% --- Raw bytecode ---
  \node[bytebox] (raw) {%
    {\ttfamily\scriptsize 85 00 00 00 0e 00 00 00}%
    \quad{\scriptsize call 14}\\[1pt]
    {\ttfamily\scriptsize 77 00 00 00 20 00 00 00}%
    \quad{\scriptsize r0 {>}{>}= 32}%
  };
  \node[stagelabel, above=0.05cm of raw]
    {Raw eBPF bytecode (16 bytes, little-endian)};

  %% --- Instruction decode ---
  \node[stagebox, below=0.55cm of raw] (decode) {%
    {\small\textbf{Insn 1:} opcode {\ttfamily 0x85} $=$
      $\underbrace{\texttt{1000}}_{\textsf{call}}$\,%
      $\underbrace{\texttt{0}}_{\textsf{K}}$\,%
      $\underbrace{\texttt{101}}_{\textsf{JMP64}}$,
      imm $=$ 14}\\[3pt]
    {\small\textbf{Insn 2:} opcode {\ttfamily 0x77} $=$
      $\underbrace{\texttt{0111}}_{\textsf{rsh}}$\,%
      $\underbrace{\texttt{0}}_{\textsf{K}}$\,%
      $\underbrace{\texttt{111}}_{\textsf{ALU64}}$,
      dst $=$ R0, imm $=$ 32}%
  };
  \node[stagelabel, left=0.15cm of decode.north west, anchor=south west]
    {Lifter decode};
  \draw[arr] (raw) -- (decode);

  %% --- VEX IR ---
  \node[stagebox, below=0.55cm of decode] (vex) {%
    {\small\textbf{Insn 1} $\to$ VEX IR:}\\
    \quad{\small\ttfamily PUT(syscall) = 0x0e}\\
    \quad{\small\ttfamily exit Ijk\_Sys\_syscall}\\[3pt]
    {\small\textbf{Insn 2} $\to$ VEX IR:}\\
    \quad{\small\ttfamily t0 = GET(R0)}\\
    \quad{\small\ttfamily t1 = Shr64(t0, 0x20)}\\
    \quad{\small\ttfamily PUT(R0) = t1}%
  };
  \node[stagelabel, left=0.15cm of vex.north west, anchor=south west]
    {VEX IR generation};
  \draw[arr] (decode) -- (vex);

  %% --- SimOS dispatch ---
  \node[stagebox, below=0.55cm of vex] (simos) {%
    \usebox{\helperbox}%
  };
  \node[stagelabel, left=0.15cm of simos.north west, anchor=south west]
    {Helper dispatch};
  \draw[arr] (vex) -- (simos);

  %% --- Result ---
  \node[resultbox, below=0.55cm of simos] (result) {%
    {\small\textbf{R0} $=$ {\ttfamily input\_pid\_tgid {>}{>} 32}
      \quad($=$ tgid, symbolic)}\\
    {\small\textbf{Constraints:} $\mathrm{pid} > 0$,\;
      $\mathrm{tgid} > 0$}%
  };
  \node[stagelabel, left=0.15cm of result.north west, anchor=south west]
    {Symbolic state};
  \draw[arr] (simos) -- (result);

\end{tikzpicture}%
}%
\end{minipage}
\caption{End-to-end trace of two eBPF instructions through our backend that
  highlights instruction lifting, VEX IR generation, helper dispatch, and the
  resulting symbolic state.}
\label{fig:e2e-example}
\end{figure}
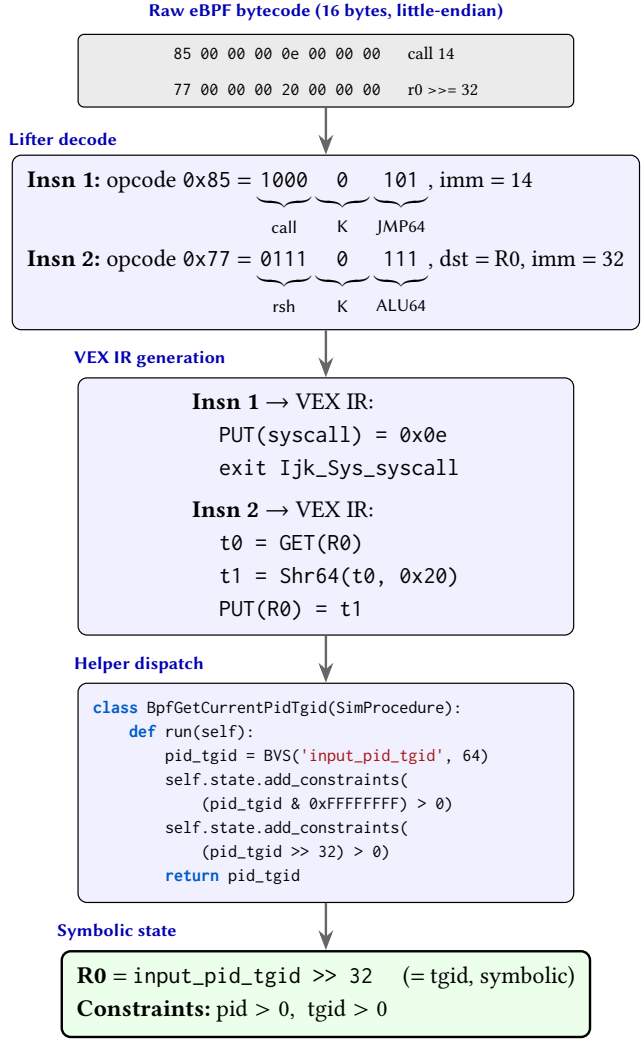

\Cref{fig:e2e-example} traces two eBPF instructions through the full symbolic
execution pipeline, from raw bytecode to symbolic formula.  The pattern
\texttt{bpf\_get\_current\_pid\_tgid()~>>~32} (extracting the thread group ID)
is ubiquitous in tracing programs, so it serves as a compact example of
lifting, helper dispatch, and symbolic state construction.

\begin{lrbox}{\helperbox}
\begin{minipage}{5.45cm}
\lstset{style=mypython}
\begin{lstlisting}
class BpfGetCurrentPidTgid(SimProcedure):
    def run(self):
        pid_tgid = BVS('input_pid_tgid', 64)
        self.state.add_constraints(
            (pid_tgid & 0xFFFFFFFF) > 0)
        self.state.add_constraints(
            (pid_tgid >> 32) > 0)
        return pid_tgid
\end{lstlisting}
\end{minipage}
\end{lrbox}

\section{Additional Implementation Details of the Equivalence Checker}
\label{sec:additional_equiv_check}

This section contains additional implementation details of the equivalence checker, including other structural invariants that are enforced to complement Z3's checks.

\paragraph{Variable unification.}
Our equivalence checker renames symbolic variables into a unified namespace with two
categories: \emph{shared} variables (kernel state such as
\texttt{input\_pid\_tgid} and \texttt{input\_ktime}, initial map contents,
and key-existence conditions) are mapped to the same Z3 variable for both
programs and \emph{distinct} variables (return values, final map state) are given per-program suffixes (e.g., \texttt{output\_r0\_c} vs.\
\texttt{output\_r0\_rust}).  This ensures both programs are evaluated under identical kernel state while allowing their outputs to possibly diverge.

\paragraph{Map type support.}
Our equivalence checker supports 20 eBPF map types across four categories.
\emph{Hash-like maps} (8 types: \texttt{hash}, \texttt{lru\_hash},
\texttt{percpu\_hash}, \texttt{lru\_percpu\_hash}, \texttt{stack\_trace},
\texttt{devmap\_hash}, \texttt{lpm\_trie}, \texttt{sockhash}) take symbolic
keys, so each lookup must reason about aliasing against every prior entry.
\emph{Array-like maps} (8 types: \texttt{array}, \texttt{percpu\_array},
\texttt{cgroup\_array}, \texttt{xskmap}, \texttt{devmap}, \texttt{cpumap},
\texttt{sockmap}, \texttt{prog\_array}) are the integer-keyed special case,
so a lookup is just a bounds check against the array length.
\emph{Output sinks} (2 types: \texttt{perf\_event\_array}, \texttt{ringbuf})
record the sequence of emitted events for equivalence comparison even
though the data leaves the kernel.
\emph{Map-of-maps} (2 types: \texttt{array\_of\_maps}, \texttt{hash\_of\_maps})
resolve outer lookups to a canonical inner map, which subsequent helper
calls then index using the inner key.

% \paragraph{Counterexample feedback.}
% When Z3 returns \texttt{SAT}, the satisfying assignment provides concrete values
% for the shared program context, initial map contents, and helper returns. The
% checker evaluates both programs under those values and classifies the observed
% difference as a sign extension, truncation, missing write, extra write, or another
% supported category. It then presents the LLM with both outputs and the program
% locations associated with the difference. The LLM first performs the bug-triage
% decision from Stage~6. A non-bug verdict feeds the same evidence into targeted
% repair, forming a loop analogous to counter-example-guided inductive synthesis
% (CEGIS)~\cite{solar2013program,alur2013syntax}. 
% The LLM proposes the next
% translation and Z3 checks it, while human review remains limited to terminal
% \textit{Bug} claims.
% ~\cite{jha2017theory,abate2018counterexample}
% The C formula is generated once and cached across all repair iterations, so only the Rust formula is regenerated each time.

\paragraph{Supplementary checks.}

Beyond the core Z3 equivalence checking, we enforce several additional structural
invariants.
During Atomic operation equivalence, symbolic execution treats atomic and
non-atomic memory operations identically on a single path, so a translation
that drops atomicity guarantees would appear equivalent under Z3.  To detect
this, we augment the checker with a static bytecode scan that counts atomic
opcodes in each binary's entry symbol section.
If the C program uses atomic
operations but the Rust translation does not (or uses fewer), the \texttt{UNSAT} result is overridden and reported as a mismatch with concrete feedback on the missing atomic operations and their Aya equivalents.
The checker also pairs mutable globals across the two binaries (matching
by name where possible, by size otherwise) and includes their final values
in the equivalence query alongside map state and return values.
Finally, to ensure entry-point type compatibility, the checker extracts the BPF program
type from each entry point's ELF section name and rejects pairs with
incompatible types before symbolic execution begins, preventing false
equivalence results from comparing programs with different context layouts.

%% ================================================================
\section{End-to-End ITE-Chains and Formula Generation Example}
\label{sec:worked-example}

We illustrate the complete symbolic execution and equivalence checking pipeline
on a minimal eBPF program, instantiating the encoding of
\Cref{sec:ite-map-encoding} (\Cref{eq:hash-ite-path}) on a concrete two-path
program.  \Cref{lst:worked-c} shows a kprobe that maintains a per-PID hit
counter in a hash map: if the PID's entry exists, it increments the value;
otherwise, it inserts~1.

\begin{lstlisting}[style=cstyle, caption={Minimal eBPF program:
  per-PID counter in a hash map.}, label=lst:worked-c]
#include "vmlinux.h"
#include <bpf/bpf_helpers.h>

struct {
    __uint(type, BPF_MAP_TYPE_HASH);
    __type(key, __u32);
    __type(value, __u32);
    __uint(max_entries, 256);
} counters SEC(".maps");

SEC("kprobe/__x64_sys_read")
int count_reads(struct pt_regs *ctx) {
    __u32 pid = bpf_get_current_pid_tgid() >> 32;
    __u32 *val = bpf_map_lookup_elem(
                     &counters, &pid);
    if (val)
        *val += 1;
    else {
        __u32 one = 1;
        bpf_map_update_elem(
            &counters, &pid, &one, 0);
    }
    return 0;
}
\end{lstlisting}

\paragraph{Symbolic execution output.}

Our angr-based engine explores two paths, corresponding to the map lookup
returning a hit or a miss.  Let $p = \texttt{input\_pid\_tgid}$ denote the
symbolic 64-bit value from \texttt{bpf\_get\_current\_pid\_tgid()},
$e_0 = \texttt{key\_exists\_counters\_v0}$ the symbolic 1-bit existence
condition for map slot~0 prior to execution, and $v_0 =
\texttt{init\_map\_counters\_v0}$ the symbolic initial value at that slot.
The lookup key is $k = \texttt{zext}_{64}(p[63{:}32])$, i.e., the upper
32~bits of $p$ zero-extended to 64~bits.  We write $v_{\mathit{init}}$ and
$p_{\mathit{init}}$ for the shared initial value/presence arrays of the
\texttt{counters} map.  Each path performs exactly one write at key $k$
(write\_seq $= 1$) with post-state existence bit $e_1 = 1$ (no path
deletes).

\medskip
\noindent\textbf{Path~0} (lookup hit $\to$ in-place increment via pointer):
\begin{align*}
  \phi_0 &= (e_0 = 1) \\
  r_0 &= 0 \\
  \text{write trace}: \quad &\langle k_1{=}k,\; v_1{=}v_0[31{:}0] + 1,\; e_1{=}1 \rangle
\end{align*}

\noindent\textbf{Path~1} (lookup miss $\to$ \texttt{update\_elem} insert):
\begin{align*}
  \phi_1 &= (e_0 = 0) \\
  r_1 &= 0 \\
  \text{write trace}: \quad &\langle k_1{=}k,\; v_1{=}1,\; e_1{=}1 \rangle
\end{align*}

Both paths return $R0 = 0$, and $\phi_0 \vee \phi_1 = \top$ partitions the
input space on the symbolic existence of the key prior to execution.

\paragraph{Per-path ITE chains.}
Instantiating \Cref{eq:hash-ite-path} on each one-write trace and
substituting $e_1 = 1$ (no deletes):
\begin{align*}
  V_{\text{ctr}}^{\pi_0}(k_q) &\triangleq \textsc{Ite}\big( k_q = k \wedge e_1 = 1,\; v_0[31{:}0] + 1,\; v_{\mathit{init}} \big) \\
  &= \textsc{Ite}\big( k_q = k,\; v_0[31{:}0] + 1,\; v_{\mathit{init}} \big) \\[2pt]
  V_{\text{ctr}}^{\pi_1}(k_q) &\triangleq \textsc{Ite}\big( k_q = k,\; 1,\; v_{\mathit{init}} \big).
\end{align*}
The parallel presence chains follow the inner shape from
\Cref{sec:ite-map-encoding} (branch on $k_q = k_i$ alone, return $e_i$):
\begin{align*}
  P_{\text{ctr}}^{\pi_0}(k_q) &\triangleq \textsc{Ite}\big( k_q = k,\; \textsc{Ite}(e_1 = 1,\, 1,\, 0),\; p_{\mathit{init}} \big) \\
  &= \textsc{Ite}\big( k_q = k,\; 1,\; p_{\mathit{init}} \big) \\[2pt]
  P_{\text{ctr}}^{\pi_1}(k_q) &\triangleq \textsc{Ite}\big( k_q = k,\; 1,\; p_{\mathit{init}} \big).
\end{align*}

\paragraph{Per-program ITE chains.}
Combining the per-path chains under their respective path predicates yields
the per-program chains:
\begin{align*}
  V_{\text{ctr}}(k_q) &\triangleq \textsc{Ite}\big(\phi_0,\; V_{\text{ctr}}^{\pi_0}(k_q),\;
                          \textsc{Ite}\big(\phi_1,\; V_{\text{ctr}}^{\pi_1}(k_q),\; V_{\mathit{init}}\big)\big) \\
  P_{\text{ctr}}(k_q) &\triangleq \textsc{Ite}\big(\phi_0,\; P_{\text{ctr}}^{\pi_0}(k_q),\;
                          \textsc{Ite}\big(\phi_1,\; P_{\text{ctr}}^{\pi_1}(k_q),\; P_{\mathit{init}}\big)\big).
\end{align*}
Per Definition \ref{def:map-output},
$\mathit{map}_{\mathcal{P}}^{\text{ctr}}(\vec{x}, k_q) = \bigl(P_{\text{ctr}}(k_q),\, V_{\text{ctr}}(k_q)\bigr)$,
with dependence on $\vec{x}$ carried implicitly through $\phi_0(\vec{x}),
\phi_1(\vec{x})$.

\paragraph{Return-value output.}
Both paths return zero, so the output predicate from Definition \ref{def:output} is
\[
  \mathit{out}_{\mathcal{P}}(\vec{x}) \triangleq
  \bigl(\phi_0 \wedge \mathcal{R}=0\bigr)
  \vee
  \bigl(\phi_1 \wedge \mathcal{R}=0\bigr).
\]
Since $\phi_0 \vee \phi_1 = \mathsf{true}$, this simplifies to
$\mathcal{R}=0$.

\paragraph{Equivalence check.}
Given the Rust translation's analogous ITE chains
$V_{\text{ctr}}^R(k_q)$ and $P_{\text{ctr}}^R(k_q)$, and its return output
$\mathcal{R}^R$, the solver checks satisfiability of the negated equivalence
(\Cref{eq:mismatch}) for the single map \texttt{counters}:
\begin{multline*}
    \exists\, p, e_0, v_0, k_q:\;
  \mathcal{R}^C \neq \mathcal{R}^R
  \;\vee\; \\
  V_{\text{ctr}}^C(k_q) \neq V_{\text{ctr}}^R(k_q)
  \;\vee\;
  P_{\text{ctr}}^C(k_q) \neq P_{\text{ctr}}^R(k_q).
\end{multline*}
% \[
%   \exists\, p, e_0, v_0, k_q:\;
%   \mathcal{R}^C \neq \mathcal{R}^R
%   \;\vee\;
%   V_{\text{ctr}}^C(k_q) \neq V_{\text{ctr}}^R(k_q)
%   \;\vee\;
%   P_{\text{ctr}}^C(k_q) \neq P_{\text{ctr}}^R(k_q).
% \]
If \texttt{UNSAT}, the programs are equivalent for all inputs and all
query keys.  If \texttt{SAT}, the model assigns concrete values to $p$,
$e_0$, $v_0$, $k_q$ that witness a divergence in either the value or
presence chain.

\section{Static Analysis Safety Engine}
\label{sec:safety_engine_full}

The safety engine is a pattern-based static analyzer that scans the
LLM-produced Rust code for unsafe constructs that silently
corrupt verifier-passing translations.  Each rule is a string match,
regex, or cross-pattern check. A hit rejects the translation from the
safety stage and emits a structured violation (rule name, message, source
span) that is fed back to the LLM for repair.  We enforce the following policies.

\begin{enumerate}
\item \textbf{No \texttt{mem::transmute} for helper invocation.} Casting
raw helper IDs through \texttt{transmute} bypasses Aya's typed wrappers
and re-introduces C-level ABI risk on every helper call.

\item \textbf{No hand-rolled \texttt{unsafe extern "C" fn} helper
trampolines.} Translations must use the helpers provided in the Aya interface. Similarly, declaring a private extern binding for
a helper is disallowed. It carries the same ABI/signature risk as transmute and
side-steps Aya's safe surface.

\item \textbf{No direct calls to \texttt{aya\_ebpf::helpers::generated::*}.}
Translations must use the typed safe wrappers
(\texttt{bpf\_probe\_read\_kernel<T>},
\texttt{bpf\_probe\_read\_user\_str\_bytes}, \texttt{RingBuf::reserve},
etc.) so that bounds, alignment, and \texttt{Result} handling stay
uniform across the translation.

\item \textbf{Helper \texttt{Result}s must be handled.} We do not allow translations to discard a failable helper's \texttt{Result} via \texttt{let \_ = ...} or
\texttt{.ok()}. The translation must branch on
\texttt{Ok}/\texttt{Err} or propagate with \texttt{?}.

\item \textbf{Ringbuf entries must be zero-initialized before submit.}
\texttt{RingBuf::reserve} returns memory containing whatever was last
written to that slot.  Submitting without an intervening
\texttt{write\_bytes(0)} or full structured initialization leaks stale
kernel-stack bytes to userspace.

% \item \textbf{No non-atomic raw-pointer field writes into HashMap entries.}
% We require value mutations on HashMap-family maps to go through
% \texttt{get(\&k) \(\to\) stack copy \(\to\) mutate \(\to\) insert(\&k,
% \&copy, 0)}. Atomic RMW via
% \texttt{AtomicU*::from\_ptr(get\_ptr\_mut(\&k))\allowbreak.fetch\_add(.., Relaxed)}
% is allowed when atomicity is required. Aya's API permits non-atomic
% \texttt{get\_ptr\_mut(\&k)} plus \texttt{unsafe \{ (*p).field = \ldots \}}
% and pushes concurrency reasoning onto the caller. We forbid that form
% because it silently re-creates C's data-race semantics on per-CPU maps
% and breaks the lookup-then-mutate invariant the verifier relies on.

\item \textbf{No non-atomic raw-pointer field writes into HashMap entries.}
On HashMap-family maps (\texttt{HashMap}, \texttt{LruHashMap},
\texttt{PerCpuHashMap}, \texttt{LruPerCpuHashMap}), a stored value must be
mutated either by (a) \texttt{get(\&k)} \(\to\) stack copy \(\to\) mutate
\(\to\) \texttt{insert(\&k, \&copy, 0)}, or (b) an atomic read-modify-write,
\path{AtomicU*::from_ptr(get_ptr_mut(&k)).fetch_add(.., Relaxed)}, when
in-place atomicity is required. We forbid the third form that
takes the raw \texttt{*mut} from \texttt{get\_ptr\_mut(\&k)} (or casting
\texttt{get(\&k)} to \texttt{*mut}) and writes a field directly with
\texttt{unsafe \{ (*p).field = \ldots \}}. This compiles and passes the
verifier, but it merely transliterates the C source's raw-pointer mutation
and bypasses Aya's typed \texttt{insert} wrapper. 
\texttt{Array} and
\texttt{PerCpuArray} are exempt since the version of Aya we use does not expose \texttt{insert}, so
\texttt{get\_ptr\_mut} plus an \texttt{unsafe} deref is the prescribed
pattern there.

\item \textbf{No \texttt{read\_volatile} loads from immutable global
\texttt{static}s.} Aya already provides
\texttt{Global<T>::load()} for loader-initialized globals. Falling back
to raw \texttt{read\_volatile} loses the load barrier and the
type-checked Aya surface.

\item \textbf{No read-only \texttt{static mut} accessed via
\texttt{read\_volatile}.} A \texttt{static mut} that is never written to
is by definition immutable. A volatile read on it almost always indicates
a missing loader-initialized binding that should have been a regular
\texttt{Global<T>} (rule~7).

\item \textbf{No untyped ringbuf reservations.} Reserving a ringbuf
entry as a byte array (\texttt{RingBuf::reserve::<[u8;\,N]>(0)} or
declaring \texttt{RingBufEntry<[u8;\,N]>}) forces the translation to
populate the slot via raw-pointer arithmetic, which replicates the
C-style partial-init bug class that rule~5 is designed to close.
Reservations must be typed: define a \texttt{\#[repr(C)]} struct with
named fields and reserve as \texttt{EVENTS.reserve::<MyEvent>(0)} so
the type system enforces field-level initialization.

\item \textbf{No inline-assembly atomic bypass.} Atomic updates on map
values or BSS globals must use
\path{core::sync::atomic::AtomicU*::from\_ptr(ptr).fetch\_add(..,
Ordering::Relaxed)} on a valid \texttt{*mut} obtained via
\texttt{HashMap::get\_ptr\_mut(\&k)} or
\texttt{core::ptr::addr\_of\_mut!(STATIC)}. We forbid the equivalent
hand-rolled escape via \texttt{core::arch::asm!}, the
\texttt{feature(asm\_experimental\_arch)} flag that enables it, and
\texttt{\#[allow(invalid\_reference\_casting)]} silencers used to feed
inline asm a \texttt{\&T} cast as \texttt{*mut T}, since these bypass
Aya's safe-API surface and the compiler's aliasing rules.

% \item \textbf{No folding of \texttt{get\_stackid}'s \texttt{Err} arm
% into the success-typed slot.} On \texttt{StackTrace::get\_stackid}
% sites, an \texttt{Err} arm that writes the error code back into the
% 4-byte \texttt{i32}/\texttt{u32} struct field used as a HashMap key
% (\texttt{Err(e) => e}, \texttt{Err(e) => e as i32}, or
% \texttt{.unwrap\_or\_else(|e| e)}) preserves the C-side
% spurious-key-on-helper-failure bug bit-for-bit at the bytecode level.
% Prescribed shapes are an early-return on the \texttt{Err} arm,
% \texttt{?}-propagation, or any pattern that does not feed the error
% code into the success-typed slot.

\item \textbf{No fixed-size buffer reads from string sources.}
\texttt{bpf\_probe\_read\_kernel\_buf} and
\texttt{bpf\_probe\_read\_user\_buf} copy the complete destination
length and do not stop at a terminating NUL. Using these helpers for a
kernel or user string, such as a \texttt{qstr} name, dentry name, file
path, or filename, can read memory beyond the source string and copy it
into an observable buffer. This pattern reproduces the over-read in
\texttt{filetop}. Translations must use
\texttt{bpf\_probe\_read\_kernel\_str\_bytes} or
\texttt{bpf\_probe\_read\_user\_str\_bytes} with a zero-initialized
destination buffer.
\end{enumerate}

A program that passes the kernel verifier but violates any of these
rules is still rejected.

\section{Differential Testing of the Trusted Computing Base}
\label{sec:differential-testing}

A \emph{Verified} verdict \toolname~produces is only as trustworthy as
the machinery that produces it. The symbolic-execution engine that lifts eBPF
bytecode into logical formulas and the equivalence checker that compares them
form the trusted computing base (TCB) of our framework: a single mis-lifted
opcode, or an unsound comparison, could silently certify a non-equivalent
translation as correct. We validate the TCB from three independent angles, each
against an oracle that shares no implementation with our own. At the instruction
level, we test the engine against \texttt{rbpf},\footnote{\url{https://github.com/qmonnet/rbpf}} a Rust-based virtual machine for eBPF programs. For each of $43$ opcodes we assemble a minimal program
that loads operands, executes the instruction under test, and returns the
result; we run it on $200$ random inputs through both the engine and the
interpreter and treat any disagreement as a candidate semantics bug, adjudicated
against the eBPF ISA specification (RFC~9669). All opcodes agree. In fact, this process resurfaced a bug in \texttt{rbpf} itself, which was fixed in a later version upstream.\footnote{\url{https://github.com/qmonnet/rbpf/commit/2f9986531c0d43abbf98fbacf20f325dfb4f72e0}} This finding attests to the quality of \toolname's instruction lifter.
% All opcodes agree. This process
% also uncovered a genuine ISA bug in the reference interpreter itself: its
% $32$-bit ALU operations failed to zero-extend their result into the $64$-bit
% destination register as the specification requires, sign-extending a negative
% result and leaking ones into the upper word. We confirmed the bug against
% RFC~9669, and it has since been fixed upstream\footnote{\url{https://github.com/qmonnet/rbpf/commit/2f9986531c0d43abbf98fbacf20f325dfb4f72e0}}.

We next validate whole programs against the ultimate oracle, the Linux kernel
itself.\footnote{We restrict this whole-program analysis to XDP programs because
\texttt{BPF\_PROG\_TEST\_RUN} does not support every eBPF program type. Program
types that cannot be driven from userspace this way are covered by the
instruction-level and checker-level tests above.} We load each of ten XDP
programs into the kernel, drive it with
\texttt{BPF\_PROG\_TEST\_RUN} on $200$ generated packets, and compare the
kernel's return value against the engine concretized on the same packet. We
repeat this both with empty maps and with pre-seeded maps, so that the
map-lookup hit paths---and the symbolic case analysis the engine builds to model
map state---are exercised against real kernel behavior rather than left dormant.
The engine and the kernel agree on every program and every packet.

Finally, we test the equivalence checker itself along two axes. For soundness, we
confirm that every program in our corpus is judged equivalent to itself
($359$ such self-checks, with no failures) and that a curated set of
deliberately divergent program pairs is always flagged. For sensitivity, we
apply mutation testing: we perturb a program by a single bytecode field, use the
interpreter to confirm the mutation actually changes behavior, and require the
checker to catch it. Of the $119$ behavior-changing mutants generated this way,
the checker caught all $119$, with no unsound misses. Taken together, these three
angles---instruction, whole-program, and checker---validate the TCB against three
independent oracles and give us confidence that an \emph{Verified} verdict
reflects real equivalence.

\section{Bug-Finding Evaluation on a Synthetic Dataset}
\label{sec:synthetic-eval}

When we ran \toolname{} on the 115-program dataset, it reported eight bugs,\footnote{Katran is not included in our dataset, and the bug was discovered in an independent, standalone analysis of Katran where we tested function-level equivalence with \toolname.}
all of which were confirmed to be genuine bugs. This suggests that \toolname{}
has a low false-positive rate. However, since we do not have the ground truth
for the real-world programs, it does not indicate the true-positive rate and
$F_1$ score. To better understand \toolname{}'s bug-finding capabilities, we
create a synthetic dataset of 34 programs into which we manually inject some
bugs. The ground truth for all these programs is known.

The dataset comprises 16 buggy programs and 18 correct (bug-free) programs.
Each buggy program contains a single injected defect drawn from the
source-level bug classes of \Cref{tab:bug-classes}. The correct
programs are faithful, bug-free eBPF tools.
We run each program through the full \toolname{} pipeline and
record a binary outcome. We label it \emph{Bug Found} when \toolname{} reports a bug that results in a divergence, and \emph{Verified} when it proves the translation
equivalent. We then compare these outcomes against the ground-truth labels.

\Cref{tab:synthetic-results} reports the results. \toolname{} flags 13 of the
16 buggy programs and verifies all 18 correct programs, giving a true-positive
rate of $0.81$, a false-positive rate of $0.00$, a precision of $1.00$, and an
$F_1$ score of $0.90$. The perfect precision corroborates the low
false-positive rate observed on the real-world dataset, while the
injected-bug ground truth additionally establishes recall. The three misses
share one shape. In each, the injected fault produces no observable divergence
for the equivalence check. Two are map key/value schema-confusion cases whose
faulty value reaches only a write-only sink or is never read back. The third is
an output-size pointer-size leak (\Cref{lst:output-size-leak-full}) that the
agent reproduced faithfully, which we examine below. Every other bug class is
detected in full.

\begin{table}[t]
\centering
\small
\setlength{\tabcolsep}{6pt}
\caption{Bug-detection results on the synthetic dataset ($16$ buggy, $18$
correct programs). \toolname{} attains TPR${}=0.81$, FPR${}=0.00$,
precision${}=1.00$, and $F_1=0.90$.}
\label{tab:synthetic-results}
\begin{tabular}{@{}lcc@{}}
\toprule
\textbf{Category} & \textbf{\#} & \textbf{\#\,\emph{Bug Found}} \\
\midrule
Uninitialized / partial-init output & 3  & 3 \\
Pointer-size mismatch            & 2  & 1 \\
Map type confusion                  & 1  & 1 \\
Map key / value schema confusion    & 4  & 2 \\
Hook / context mismatch             & 4  & 4 \\
Endianness confusion                & 2  & 2 \\
\midrule
Buggy (total)                       & 16 & 13\ \ (TP) \\
Correct (no bug)                    & 18 & 0\ \ (FP) \\
\bottomrule
\end{tabular}
\end{table}

The one missed leak illustrates how bug finding draws on both the agent and the
equivalence checker. \toolname{} reports a bug only when the agent's translation
diverges from the C and the checker then witnesses that divergence, and neither
part suffices alone. Here the C names the public prefix with the output pointer
but sizes the transfer by the enclosing object.

\begin{lstlisting}[style=cstyle,
caption={The output pointer names the prefix; the size names the enclosing
object.},
label=lst:output-leak-c]
/* BUG: prefix pointer, enclosing-object size. */
bpf_perf_event_output(ctx, &events, flags,
                      &evt.pub, sizeof(evt));
\end{lstlisting}

Aya's typed output takes no separate length, since the size follows from the
payload type, so the agent's only decision is which type to emit.

\begin{lstlisting}[style=cstyle,
caption={The payload type, not a length, decides whether the leak survives.},
label=lst:output-leak-rust]
// Emit the public prefix -> 16 B, diverges from C.
static EVENTS: PerfEventArray<PublicEvent> = ...;
EVENTS.output(&ctx, &evt.pub_event, 0);

// BUG: emit the enclosing struct -> 32 B, reproduces
//      the leak (private task_ptr/ip_ptr are exported).
static EVENTS: PerfEventArray<FullEvent> = ...;
EVENTS.output(&ctx, &evt, 0);
\end{lstlisting}

The correct choice is \texttt{PerfEventArray<PublicEvent>} with
\texttt{\&evt.pub\_event}, from which Aya derives a 16-byte length. This matches
the intent the C pointer \texttt{\&evt.pub} already names and leaves
\texttt{task\_ptr} and \texttt{ip\_ptr} in the kernel. The 16-byte record then
diverges from the C's 32 bytes, so \toolname{} reports the leak. The agent
instead emitted the enclosing \texttt{FullEvent}, which reproduces the exact
bytes and passes the equivalence check.

Note that the C pattern itself, a prefix pointer paired with an
enclosing-object length, has no direct equivalent in typed Aya Rust. Had the
agent tried to replicate it literally, \toolname{} would have rejected the
translation, since a raw output with an independent byte count violates the
safety policy. The agent is therefore left with the two valid, typed forms
above, both of which compile and pass the safety checks. Deciding that a bug is
present, and that only the prefix form is correct, rests on the agent's
reasoning about the program's intent. This miss is not a failure of
\toolname{}'s equivalence checker or safety policy. Both behaved correctly, as
the emitted Rust contains no unsafe construct and is genuinely equivalent to the
C. The bug went unreported only because the agent chose the faithful form, which
carries the leak into the Rust rather than surfacing it as a divergence.

Because the translation depends on the agent's reasoning, the set of bugs found
can vary with the underlying LLM, and this is expected. The equivalence checker
is what keeps such findings sound. It confirms that a reported bug is a genuine
C-vs-Rust divergence rather than a translation artifact, so precision holds
regardless of the model.

The other two misses are map key/value schema-confusion cases that fail for a
related reason. \Cref{lst:map-schema-confusion-full} is one such case, where the
C writes a 16-byte struct into an 8-byte map value and reads it back through an
incompatible type, yet the only observable effect is a single printed integer.
Aya's typed maps cannot express the mismatched write, so the agent keeps the
declared value type and reproduces the observed value by hand.

\begin{lstlisting}[style=cstyle,
caption={The agent keeps the typed map and hand-packs the punned value.},
label=lst:map-schema-rust]
static DATA: Array<Conn> = ...;      // declared 8-byte value
if let Some(p) = DATA.get_ptr_mut(0) {
    *p = Conn { src_ip: 1, dst_ip: 2 }; // oversized struct dropped
}
let s = unsafe { &*DATA.get_ptr(0).unwrap() };
// BUG: hand-packs the fields into the u64 the C read,
//      so the printed value matches and no divergence shows.
let bytes = s.src_ip as u64 | ((s.dst_ip as u64) << 32);
\end{lstlisting}

The output matches the C exactly, so the checker reports no divergence. The
second case is analogous. It drops the oversized struct and stores only the
correctly typed field. As with the leak, this is not a checker or safety-policy
failure. The emitted Rust is well-typed and genuinely equivalent, and the schema
confusion survives only in an erased intermediate that never reaches an
observable output.

\section{Sample Rust Translations}
\label{sec:sample-translations}

This appendix reproduces three representative Rust translations that
\toolname{} (Agentic) with Opus 4.6 produced and the pipeline verified as
equivalent to their C originals.  Each listing is reproduced verbatim, including
attributes, safety comments, and boilerplate, exactly as the pipeline emitted
it.  The three cover distinct hook types and Aya abstractions: a
\texttt{uretprobe} with a perf-event array (\Cref{lst:rust-bashreadline}), an
\texttt{xdp} program with a per-CPU hash map (\Cref{lst:rust-xdp-map}), and a
multi-hook \texttt{tracepoint} program (\Cref{lst:rust-syncsnoop}). We note that every unsafe present in the program is inevitable and is sanctioned by our safety policy. 

\Cref{lst:rust-bashreadline} is the \texttt{bashreadline} translation discussed
in \Cref{sec:verifier-gaps}.  It closes the KASLR leak by zero-initializing the
complete \texttt{StrT} event before use (\texttt{str\_buf: [0u8;
MAX\_LINE\_SIZE]}) and by propagating the \texttt{bpf\_probe\_read\_user\_str}
result through the \texttt{?} operator, so an unreadable user pointer returns
early instead of emitting stale bytes.\texttt{bpf\_probe\_read\_user\_str} is \texttt{unsafe} since Aya exposes
every probe-read of an untrusted user or kernel pointer as an \texttt{unsafe fn}
because the source pointer can only be validated at run time, so no safe wrapper
exists.

\begin{lstlisting}[style=ruststyle,
  caption={Verified Rust translation of \texttt{bashreadline}. The event is
  zero-initialized and the user-string read is checked, closing both the
  uninitialized-memory leak and the unchecked-helper bug.},
  label=lst:rust-bashreadline]
#![no_std]
#![no_main]
#![deny(clippy::multiple_unsafe_ops_per_block)]
#![deny(clippy::undocumented_unsafe_blocks)]
#![deny(unused_unsafe)]

use aya_ebpf::macros::*;
use aya_ebpf::maps::*;
use aya_ebpf::helpers::*;
use aya_ebpf::programs::*;
use aya_ebpf::cty::*;

const MAX_LINE_SIZE: usize = 80;

#[repr(C)]
struct StrT {
    pid: u32,
    str_buf: [u8; MAX_LINE_SIZE],
}

#[map(name = "events")]
static EVENTS: PerfEventArray<StrT> = PerfEventArray::new(0);

#[uretprobe]
pub fn printret(ctx: RetProbeContext) -> i32 {
    match try_printret(&ctx) {
        Ok(ret) => ret,
        Err(_) => 0,
    }
}

fn try_printret(ctx: &RetProbeContext) -> Result<i32, i32> {
    let ret: u64 = ctx.ret();
    if ret == 0 {
        return Ok(0);
    }

    let comm = bpf_get_current_comm().map_err(|_| 0i32)?;
    if comm[0] != b'b' || comm[1] != b'a' || comm[2] != b's' || comm[3] != b'h' || comm[4] != 0 {
        return Ok(0);
    }

    let pid = (bpf_get_current_pid_tgid() >> 32) as u32;
    let mut data = StrT {
        pid,
        str_buf: [0u8; MAX_LINE_SIZE],
    };

    // SAFETY: reading user string from return value pointer
    unsafe { bpf_probe_read_user_str(ret as *const u8, &mut data.str_buf) }.map_err(|_| 0i32)?;

    EVENTS.output(ctx, &data, 0);

    Ok(0)
}

#[cfg(not(test))]
#[panic_handler]
fn panic(_info: &core::panic::PanicInfo) -> ! {
    loop {}
}

#[no_mangle]
#[link_section = "license"]
pub static LICENSE: [u8; 4] = *b"GPL\0";
\end{lstlisting}

\Cref{lst:rust-xdp-map} translates an XDP program that maintains a per-CPU
packet counter.  Aya's typed \texttt{XdpContext} exposes \texttt{data} and
\texttt{data\_end} for explicit bounds checks, and the per-CPU hash map is
updated in place through a checked pointer returned by the lookup. Both \texttt{unsafe} blocks are inevitable. The packet field is read through a
raw pointer into the \texttt{data}\ldots\texttt{data\_end} window, which Aya models
as bare addresses with no typed accessor, and the counter is incremented through
the \texttt{*mut} from \texttt{get\_ptr\_mut} because the hash-map API offers no
safe value read (\texttt{get} is itself an \texttt{unsafe fn} and
\texttt{get\_ptr}/\texttt{get\_ptr\_mut} return raw pointers), so an in-place
read-modify-write needs exactly one dereference.

\begin{lstlisting}[style=ruststyle,
  caption={Verified Rust translation of an XDP program with a per-CPU hash map
  and explicit packet-bounds checks.},
  label=lst:rust-xdp-map]
#![no_std]
#![no_main]
#![deny(clippy::multiple_unsafe_ops_per_block)]
#![deny(clippy::undocumented_unsafe_blocks)]
#![deny(unused_unsafe)]

use aya_ebpf::macros::{map, xdp};
use aya_ebpf::maps::PerCpuHashMap;
use aya_ebpf::programs::XdpContext;
use aya_ebpf::bindings::xdp_action;

#[repr(C)]
#[derive(Clone, Copy)]
struct DummyKey {
    key: u8,
}

#[map(name = "rxcnt")]
static RXCNT: PerCpuHashMap<DummyKey, i64> = PerCpuHashMap::with_max_entries(256, 0);

#[xdp]
pub fn xdp_prog1(ctx: XdpContext) -> u32 {
    match try_xdp_prog1(&ctx) {
        Ok(ret) => ret,
        Err(ret) => ret,
    }
}

fn try_xdp_prog1(ctx: &XdpContext) -> Result<u32, u32> {
    let data = ctx.data();
    let data_end = ctx.data_end();
    let rc: u32 = xdp_action::XDP_DROP;

    // ethhdr is 14 bytes
    let nh_off: usize = 14;
    if data + nh_off > data_end {
        return Ok(rc);
    }

    // Read h_proto (not used for branching, but matches C behavior)
    // SAFETY: bounds checked above, reading u16 at offset 12 within ethhdr
    let _h_proto: u16 = unsafe { *((data + 12) as *const u16) };

    let key = DummyKey { key: 23 };
    let value = RXCNT.get_ptr_mut(&key);
    if let Some(v) = value {
        // SAFETY: pointer from map lookup is valid for program lifetime
        unsafe { *v += 1 };
    } else {
        let dummy_value: i64 = 1;
        let _ = RXCNT.insert(&key, &dummy_value, 0);
    }

    Ok(rc)
}

#[cfg(not(test))]
#[panic_handler]
fn panic(_info: &core::panic::PanicInfo) -> ! {
    loop {}
}

#[no_mangle]
#[link_section = "license"]
pub static LICENSE: [u8; 4] = *b"GPL\0";
\end{lstlisting}

\Cref{lst:rust-syncsnoop} translates \texttt{syncsnoop}, which attaches the same
handler to eight syscall tracepoints and emits a zero-initialized event through
a perf-event array.
Its two \texttt{unsafe} uses are inevitable since \texttt{core::mem::zeroed()}
clears the struct's alignment padding so no uninitialized bytes leak through the
perf buffer (a field-wise literal would leave that padding undefined), and
\texttt{bpf\_ktime\_get\_ns()} is available only as a generated FFI binding with
no safe Aya wrapper.

\begin{lstlisting}[style=ruststyle,
  caption={Verified Rust translation of \texttt{syncsnoop}, a multi-hook
  tracepoint program that shares one handler across eight syscalls.},
  label=lst:rust-syncsnoop]
#![no_std]
#![no_main]
#![deny(clippy::multiple_unsafe_ops_per_block)]
#![deny(clippy::undocumented_unsafe_blocks)]
#![deny(unused_unsafe)]

use aya_ebpf::macros::*;
use aya_ebpf::maps::*;
use aya_ebpf::helpers::*;
use aya_ebpf::programs::TracePointContext;
use aya_ebpf::cty::*;

const TASK_COMM_LEN: usize = 16;

#[repr(C)]
struct Event {
    comm: [u8; TASK_COMM_LEN],
    ts_us: u64,
    sys: i32,
}

#[map(name = "events")]
static EVENTS: PerfEventArray<Event> = PerfEventArray::new(0);

#[inline(always)]
fn __syscall(ctx: &TracePointContext, sys: i32) {
    // SAFETY: zeroing repr(C) struct with valid all-zero representation
    let mut event: Event = unsafe { core::mem::zeroed() };

    if let Ok(comm) = bpf_get_current_comm() {
        event.comm = comm;
    }
    // SAFETY: calling BPF helper to get current time
    event.ts_us = unsafe { bpf_ktime_get_ns() } / 1000;
    event.sys = sys;
    EVENTS.output(ctx, &event, 0);
}

#[tracepoint(category = "syscalls", name = "sys_enter_sync")]
pub fn tracepoint__syscalls__sys_enter_sync(ctx: TracePointContext) -> i32 {
    __syscall(&ctx, 1);
    0
}

#[tracepoint(category = "syscalls", name = "sys_enter_fsync")]
pub fn tracepoint__syscalls__sys_enter_fsync(ctx: TracePointContext) -> i32 {
    __syscall(&ctx, 2);
    0
}

#[tracepoint(category = "syscalls", name = "sys_enter_fdatasync")]
pub fn tracepoint__syscalls__sys_enter_fdatasync(ctx: TracePointContext) -> i32 {
    __syscall(&ctx, 3);
    0
}

#[tracepoint(category = "syscalls", name = "sys_enter_msync")]
pub fn tracepoint__syscalls__sys_enter_msync(ctx: TracePointContext) -> i32 {
    __syscall(&ctx, 4);
    0
}

#[tracepoint(category = "syscalls", name = "sys_enter_sync_file_range")]
pub fn tracepoint__syscalls__sys_enter_sync_file_range(ctx: TracePointContext) -> i32 {
    __syscall(&ctx, 5);
    0
}

#[tracepoint(category = "syscalls", name = "sys_enter_sync_file_range2")]
pub fn tracepoint__syscalls__sys_enter_sync_file_range2(ctx: TracePointContext) -> i32 {
    __syscall(&ctx, 6);
    0
}

#[tracepoint(category = "syscalls", name = "sys_enter_arm_sync_file_range")]
pub fn tracepoint__syscalls__sys_enter_arm_sync_file_range(ctx: TracePointContext) -> i32 {
    __syscall(&ctx, 7);
    0
}

#[tracepoint(category = "syscalls", name = "sys_enter_syncfs")]
pub fn tracepoint__syscalls__sys_enter_syncfs(ctx: TracePointContext) -> i32 {
    __syscall(&ctx, 8);
    0
}

#[cfg(not(test))]
#[panic_handler]
fn panic(_info: &core::panic::PanicInfo) -> ! {
    loop {}
}

#[no_mangle]
#[link_section = "license"]
pub static LICENSE: [u8; 4] = *b"GPL\0";
\end{lstlisting}

\section{LLM Prompts}
\label{sec:llm_prompts}

This appendix lists the prompts used by both \toolname{} pipelines.  The
deterministic pipeline drives the LLM stateless\-ly through one prompt per
stage of \Cref{fig:pipeline}, with placeholders in
\texttt{\{\{...\}\}} substituted at run time.  The agentic pipeline gives
the agent a translation protocol document plus a per-program dispatch
prompt from a shell script; the agent decides on its own which tools to
invoke at each stage.

The full prompts are included in the artifact. For brevity and
readability, the listings below show shortened versions that retain only
each prompt's high-level role.

\subsection{\toolname{} (Deterministic)}
\label{sec:appendix-prompts-deterministic}

The deterministic pipeline uses four stage-specific prompts plus a
condensed-rules block that is prepended to the equivalence-fix prompt as
a cached prefix.

\subsubsection{Initial translation prompt.}
Sent on the first attempt with the C source substituted for
\texttt{\{\{target\_c\_code\}\}} and (optionally) Aya documentation
substituted for \texttt{\{\{aya\_api\_docs\}\}}.

\begin{lstlisting}[style=promptstyle,
  caption={Shortened prompt for initial translation (\texttt{first\_prompt}).},
  label=lst:prompt-first]
You are translating a C/libbpf eBPF program to Rust using Aya.

Inputs:
- C source: {{target_c_code}}
- Aya API documentation, when available: {{aya_api_docs}}

Produce an idiomatic, functionally equivalent Aya program.
- Translate every SEC() entry point and preserve its name and hook type.
- Preserve map names, data layouts, control flow, return values, and
  observable effects.
- Use typed aya_ebpf contexts, maps, and helper APIs.
- Minimize and document unavoidable unsafe operations, and handle
  fallible helpers explicitly.
- Include the required eBPF boilerplate.

The result must compile and pass kernel verification, the safety policy,
and symbolic equivalence checking. Output only the translated Rust code.
\end{lstlisting}

\subsubsection{Compile-error fix prompt.}
Sent on every compile retry.  The previous Rust source is substituted
for \texttt{\{\{rust\_code\}\}} and the compiler error log for
\texttt{\{\{error\_log\}\}}.

\begin{lstlisting}[style=promptstyle,
  caption={Shortened prompt for compile-error repair (\texttt{error\_prompt}).},
  label=lst:prompt-error]
The following Rust eBPF program failed to compile.

Rust source:
{{rust_code}}

Compiler diagnostics:
{{error_log}}

Fix all compilation errors and warnings while preserving the program's
behavior, entry points, hook types, and maps. Output only the corrected
Rust code.
\end{lstlisting}

\subsubsection{Safety-policy fix prompt.}
Sent when the code compiles but the static safety analyzer reports a
blocking violation.  The current Rust source is substituted for
\texttt{\{\{rust\_code\}\}} and the analyzer report for
\texttt{\{\{safety\_report\}\}}.

\begin{lstlisting}[style=promptstyle,
  caption={Shortened prompt for safety-policy repair (\texttt{safety\_fix\_prompt}).},
  label=lst:prompt-safety]
The following Rust eBPF program compiles but violates the safety policy.

Rust source:
{{rust_code}}

Safety report:
{{safety_report}}

Fix every blocking violation without changing program semantics.
- Apply the analyzer's recommended idiomatic Aya wrappers and typed APIs.
- Do not bypass safe APIs with raw/generated helper calls or trampolines.
- Keep each unavoidable unsafe operation minimal and documented.
- Handle helper failures explicitly rather than treating errors as data.

Preserve entry points, hook types, maps, and observable behavior.
Output only the corrected Rust code.
\end{lstlisting}

\subsubsection{Equivalence-fix prompt.}
Sent when the code compiles, kernel-verifies, and passes safety, but the
symbolic equivalence checker reports a divergence.  The C source, the
current Rust source, and the formatted counter-example are substituted
for the three placeholders.  A condensed-rules block (below) is
prepended as a cached prefix so the model retains the translation rules
across the equivalence-fix loop without re-shipping the full first
prompt.

\begin{lstlisting}[style=promptstyle,
  caption={Shortened prompt prefix for equivalence repair (\texttt{CONDENSED\_TRANSLATION\_RULES}).},
  label=lst:prompt-condensed]
When repairing an Aya translation:
- Preserve every C entry point, hook type, map name, and observable effect.
- Match C data layouts, integer widths and signedness, unions, atomics,
  control flow, return codes, and helper-error behavior.
- Use typed aya_ebpf APIs and the required eBPF boilerplate.
- Keep unsafe operations minimal and documented, and introduce no new
  safety-policy violation.
\end{lstlisting}

\begin{lstlisting}[style=promptstyle,
  caption={Shortened prompt for equivalence repair (\texttt{equivalence\_fix\_prompt}).},
  label=lst:prompt-equiv]
The Rust translation compiles, kernel-verifies, and passes the safety
policy, but it is not semantically equivalent to the C program.

Original C source:
{{c_source}}

Current Rust source:
{{rust_source}}

Symbolic counter-example and diagnosis:
{{counter_example}}

Use the concrete input and divergent outputs to locate the first semantic
difference, then repair its root cause. Preserve entry points, hook types,
maps, and the safety-policy guarantees. Output only the corrected Rust
code.
\end{lstlisting}

\subsection{\toolname{} (Agentic)}
\label{sec:appendix-prompts-agentic}

The agentic pipeline gives the agent two inputs: a translation
\emph{protocol document} (\Cref{lst:prompt-agentic-protocol}) that
codifies the rules and the per-stage commands of \Cref{fig:pipeline},
and a per-program \emph{dispatch prompt}
(\Cref{lst:prompt-agentic-dispatch}) that fixes the source/binary paths,
the entry-symbol set, and the output directory for the current
translation.  The agent decides at every step which tool to invoke;
\toolname{} only enforces the success criterion (compile, kernel
verify, safety, and equivalence all pass) and the artifact layout.

\subsubsection{Translation protocol document.}
Saved alongside the agent's working directory and referenced by the
dispatch prompt as \texttt{AGENTS.md}.  The agent reads it
before producing any Rust.

\begin{lstlisting}[style=promptstyle,
  caption={Shortened prompt for the agentic translation protocol (\texttt{AGENTS.md}).},
  label=lst:prompt-agentic-protocol]
# C eBPF to Aya Rust Translation Protocol

Goal: produce an idiomatic Aya translation that preserves the observable
behavior of every entry point in the C object.

Workflow:
1. Read the C source and inventory entry points, hooks, maps, helpers,
   data layouts, and control flow.
2. Translate every entry point using typed aya_ebpf APIs.
3. Compile; use diagnostics to repair the source until it builds.
4. Run the kernel verifier; repair and repeat after any rejection.
5. Run the safety analyzer; replace violations with idiomatic APIs.
6. Check symbolic equivalence for every entry point; use any
   counter-example to repair the translation and repeat all checks.
7. Save the final source, object, result, and attempt log.

Preserve names, layouts, helper semantics, map effects, return values, and
atomic operations. Handle fallible helpers, minimize and document unsafe,
and do not bypass the safety policy. A translation succeeds only after
compilation, kernel verification, safety analysis, and all equivalence
checks pass.
\end{lstlisting}

\subsubsection{Per-program dispatch prompt.}
A small wrapper script generates this prompt for every (C source, C
binary) pair before invoking the agent.  It pins the file paths, the
map list, and the set of entry symbols extracted from the C
\texttt{.o}, and it specifies the artifact layout the runner expects on
disk.  Placeholders shown in angle brackets are substituted by the
wrapper.

\begin{lstlisting}[style=promptstyle,
  caption={Shortened prompt for per-program agentic dispatch.},
  label=lst:prompt-agentic-dispatch]
Translate this eBPF program from C to Aya Rust by following AGENTS.md.

Program:
- C source: <C_DIR>/<C_SOURCE>
- C object: <OBJ_PATH>
- Maps: <MAPS>
- Required entry points: <C_ENTRIES>
- Entry-point sections: <C_ENTRY_SECTIONS>
- Output directory: <PROG_DIR>

Translate every required entry point. Iterate through compilation,
entry-symbol auditing, kernel verification, safety analysis, and symbolic
equivalence for every entry. Repair failures and rerun all affected checks.

Log every attempt. Save the final Rust source, compiled object, and result
in <PROG_DIR>. Report EQUIVALENT only when every entry passes; otherwise
record the verified and unverified entries or the failure reason. Do not
modify files outside the translation workspace and output directory.
\end{lstlisting}

\end{document}